\documentclass{aa}
\usepackage{courier}
\usepackage[T1]{fontenc}
\usepackage{mathtools}
\usepackage{longtable}
\usepackage[para]{threeparttable}

\usepackage{xspace}
\usepackage{gensymb}
\newcommand{\swift}{\textit{Swift}}
\newcommand{\myt}{\texttt{MYTorus}\xspace}
\newcommand{\bor}{\texttt{borus02}\xspace}
\newcommand{\uxc}{\texttt{UXCLUMPY}\xspace}
\newcommand{\xmm}{\textit{XMM-Newton}\xspace}
\newcommand{\nustar}{\textit{NuSTAR}\xspace}
\newcommand{\chandra}{\textit{Chandra}\xspace}

\begin{document}

\title{Hydrogen Column Density  Variability in a Sample of Local Compton-Thin AGN}

%\subtitle{}

   \author{N. Torres-Alb\`{a}\inst{1}
          \and
          S. Marchesi\inst{1}\inst{2}
          \and
          X. Zhao\inst{3}
          \and
          I. Cox\inst{1}
          \and
          A. Pizzetti\inst{1}
          \and
          M. Ajello\inst{1}
          \and
          R. Silver\inst{1}
          }

   \institute{Department of Physics and Astronomy, Clemson University,  Kinard Lab of Physics, Clemson, SC 29634, USA\\
              \email{nuriat@clemson.edu}
         \and
            INAF - Osservatorio di Astrofisica e Scienza dello Spazio di Bologna, Via Piero Gobetti, 93/3, 40129, Bologna, Italy\\
         \and
            Center for Astrophysics | Harvard \& Smithsonian, 60 Garden Street, Cambridge, MA 02138, USA
             }

\date{Received September 15, 1996; accepted March 16, 1997}

 \abstract{
We present the analysis of multiepoch observations of a set of 12 variable, Compton-thin, local (z<0.1) active galactic nuclei (AGN) selected from the 100-month BAT catalog. We analyze all available X-ray data from \chandra, \xmm, and \nustar, adding up to a total of
53 individual observations. This corresponds to between 3 and 7 observations per source, probing variability timescales between a few days and $\sim 20$~yr. All sources have at least one \nustar observation, ensuring high-energy coverage, which allows us to disentangle the line-of-sight and reflection components in the X-ray spectra. For each source, we model all available spectra simultaneously, using the physical torus models \myt, \bor, and \uxc. The simultaneous fitting, along with the high-energy coverage, allows us to place tight constraints on torus parameters such as the torus covering factor, inclination angle, and torus average column density. We also estimate the line-of-sight column density ($N_{\rm H}$) for each individual observation. Within the 12 sources, we detect clear line-of-sight $N_{\rm H}$ variability in 5, non-variability in 5, and for 2 of them it is not possible to fully disentangle intrinsic-luminosity and $N_{\rm H}$ variability. We observe large differences between the average values of line-of-sight $N_{\rm H}$ (or $N_{\rm H}$ of the obscurer) and the average $N_{\rm H}$ of the torus (or $N_{\rm H}$ of the reflector), for each source, by a factor between $\sim2$ to $>100$. This behavior, which suggests a physical disconnect between the absorber and the reflector, is more extreme in sources that present $N_{\rm H}$ variability. $N_{\rm H}$-variable AGN also tend to present larger obscuration and broader cloud distributions than their non-variable counterparts. We observe that large changes in obscuration only occur at long timescales, and use this to place tentative lower limits on torus cloud sizes.}


   \maketitle

\section{Introduction} 

Active galactic nuclei (AGN) are powered by accreting supermassive black holes (SMBHs), surrounded by a torus of obscuring material. According to the unification theory \citep{Urry1995}, this torus is uniform and obscures certain lines of sight, preventing us from observing the broad line region (BLR, composed of gas clouds closely orbiting the black hole) from certain lines of sight. However, more recent studies based on infrared (IR) spectral energy distributions (SEDs) favor a scenario in which this torus is clumpy or patchy, rather than uniform \citep[e.g.][]{Nenkova2002,Ramos-Almeida2014}. This has been further confirmed by direct observations of changes in the line-of-sight (l.o.s.) obscuration ($N_{\rm H,los}$) in the X-ray spectra of nearby AGN \citep[e.g.][]{Risaliti2002}.

Obscuration variability in X-rays has been detected in a large range of timescales, from $\lesssim 1$ day \citep[e.g.][]{Elvis2004,Risaliti2009} to years \citep[e.g.][]{Markowitz2014}. Similarly, a large range of obscuring density variations have been observed: from small variations of $\Delta(N_{\rm H,los})\sim10^{22}$ cm$^{-2}$ \citep[e.g.][]{Laha2020} to the so-called `Changing-Look' AGN, which transition between Compton-thin ($N_{\rm H,los}<10^{24}$ cm$^{-2}$) and Compton-thick ($N_{\rm H,los}>10^{24}$ cm$^{-2}$) states \citep[e.g.][]{Risaliti2005,Bianchi2009,Rivers2015}.

Despite the multiple works that detect a $\Delta(N_{\rm H,los})$ between two different observations of the same source, very few have observations covering a complete eclipsing event \citep[e.g.][]{Maiolino2010,Markowitz2014}. This is because oserving the ingress and egress of single clouds across the line of sight may require daily observations across years. In fact, the most extensive statistical study of $N_{\rm H,los}$
variability to date is the result of frequent monitoring of 55 sources, spanning a total of 230 years of equivalent observing time with RXTE \citep{Markowitz2014}. And it resulted in the detection of variability in only 5 Seyfert 1 (Sy1) and 3 Seyfert 2 (Sy2) galaxies, with a total of 8 and 4 eclipsing events respectively. This study has been used to calibrate the most recent X-ray emission models based on clumpy tori \citep[e.g.][]{Buchner2019}.

While it is clear that further studies such as the one mentioned are not possible with the current X-ray telescopes, due to time constraints of pointed observations, studies including large samples of sources with sporadic observations can still be particularly helpful in understanding the torus structure. The $\Delta(N_{\rm H,los})$ between two different observations, separated by a given $\Delta t$, has been used to set upper limits to cloud sizes and/or their distances to the SMBH \citep[e.g.][]{Risaliti2002,Risaliti2005,Pizzetti2022,Marchesi2022}.

Recently, \citet{Laha2020} studied the variability of 20 Sy2s and found that only 7/20 sources showed changes in $N_{\rm H,los}$ over timescales from months to years.  A particularly interesting source also showed an increase of $N_{\rm H,los}$ over a period of 3.5 yr, and then remained seemingly constant for $\sim$11 yr. \citet{Laha2020} further argued that obscured AGN in which $N_{\rm H,los}$ variability is not present, or is only present on $\sim$yearly timescales, are difficult to reconcile with a simple clumpy torus scenario. The presence of a two-phase medium \citep[e.g.][]{Siebenmorgen2015}, or important contributions of larger-scale structures in the galaxy (e.g. gas lines or filaments) have been suggested as possible alternatives to obscuration in such cases. 

Even now, the number of well-studied sources in the literature still remains small. In particular, very few works exist dedicated to analyzing larger samples of AGN with multiepoch X-ray observations. Even in such cases, they tend to use phenomenological models \citep[e.g.][]{Markowitz2014,Laha2020}, which do not allow for a comparison between the $N_{\rm H,los}$ variability and general torus properties.

Recently, a variety of self-consistent physical torus models aiming to better-fit the reflection component of AGN X-ray spectra have been developed. Some are based on a uniform torus assumption, such as \myt \citep{Murphy2009} or \bor \citep{Balokovic2018}, and have been widely tested. Others, while more recent and perhaps not as robustly tested, include the option of a clumpy or patchy torus, such as \uxc \citep{Buchner2019} and \texttt{XCLUMPY} \citep{Tanimoto2019}. All these models, both uniform and patchy, take advantage of the high-energy coverage of telescopes such as the Nuclear Spectroscopic Telescope Array \citep[hereafter \nustar,][]{Harrison2013} to accurately model the reprocessed emission of the torus (i.e. the reflection component). Through this process, quantities such as the torus covering factor, the inclination angle, and the average torus column density can be estimated.

In this work, we aim to analyze a sample of 12 likely-variable AGN that have multiple X-ray observations, covering timescales of weeks to decades. These have been selected from a parent sample of BAT-detected, Compton-thin AGN at low ($z<0.1$) redshift, which have archival \nustar observations. We use three different physical torus models, with the objective of comparing our results on $N_{\rm H,los}$ variability to various torus properties.

The sample selection and data reduction processes are discussed in Sect. \ref{Sample}. In Sect. \ref{Analysis} we discuss the physical torus models used to model the spectra of the sources, and the various torus properties that can be derived from each of them. In Sect. \ref{Variability} we discuss the methods we use to classify a source as $N_{\rm H,los}$-variable, or non-variable. And finally, our results and a discussion on those are provided in Sects. \ref{Results} and \ref{Discussion}, respectively. We add a conclusion in Sect. \ref{Conclusions}. Further information, such as tables listing fit parameters, images of the spectra, and comments on individual sources can be found in Appendixes \ref{App:x-ray}, \ref{App:spectra}, and \ref{App:sources}, respectively.

\section{Sample Selection and Data Reduction}\label{Sample}

The sample in this work has been selected from \citet{Zhao2020}, a work performing a broadband X-ray spectral analysis of an unbiased sample of 93 heavily obscured AGN (with line-of-sight column density 23$\leq$log($N_{\rm H}$)$\leq$24; i.e. Compton-thin AGN) in the nearby Universe, for which high-quality archival \nustar data are available. This sample, derived from the Swift-BAT catalog \citep[Burst Alert Telescope, observing in the 15-150 KeV range,][]{Oh2018} is the largest \nustar dataset analyzed to date. \citet{Zhao2020} estimated torus geometry and $N_{\rm H,los}$ for the whole sample by jointly fitting a \nustar observation and a non-simultaneous soft X-ray observation, from either \xmm, \chandra, or \swift.

It is an ideal starting sample, first of all because a BAT detection already guarantees that the sources are X-ray bright and are typically at low redshift ($z<0.12$). Secondly, all sources analyzed already have one \nustar observation, which is essential in breaking the degeneracy between reflection and line-of-sight components, allowing us to constrain torus parameters. On top of that, it is a sample of Compton-thin AGN. These are obscured enough to let the reflection component shine through, allowing us to study the torus geometry, while being unobscured enough to allow us to constrain $N_{\rm H,los}$ with low uncertainty (compared to e.g. Compton-thick AGN).

Through a preliminary study performed in their analysis of the sample, \citet{Zhao2020} found that at least 31\footnote{We note that 22 out of the 93 sources analyzed in \citet{Zhao2020} have simultaneous \nustar and soft X-ray observations. Moreover, 13 additional sources were analyzed using \swift-XRT data, which typically has very low signal-to-noise ratio. It is therefore more accurate to say that 31 out of 58 sources presented some form of variability.} of the sources presented variability (either in $N_{\rm H,los}$ or flux). Flux variability can often be confused with $N_{\rm H,los}$ variability when the data quality is low; therefore we consider all these sources possible candidates to perform an in-depth study of $N_{\rm H,los}$ variability.

Out of the mentioned 31 sources, only 18 had additional archival data to that analyzed by \citet{Zhao2020}\footnote{As of January 2021}. Out of those, NGC 7479 was analyzed and published as a pilot project \citep{Pizzetti2022}, and Mrk 477 is currently the subject of a monitoring campaign (Torres-Alb\`a et al. in prep.). ESO 201-IG004 is part of a double system, which is not clearly resolved in the \nustar data, and was therefore removed from our sample, given the sensitivity required of the proposed analysis. 4C+73.08 was also removed as the \xmm observations \citep[additional to the one used by][]{Zhao2020} were corrupted by flares.  NGC 7582 and NGC 6300 both have a large number of observations, and have been studied in depth in previous works \citep[e.g.][respectively]{Rivers2015,Jana2020} regarding $N_{\rm H,los}$ variability. Both sources require a more careful comparison with previous results, which is beyond the scope of this work. In order to complete a self-consistent analysis of the whole sample, we will present their in-depth analysis in future works (Sengupta et al. in prep., Torres-Alb\`a et al. in prep.)

This leaves us with 12 sources, with a total of 54 observations. These are listed in Table \ref{tab:Sample}.

\onecolumn
\begin{small}
\renewcommand*{\arraystretch}{1.1}
\begin{longtable}{ccccccccc} \\
\hline \hline
 & {Source Name} & {R.A.} & {Decl.} &{z} & {Telescope} & {Obs ID} &{Exp. Time} & {Obs Date}  \\ \ & \ & {[deg (J2000)]} & {[deg (J2000)]} & \ & \ & \ & {[ks]} & \  \\  & {(1)} & {(2)} & {(3)} &{(4)} & {(5)} & {(6)} & {(7)} & {(8)} \\
\hline
& NGC 612 & 01 33 57.75 &  -36 29 35.80   & 0.0299 & \xmm & 0312190201 & 9.5 & June 26 2006  \\
&  & & & & \nustar & 60061014002 & 16.7 & September 14 2012 \\ 
&  & & & & \chandra 1 & 16099 & 10.9 & December 23 2014  \\
&  & & & & \chandra 2 & 17577  & 25.1 & February 2 2015\\
\hline
& NGC 788 & 02 01 06.46 & -06 48 57.15  & 0.0136 & \chandra & 11680 & 15.0 & September 6 2009 \\
& & & & & \xmm & 0601740201 & 15.6 &  January 15 2010\\
& & & & & \nustar & 60061018002 & 15.4 & January 28 2013\\ \hline

& NGC 835/833 & 02 09 24.61 & -10 08 09.31 & 0.0139 & \xmm & 0115810301 & 28.5 & January 1 2000  \\
& & &  &  & \chandra 1 & 923 & 12.7  & November 16 2000\\
& & &  &  & \chandra 2 & 10394 & 14.2  & November 23 2008 \\
& & &  &  & \chandra 3 & 15181 & 50.1  & July 16 2013 \\ 
& & &  &  & \chandra 4 & 15666 & 30.1  & July 18 2013 \\
& & &  &  & \chandra 5 & 15667 & 59.1  & July 21 2013 \\
& & &  &  & \nustar & 60061346002 & 20.7  & September 13 2015 \\ \hline
& 3C 105 & 04 07 16.44 & +03 42 26.33 & 0.1031 & \chandra & 9299 & 8.2 & December 17 2007\\
&  & & & & \xmm & 0500850401 & 4.2 & February 25 2008\\
&  & & & & \nustar 1 & 60261003002 & 20.7 & August 21 2016\\
&  & & & & \nustar 2 & 60261003004 & 20.7 & March 14 2017\\ \hline
& 4C+29.30 & 08 40 02.34 & +29 49 02.73  & 0.0648 & \chandra 1 & 2135 & 8.5  & April 8 2001 \\
& & &  &  & \xmm & 0504120101 & 18.0 & April 11 2008 \\
& & &  &  & \chandra 2 & 12106 & 50.5  & February 18 2010 \\
& & &  &  & \chandra 3 & 11688 & 125.1  & February 19 2010 \\
& & &  &  & \chandra 4 & 12119 & 56.2  & February 23 2010 \\
& & &  &  & \chandra 5 & 11689 & 76.6  & February 25 2010 \\
& & &  &  & \nustar & 60061083002 & 21.0 & November 8 2013 \\ \hline
& NGC 3281 & 10 31 52.09 & -34 51 13.40  &	0.0107 & \xmm & 0650591001 & 18.5 &  January 5 2011 \\
&  & & & & \nustar 1 & 60061201002 & 20.7 & January 22 2016\\ 
& & & & & \chandra & 21419 & 10.1 & January 24 2019\\
\hline
& NGC 4388 & 12 25 46.82 & +12 39 43.45  & 0.0086 & \chandra 1 & 1619 & 20.2 & June 8 2001 \\
&  &  &   &  & \xmm 1 & 0110930701 & 6.6 & December 12 2002 \\
&  &  &   &  & \chandra 2 & 12291 & 28.0 & December 7 2011 \\ 
&  &  &   &  & \xmm 2 & 0675140101 & 20.6 & June 17 2011 \\
&  &  &   &  & \nustar 1 & 60061228002 & 21.4 & December 27 2013 \\
&  &  &   &  & \xmm 3 & 0852380101 & 17.8 & December 25 2019 \\
&  &  &   &  & \nustar 2 & 60061228002 & 50.4 & December 25 2019 \\ \hline
& IC 4518 A & 14 57 40.42 & -43 07 54.00  & 0.0166 & \xmm 1 & 0401790901 & 7.4 & August 07 2006 \\
&  & & & & \xmm 2 & 0406410101 & 21.2 & August 15 2006\\
&  & & & & \nustar & 60061260002 & 7.8 & August 2 2013\\ \hline
& 3C 445 &	22 23 49.54 & -02 06 12.90  & 0.0564 & \xmm & 0090050601  & 15.4 & June 12 2001 \\
&  &  &   &  & \chandra 1 & 7869 & 46.2 & October 18 2007 \\
 &  &  &   &  & \nustar & 60160788002  & 19.9 & May 15 2016 \\ 
&  &  &   &  & \chandra 2 & 21506 & 31.0 & September 9 2019 \\
&  &  &   &  & \chandra 4 & 22842  & 55.1 & September 12 2019 \\
&  &  &   &  & \chandra 3 & 21507 & 45.1 & December 31 2019 \\ 
&  &  &   &  & \chandra 5 & 23113 & 44.2 & January 1 2020 \\
\hline
& NGC 7319 & 22 36 03.60 & +33 58 33.18  & 0.0228 & \xmm & 0021140201 & 32.7 & July 7 2001  \\
&  & & & & \chandra 1 & 789 & 20.0 & July 19 2001  \\
&  & & & & \chandra 2 & 7924 & 94.4 & August 20 2008\\
&  & & & & \nustar 1 & 60061313002 & 14.7 & November 9 2011 \\
&  & & & & \nustar 2 & 60261005002 & 41.9 & September 27 2017 \\ \hline
& 3C 452 & 22 45 48.787 & +39 41 15.36  & 0.0811 & \chandra & 2195 & 80.9 & August 21 2001 \\
& & & & & \xmm & 0552580201 & 54.2 &  November 30 2008\\
& & & & & \nustar & 60261004002 & 51.8 & May 1 2017\\ \hline
\caption{\textbf{Notes:} (1): Source name. (2) and (3): R.A. and decl. (J2000 Epoch). (4): Redshift. (5): Telescope used in the analysis. (6): Observation ID. (7): {Exposure time, in ks. \textit{XMM-Newton} values are reported for EPIC-PN, after cleaning for flares.} (8): Observation date.}
\label{tab:Sample}
\end{longtable}
\end{small}
%\end{table} 
\twocolumn

\subsection{Data reduction}

The data retrieved for both \nustar Focal Plane Modules \citep[FPMA and FPMB;][]{Harrison2013} were processed using the NuSTAR Data Analysis Software (NUSTARDAS) v1.8.0. The event data files were calibrated running the \texttt{nupipeline} task using the response file from the Calibration Database (CALDB) v. 20200612. With the nuproducts script, we generated both the source and background spectra, and the ancillary and response matrix files. For both focal planes, we used a circular source extraction region with a 50$\arcsec$ diameter centered on the target source. For the background, we used an annular extraction region (inner radius 100$\arcsec$, outer radius 160$\arcsec$) surrounding the source, excluding any resolved sources. The \nustar spectra have then been grouped with at least 20 counts per bin.

We reduced the \xmm data using the SAS v18.0.0 after cleaning for flaring periods, adopting standard procedures. The source spectra were extracted from a 30$\arcsec$ circular region, while the background spectra were obtained from a circle that has a radius 45$\arcsec$ located near the source (avoiding contamination by nearby objects). All spectra have been binned with at least 15 counts per bin.

The \chandra data was reduced using CIAO v4.12 \citep{Fruscione2006}. The source spectra were extracted from a 5$\arcsec$ circular region centered around the source, while the background spectra were obtained using an annulus (inner radius 6$\arcsec$, outer radius 15$\arcsec$) surrounding the source, excluding any resolved sources. All spectra have been binned with at least 15 counts per bin.

All spectrum extracting regions have sizes and characteristics as specified above unless otherwise stated in the source comments in Appendix \ref{App:sources}. Likewise, any exceptions on the mentioned minimum counts per bin (which ensure good usage of $\chi^2$ statistics) are mentioned in the same appendix.

We fitted our spectra using the XSPEC software \citep[][in HEASOFT version 6.26.1]{Arnaud1996}, taking into account the Galactic absorption measured by \citet{Kalberla2005}. We used \citet{Anders1989} cosmic abundances, fixed to the solar value, and the \citet{Verner1996} photoelectric absorption cross-section. The luminosity distances are computed assuming a cosmology with H$_0$=70 km s$^{-1}$ Mpc$^{-1}$, and
$\Omega_{\rm \Lambda}$=0.73. {We used $\chi ^2$ as the fitting statistic unless otherwise mentioned.}

\section{X-ray Spectral Analysis}\label{Analysis}

All sources are fit using a physically-motivated torus model, with the addition of a soft component, generally of thermal origin. Three torus models, responsible for the reflection of the AGN emission in the spectra, are used (and described below): \myt \citep{Murphy2009}, \bor \citep{Balokovic2018} and \uxc \citep{Buchner2019}. To account for the soft excess present in most galaxies, we use the thermal emission model \texttt{apec} \citep{Smith2001}. In multiple occasions, sources required the use of two \texttt{apec} components to accurately describe the soft excess. This has been shown to reproduce the complex thermal emission in star-forming galaxies \citep{Torres-Alba2018}\footnote{We note however that this approach is not necessarily superior to using a single thermal emission model with non-solar metalicity. In any case, thermal emission in the centers of galaxies is likely to come from a complex, multi-phase medium, and derived values should be used only as a first-order approximation. See \citep{Torres-Alba2018} for an in-depth discussion}. 

X-ray data for each source are fit simultaneously. That is, parameters that are not expected to change in the considered timescales (of up to $\sim$ 20 yr), are linked between different observations, and thus keep a constant value. As shown in previous works, this strategy can significantly reduce the error of the common parameters \citep[e.g.][]{Marchesi2022}. Parameters kept constant include the intrinsic photon index of the AGN (i.e. $\Gamma$) and torus geometry parameters (see individual torus model sections for details). Any caveats and/or implications of this approach are discussed in Sect. \ref{Discussion}.

The model used is
\begin{eqnarray}
 \label{eq:xspec_model}
    \textit{Model} = C  * phabs *  (\textit{Soft Model} + \textit{AGN Model}),
\end{eqnarray}
where C accounts for intrinsic flux variability and/or cross-calibration effects between different observations; and \texttt{phabs} is a photoelectric model that accounts for the Galactic absorption
in the direction of the source \citep{Kalberla2005}. We note that, for the purposes of this paper, we consider $N_{\rm H,los}$ free to vary at all epochs. However, this is not the case for C. In order to minimize the number of free parameters in the models\footnote{This number can be as high as $\sim25$, which results in computational difficulties.}, we do not consider intrinsic flux variability between two observations (A and B) when: 1) $\chi^2$ does not improve significantly when adding the additional free parameter (which we ensure via f-test); 2) $C_A$ and $C_B$ are compatible with each other within errors at $1\sigma$; and 3) forcing $C_A=C_B$ does not result in a source that was $N_{\rm H,los}$ variable to become non-$N_{\rm H,los}$ variable (and vice-versa).

The Soft Model can take the two following forms:
\begin{eqnarray}
 \label{eq:soft_model}
    \textit{Soft Model} = apec, \ \mathrm{or} \\ \textit{Soft Model} = apec_1 + zphabs * apec_2, 
\end{eqnarray}
and in which $kT_2>kT_1$. As mentioned above, this is a first approximation to a multiphase medium, in which the material closer to the nucleus of the galaxy is hotter, as well as more obscured \citep{Torres-Alba2018}. 

The AGN Model accounts for both line of sight and reflection components, as well as a scattered component. The latter characterizes the intrinsic powerlaw emission of the AGN that either leaks through the torus without interacting with it, or interacts with the material via elastic collisions. This component is set equal to the intrinsic powerlaw, multiplied by a constant, $F_{\rm s}$, that represent the fraction of scattered emission (typically of the order of few percent, or less).

All sources have been fit in the range from 0.6 keV to 25$-$55 keV, with the higher energy limit depending on the point in which \nustar data is overtaken by the background. For every source, all models have been consistently applied to the same energy range. Results of the X-ray spectral analysis of each source can be found in Sect. 4 and Appendix \ref{App:x-ray}. The obtained spectra along with the simultaneous \bor best-fit can be found, for all sources, in Appendix \ref{App:spectra}. Comments on the specific fitting details of each source can be found in Appendix \ref{App:sources}.

\subsection{\myt}

The \myt model \citep{Murphy2009} assumes a uniform, neutral (cold) torus with half-opening angle fixed to 60º, containing a uniform X-ray source. It is decomposed into three different components: an absorbed line-of sight emission, a reflected continuum, and a fluorescent line emission. These components are linked to each other via the same power-law normalization and torus parameters (i.e. torus absorbing column density, $N_{\rm H}$, and inclination angle $\theta_{\rm i}$). The inclination angle is measured from the axis of the torus, so that $\theta_{\rm i}$=0º represents a face-on AGN, and $\theta_{\rm i}$=90º an edge-on one.

Both the reflected continuum and line emission can be weighted via multiplicative constants, $A_{\rm S}$ and $A_{\rm L}$, respectively. When left free to vary, these can account for differences in the fixed torus geometry (i.e. metallicity or torus half-opening angle) and time delays between direct, scattered and fluorescent line photons.

We use \myt in `decoupled configuration' \citep{Yaqoob2012}, so as to better represent the emission from a clumpy torus. Generally, a better description of the data is possible when decoupling the line-of-sight emission from the reflection component \citep[e.g.][]{Marchesi2019,Torres-Alba2021}. That is, the $N_{\rm H}$ associated to absorption, $N_{H,los}$, and the $N_{\rm H}$ associated to reflection, $N_{H,av}$, are not fixed to the same value. This allows for the flexibility of having a particularly dense line of sight in a (still uniform) Compton-thin torus, or vice versa. 

In this configuration, the line of sight inclination angle is frozen to $\theta_{\rm i}=90\degree$. In order to better represent scattering, two reflection and line components are included. One set with $\theta_{\rm i}=90\degree$ (forward scattering), weighted with $A_{\rm S,L 90}$; and one set with $\theta_{\rm i}=0\degree$ (backward scattering), weighted with $A_{\rm S,L 0}$. In this configuration $\theta_{\rm i}$ is no longer a variable. We note however that the ratio between forward to backward scattering (i.e. $A_{\rm S,L 90}$/$A_{\rm S,L 0}$), can give a qualitative idea of the relative orientation of the AGN, as it indicates the predominant direction reflection comes from.

In the particular case of fitting multiple observations together, we consider that $N_{H,av}$ does not vary with time, and neither do the constants $A_{\rm S}$ and $A_{\rm L}$. All of these parameters are representative of properties of the overall torus, which is assumed to not vary in the considered timescales. However, $N_{\rm H,los}$ can change as the torus rotates and our line of sight pierces a different material. Therefore, each individual observation is associated to a different $N_{\rm H,los}$. 

In \texttt{XSPEC} this model configuration is as follows,

\begin{eqnarray}
 \label{eq:myt}
    \textit{AGN Model} = mytorus\_Ezero\_v00.fits * zpowerlw + \nonumber \\ A_{\rm S,0} * mytorus\_scatteredH500\_v00.fits + \nonumber \\ A_{\rm L,0} * mytl\_V000010nEp000H500\_v00.fits +
    \nonumber \\ A_{\rm S,90} * mytorus\_scatteredH500\_v00.fits + \nonumber \\ A_{\rm L,90} * mytl\_V000010nEp000H500\_v00.fits +
    \nonumber \\ + F_{\rm s} * zpowerlw. \ \ 
\end{eqnarray}

We fix $A_{S,90}$ = $A_{L,90}$ and $A_{S,0}$ = $A_{L,0}$, as is standard.

\subsection{BORUS02}

\texttt{borus02} \citep{Balokovic2018} is also a uniform torus model, but with a more flexible geometry: the opening angle is not fixed, and can be changed via the covering factor, $C_{\rm F}$, parameter ($C_{\rm F}\in[0.1,1]$). The model consists of a reflection component, which accounts for both the continuum and lines. Therefore, an absorbed line-of-sight component must be added.

We also use this model in a decoupled configuration, with $N_{\rm H,los}$ and $N_{\rm H,av}$ set to vary independently. In this case, however, $\theta_{\rm i}$ (with $\theta_{\rm i}\in  [18º-87º]$) can still be fitted in a decoupled configuration. \bor also includes a high-energy cutoff \citep[which we freeze at $\sim~300$~keV, consistent with the results of][on the local obscured AGN population]{Balokovic2020} and iron abundance (which we freeze at 1) as free parameters. We are not able to constrain these two parameters with the data available. 

When considering our variability analysis, we again allow $N_{\rm H,los}$ to vary between different observations, but force all torus parameters ($N_{\rm H,av}$, $C_{\rm F}$, $\theta_{\rm i}$) to remain constant.

In \texttt{XSPEC} this model configuration is as follows,
\begin{equation}
\begin{aligned}
 \label{eq:borus}
   \textit{AGN Model} =  borus02\_v170323a.fits + \\ zphabs * cabs * zpowerlw \\
    + F_{\rm s} * zpowerlaw,
    % \quad
\end{aligned}
\end{equation}
where \texttt{zphabs} and \texttt{cabs} are the photoelectric absorption and Compton scattering, respectively, applied to the line-of-sight component.

\subsection{UXCLUMPY}

\uxc is a clumpy torus model, which uses the \citet{Nenkova2008} formalism to describe the distribution and properties of clouds. Possible torus geometries are further narrowed down using known column density distributions \citep{Aird2015,Buchner2015,Ricci2015}, as well as by reproducing observed frequencies of eclipsing events \citep{Markowitz2014}. 

Clouds are set in a Gaussian distribution of width $\sigma$ (with $\sigma\in  [6º-90º]$) away from the equatorial plane. This distribution is viewed from a given inclination angle, $\theta_{\rm i}$ (with $\theta_{\rm i}\in  [0\degree-90\degree]$).

The model consists of one single component, which includes both reflection and line of sight in a self-consistent way, allowing for a high-energy cutoff, which we again freeze at $E_{\rm cut}=300$ keV. Although this model has the advantage of providing a clumpy distribution of material, it does not provide an estimate of the average column density of the torus, $N_{\rm H,av}$, which can be compared to the that provided by \myt and \bor. Therefore, $N_{\rm H,los}$ is the sole column density provided by the model.

In addition to the cloud distribution, \uxc offers the possibility of adding an inner `thick reflector' ring of material, which was shown to be needed to fit sources with strong reflection \citep{Buchner2019,Pizzetti2022}. This material has a covering factor, $C_{\rm F}$ (with $C_{\rm F}\in  [0-0.6]$). Sources with $C_{\rm F}=0$ do not require this additional inner reflector. 

When considering our variability analysis, we again allow $N_{\rm H,los}$ to vary between different observations, but force all torus parameters ($C_{\rm F}$, $\theta_{\rm i}$, $\sigma$) to remain constant.

In \texttt{XSPEC} this model configuration is as follows,
\begin{equation}
\begin{aligned}
 \label{eq:uxc}
   \textit{AGN Model} =  uxclumpy.fits + \\
    + F_{\rm s} * uxclumpy-scattered.fits,
    % \quad
\end{aligned}
\end{equation}
where \texttt{uxclumpy-scattered} is the scattered emission that leaks through the torus. \uxc however provides a more realistic version than a simple powerlaw, which includes the emission that leaks after being reflected.

\section{Variability Estimates}\label{Variability}

The main objective of this work is to measure the variability in obscuring column density, or $N_{\rm H,los}$, for the proposed sample of sources. As such, a method to determine whether sources are variable is needed. Here, we propose two estimators of source variability. A detailed explanation on the interpretation of these comparisons for each source can be found in Appendix \ref{App:sources}.

\subsection{Reduced $\chi^2$ Comparison}\label{redchi2comp}

The parameters of the best-fit models to the data are reported in Table \ref{tab:NGC612}, and Tables \ref{tab:NGC788} through \ref{tab:3C452}. The reduced $\chi^2$ ($\chi^2_{\rm red}$) of the best-fit is reported for all three models used.

As a further test for the need to introduce variability in the models, we present a comparison with $\chi^2_{\rm red}$ for the best fit under three different assumptions:
\begin{itemize}
\setlength\itemsep{1em}
    \item There is no variability, either in intrinsic flux or $N_{\rm H,los}$, at any epoch ($\chi^2_{\rm red}$ No Var).
    \item There is no intrinsic flux variability at any epoch, but $N_{\rm H,los}$ variability is allowed at all epochs ($\chi^2_{\rm red}$ No C Var.).
    \item There is no $N_{\rm H,los}$ variability at any epoch, but intrinsic flux variability is allowed at all epochs ($\chi^2_{\rm red}$ No $N_{\rm H}$ Var.).
\end{itemize}

A $\chi^2$ distribution approximates a Gaussian for large values of N (number degrees of freedom), with a variance $\sigma=1/\sqrt{N}$. $\chi^2_{\rm red}$ can then be used to compare different models to select the one that best fits the data. The $\chi^2_{\rm red}$ of the `true' model, the one with the `true' parameter values, is a Gaussian distributed around the mean value of 1 with standard deviation $\sigma$ \citep[see e.g.][]{Andrae2010}. A tension can then be defined between the proposed model and the data, as $T=\lvert 1-\chi^2_{\rm red} \rvert / \sigma $.

We consider that a model fits a source significantly better than another when the former has a $T<3\sigma$, and the latter yields  $T>5\sigma$ \citep[see e.g.,][]{Andrae2010}. We use this system to classify sources as $N_{\rm H,los}$-variable, by comparing the best-fit $T$ with the no-$N_{\rm H,los}$-variability $T$. When both models yield $T<3\sigma$ we interpret that $N_{\rm H,los}$-variability is not required to fit the data, and thus classify the source as non-variable. Disagreement between the different torus models used will result in classifying the source as `Undetermined'.

An exception to this rule is made for NGC 4388. No model fits the data with $T<3\sigma$ (see discussion in Appendix \ref{App:sources}), but the difference in significance between the best-fit (which includes $N_{\rm H,los}$ variability) and the non-variability scenarios is of $30-40\sigma$. Therefore, we consider that including $N_{\rm H,los}$ variability results in a significant improvement to the fit, and thus we classify this source as $N_{\rm H,los}$-variable.

We note that for two sources in our sample, NGC 612 and 4C+29.30, the fitting statistic used is a mix of C-stat and $\chi^2$ (due to one or more of the spectra having very few cts/bin. See Sect. \ref{Results}, and individual source comments in Appendix \ref{App:sources}). In such cases, we use $T=\lvert 1-{\rm Stat}_{\rm red} \rvert /\sigma$. However, given how this distribution does not necessarily approximate a Gaussian, the interpretation of $T$ in such cases is not straightforward. We opt to still provide this value as a reference.

%\subsection{AIC}

%We employ another statistical analysis based on the Akaike information criterion \citep[AIC,][]{Aikake1974}. AIC is designed for the purpose of model selection, and can be used to compare two different models fitting the same dataset, without the necessity of those being nested (as would be the case for e.g. f-test). $AIC$ is a quantity defined as $AIC=C+2k$, where C is the fitting statistic (either $\chi^2$ or a combination of $\chi^2$ and C-stat, see Sect. \ref{Results}) and $k$ is the number of free parameters in the model. When comparing two models, the one with lower $AIC$ is preferred. $\Delta (AIC)=AIC_1 - AIC_2$ is used to set a limit to the significance of such comparison, with $\lvert \Delta (AIC) \rvert>7$ generally chosen as a $3\sigma$ threshold \citep[e.g.][]{Yang2018,Ansh2022}.

%We use AIC to compare two scenarios: 

\subsection{P-value}\label{pvalue}

We take the derived best-fit values of $N_{\rm H,los}$ for all epochs (as depicted in Figures \ref{fig:nhvstime1} and \ref{fig:nhvstime2}) and estimate the probability that they all result from the same `true' value. Here the null-hypothesis is that no $N_{\rm H,los}$ variability was found among different observations of the source. That is, the probability that the source is not $N_{\rm H,los}$-variable. We do this via a $\chi^2$ computation, that we later convert into a p-value (probability of the hypothesis: the source is not $N_{\rm H,los}$ variable). The $\chi^2$ is generally computed as follows:
\begin{equation}\label{eq:chi2}
    \chi^2=\sum_{i=1}^{n} \frac{(N_{\rm H, los, i}-\langle N_{\rm H,los} \rangle)^2}{\delta(N_{\rm H,los,i})^2}
\end{equation}
However, in our particular scenario, the errors of the $N_{\rm H,los}$ determinations are asymmetric (i.e. not Gaussian). In order to calculate the equivalent to Equation \ref{eq:chi2} one needs to know (or, in its default, assume) the probability distribution of the error around the best-fit value. We follow the formalism detailed in \citet{Barlow2003} and opt to assume a simple scenario to describe this function: two straight lines which meet at the central value. In such a case, in order to evaluate the $\chi^2$ one needs only to assume as the error $\delta$ either $\sigma^+$ or $\sigma^-$, as appropriate.

From the obtained $\chi^2$ we obtain the probability (p-value) of the null-hypothesis. 
\begin{itemize}
\setlength\itemsep{1em}
    \item We classify a source as $N_{\rm H,los}$-variable if p-value $<0.01$ for all three models used (\texttt{MYTorus}, \texttt{borus02},\texttt{UXCLUMPY}).
    \item We classify a source as not $N_{\rm H,los}$-variable if p-value $>0.01$ for all three models used.
    \item We classify a source as `Undetermined' if p-value is above the given threshold for at least one model, and below it for the others.
\end{itemize}

\begin{table*}
\centering
\begin{threeparttable}
\label{tab:NGC612}
\caption{NGC 612 fitting results}
\renewcommand*{\arraystretch}{1.4}
\begin{tabular}{cccc}
{\bf Model} & {\bf MYTorus} & {\bf borus02} & {\bf UXCLUMPY} \\ \hline
Stat$_{\rm red}$ & 1.01  & 0.99 & 1.05  \\
Stat/d.o.f. & 271.96/268 & 265.91/268 & 281.77/268 \\
T & 0.2$\sigma$ & 0.2$\sigma$ & 0.8$\sigma$\\\hline
kT & 0.72$^{+0.11}_{-0.11}$ & 0.70$^{+0.12}_{-0.08}$ & 0.64$^{+0.12}_{-0.13}$ \\ \hline
$\Gamma$ & 1.54$^{+0.16}_{-u}$ & 1.43$^{+0.02}_{-u}$ & 1.52$^{+0.17}_{-0.14}$  \\
$N_{H,av}$ & 0.67$^{+1.63}_{-0.33}$ & 0.50$^{+0.13}_{-0.10}$ & $-$ \\
A$_{S90}$ & 0* & $-$ & $-$  \\
A$_{S0}$ & 0.12$^{+0.06}_{-0.04}$ & $-$ & $-$  \\
$C_{\rm F}$ & $-$ & 0.10$^{+0.03}_{-u}$ & 0*  \\
Cos ($\theta_{Obs}$) & $-$ & 0.05$^{+0.05}_{-u}$ & 0.00$^{+0.08}_{-u}$\\ 
$\sigma_{\rm tor}$ & $-$ & $-$ & 0.91$^{+10.82}_{-0.31}$  \\ 
F$_s$ (10$^{-3}$) & 0.84$^{+0.51}_{-0.38}$ & 1.13$^{+0.20}_{-0.19}$ & 0.15$^{+5.29}_{-0.03}$ \\
norm (10$^{-3}$) & 5.20$^{+0.42}_{-0.22}$ & 3.58$^{+0.10}_{-0.10}$ & 19.9$^{+6.47}_{-0.77}$ \\
\hline
$N_{H,xmm}$ & 0.90$^{+0.11}_{-0.10}$ & 0.89$^{+0.02}_{-0.02}$ & 0.92$^{+0.11}_{-0.13}$  \\
$N_{H,nus}$ & 0.84$^{+0.13}_{-0.11}$ & 0.81$^{+0.02}_{-0.02}$ & 0.79$^{+0.17}_{-0.08}$  \\
$N_{H,Ch1}$ & 1.29$^{+0.29}_{-0.22}$ & 1.27$^{+0.18}_{-0.13}$ & 0.93$^{+0.18}_{-0.19}$  \\
$N_{H,Ch2}$ & 1.39$^{+0.28}_{-0.22}$ & 1.55$^{+0.19}_{-0.14}$ & 1.10$^{+0.29}_{-0.14}$  \\
 \hline
$C_{xmm}$ & 1.14$^{+0.43}_{-0.33}$ & 1.22$^{+0.06}_{-0.06}$ & 2.62$^{+1.24}_{-0.77}$  \\
$C_{nus}$ & 0.68$^{+0.38}_{-0.26}$ & 0.70$^{+0.03}_{-0.02}$ & 1.37$^{+0.90}_{-0.49}$  \\ 
$C_{Ch1}$ & 1* & 1* & 1* \\
$C_{Ch2}$ & $=C_{Ch1}$ & 1.22$^{+0.16}_{-0.14}$ & 1.31$^{+0.77}_{-0.43}$  \\
\hline \hline
Stat$_{\rm red}$  No Var.  & 1.73 & 1.72 & 1.87 \\
$T$ & 12.1$\sigma$ & 11.9$\sigma$ & 14.4$\sigma$\\ \hline
Stat$_{\rm red}$   No C Var. & 1.03 & 1.02 & 1.19\\
$T$ & 0.5$\sigma$ & 0.3$\sigma$ & 3.1$\sigma$\\ \hline
Stat$_{\rm red}$   No $N_{\rm H}$ Var. & 1.09 & 1.63 & 1.07\\ 
$T$ & 1.5$\sigma$ & 10.4$\sigma$ & 1.2$\sigma$\\ \hline
%p-value 1 & 4.6e-2 & 3.1e-29 & 9.2e-1\\
p-value & 5.0e-1 & 1.42e-28 & 1.00\\ \hline \hline 
\end{tabular}
\begin{tablenotes}{\textbf{Notes:} \newline red $\chi^2$ (or Stat): reduced $\chi^2$ or total Statistic \newline $\chi^2$(or Stat)/d.o.f.: $\chi^2$ (or total Statistic) over degrees of freedom. \newline $kT:$ \texttt{apec} model temperature, in units of keV. \newline $\Gamma$: Powerlaw photon index. \newline $N_{H,av}$: Average torus column density, in units of $10^{24}$ cm$^{-2}$. \newline $A_{S90}$: Constant associated to the reflection component, edge-on. \newline $A_{S0}$: Constant associated to the reflection component, face-on. \newline $C_{\rm F}$: Covering factor of the torus. \newline cos ($\theta_{i}$): cosine of the inclination angle. cos ($\theta_{i}$)=1 represents a face-on scenario. \newline F$_s$: Fraction of scattered continuum \newline Norm: Normalization of the AGN emission. \newline $N_{\rm H, inst., num.}$: Line-of-sight hydrogen column density for a given observation, in units of $10^{24}$ cm$^{-2}$.\newline C$_{\rm inst., num.}$: Cross-normalization constant for a given observation, with respect to the intrinsic flux of the first Chandra observation. \newline
The last block shows the reduced $\chi^2$ (or Stat) of the best-fit when considering a) No variability between different observations; b) No intrinsic flux (i.e. C) variability; c) No $N_{\rm H,los}$ variability. \newline ($-u$) refers to a parameter being compatible with the hard limit of the available range.}
\end{tablenotes}
\end{threeparttable}
\end{table*}
%calculated p-value from chi2 as: http://courses.atlas.illinois.edu/spring2016/STAT/STAT200/pchisq.html

\section{Results}\label{Results}

In this section we present results on the analysis of all sources. Table \ref{tab:NGC612} is an example of the tabulated best-fit parameters for NGC 612. The table lists, for each of the three models used, the best-fit statistics (reduced $\chi^2$ and $\chi^2$/d.o.f., i.e. degrees of freedom; or a mix of $\chi^2$ and C-stat for sources with at least one spectra binned with $<$ 15 cts/bin\footnote{See Appendix \ref{App:sources} for details.}) in the first block. It also includes the tension, $T$, between the data and the obtained best-fit model, derived as described in Sect. \ref{redchi2comp}.

The second block shows parameters related to the soft emission. The third block shows the parameters corresponding to the AGN emission models. The fourth and fifth blocks refer to source variability, either of $N_{\rm H}$ or intrinsic flux ($C$, the cross-normalization constant), respectively. 

The final blocks show the best fit statistics that could be achieved when considering: a) No variability at all between observations; b) No intrinsic flux variability between observations; c) No obscuring column density variability between observations. For each of these scenarios, the tension between the data and the best-fit models is also computed, as described in Sect. \ref{redchi2comp}. Finally, we compute the probability of the source being not variable in $N_{\rm H,los}$ (p-value), as described in Sect. \ref{pvalue}.

\begin{figure}
\begin{center}
\includegraphics[scale=0.35,angle=-90]{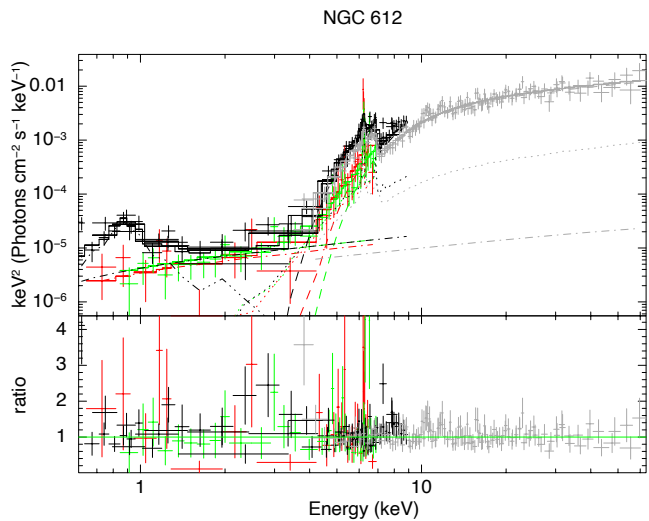}
\caption{\bor fit to the data for NGC 612. Color code is as explained in Appendix \ref{App:spectra}.}
\label{fig:NGC612}
\end{center}
\end{figure}

\begin{table*}
\centering
\begin{threeparttable}
\label{tab:var_results}
\caption{$N_{\rm H,los}$ Variability Results}
\renewcommand*{\arraystretch}{1.4}
\begin{tabular}{ccc|cc|cc|c}
{Source} & \multicolumn{2}{c}{\myt} & \multicolumn{2}{c}{\bor} & \multicolumn{2}{c}{\uxc} & Classification\\
{} & $\chi^2_{\rm red}$& P-val. & $\chi^2_{\rm red}$& P-val. & $\chi^2_{\rm red}$& P-val. \vspace{0.1cm}\\ \hline
NGC 612 & N & N & Y & Y & N & N & Undetermined\\
NGC 788 & N & N & N & N & N & Y & Not Variable\\
NGC 833 & N & N & N & N & N & N & Not Variable\\
NGC 835 & Y & Y & Y & Y & Y & Y & Variable\\
3C 105 & N & N & N & N & N & N & Not Variable\\
4C+29.30 & N & N & N & N & N & N & Not Variable\\
NGC 3281 & Y & Y & Y & Y & Y & Y & Variable\\
NGC 4388 & Y* & Y & Y* & Y & Y* & Y & Variable\\
IC 4518 A & Y & N & Y & N & Y & Y & Undetermined\\
3C 445 & N & N & N & N & N & Y & Not Variable\\
NGC 7319 & Y & Y & Y & Y & Y & Y & Variable\\
3C 452 & Y & Y & Y & Y & Y & Y & Variable\\ \hline
\end{tabular}
\begin{tablenotes}
{\textbf{Notes:} $N_{\rm H,los}$-variability determinations using the $\chi^2_{\rm red}$ and the p-value methods described in Sect. \ref{Variability}.\newline N: Not variable. Y: Variable. \newline *: See Sect. \ref{redchi2comp} and Source Notes on NGC 4388.}
\end{tablenotes}
\end{threeparttable}
\end{table*}

Tables containing the best-fit results for the rest of the sample can be found in Appendix \ref{App:x-ray}. Table \ref{tab:var_results} contains a summary of the results of applying the variability determination methods described in Sect. \ref{Variability} to all sources, for all three models used. 

We classify a source as $N_{\rm H,los}$-variable or as not $N_{\rm H,los}$-variable if at least 5 out of 6 classifications (accounting for both variability estimation methods, applied on the $N_{\rm H,los}$ determinations from all three used models) agree on the classification. If two or more determinations disagree for any source, we classify it as `Undetermined'. This is the case for only two sources within the sample: NGC 612, for which \bor results in variability according to both determinations; and IC 4518 A, for which the p-value and the $\chi^2_{\rm red}$ determinations disagree for both \myt and \bor. Further commentary on these disagreements can be found in Appendix \ref{App:sources}.

Following the method described above, out of the 12 sources analyzed in this work, 5 are not $N_{\rm H,los}$-variable, 5 are $N_{\rm H,los}$-variable, and 2 remain undetermined. It is worth noting that all sources require at least one type of variability (either $N_{\rm H,los}$ or intrinsic flux) in order to explain the data, as expected from our sample selection. This can be appreciated when comparing the best-fit $\chi^2_{\rm red}$ to the no-variability $\chi^2_{\rm red}$ in the tables presented in Appendix \ref{App:x-ray}.

Figures \ref{fig:nhvstime1} and \ref{fig:nhvstime2} show the $N_{\rm H,los}$ variability as a function of time for all the sources analyzed, considering all three physical torus models: \myt, \bor and \uxc. The dashed horizontal lines represent the best fit values for $N_{\rm H,av}$ obtained with \myt and \bor. The shaded areas correspond to the uncertainties associated to those values. All values of $N_{\rm H}$ depicted can be found in Table \ref{tab:NGC612}, and Tables \ref{tab:NGC788}$-$\ref{tab:3C452}.

\begin{figure*}[h]
\begin{center}
\hbox{
\includegraphics[scale=0.52]{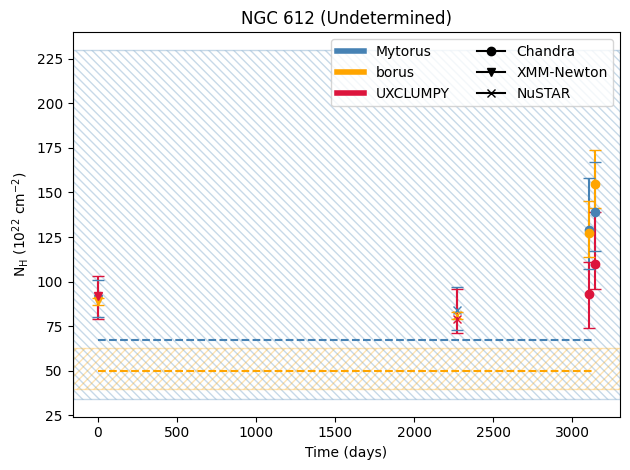}
\hspace{5mm}
\includegraphics[scale=0.52]{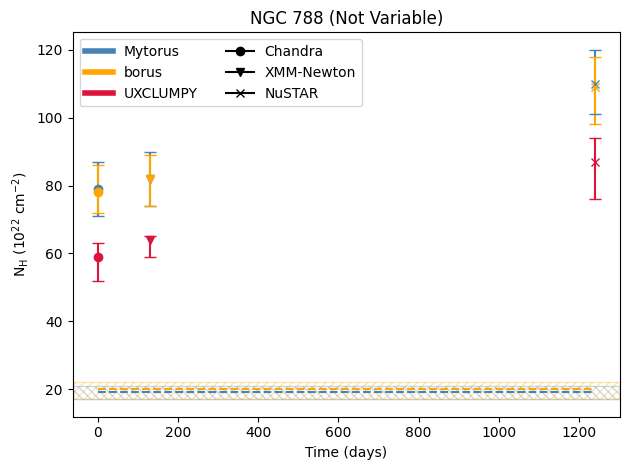}
}
\hbox{
\includegraphics[scale=0.52]{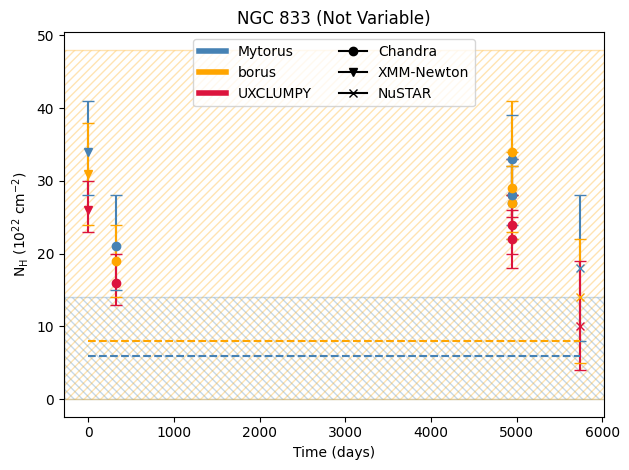}
\hspace{5mm}
\includegraphics[scale=0.52]{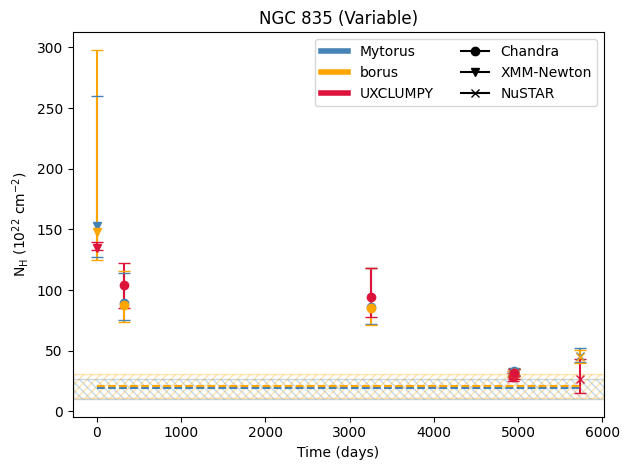}
}
\hbox{
\includegraphics[scale=0.52]{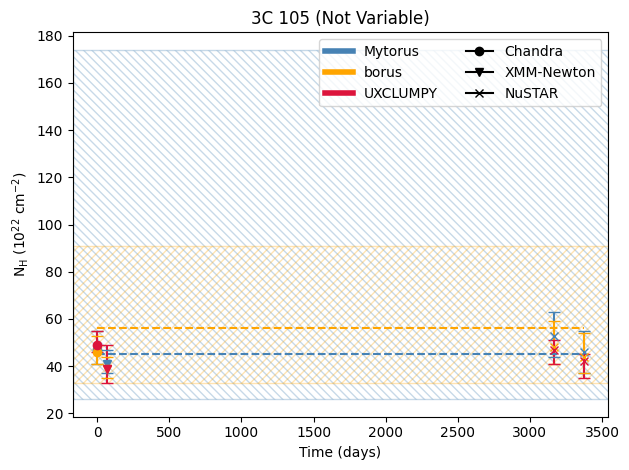}
\hspace{5mm}
\includegraphics[scale=0.52]{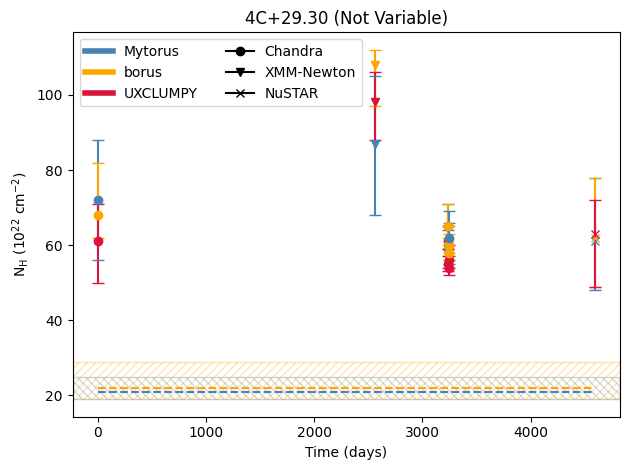}
}
\caption{$N_{\rm H,los}$ as a function of time (data points) for \myt, \bor and \uxc. Dashed horizontal lines and shaded areas correspond to the best-fit values of $N_{\rm H,av}$, and their error, respectively, for \myt and \bor. This quantity is considered constant with time.}
\label{fig:nhvstime1}
\end{center}
\end{figure*}

\begin{figure*}[h]
\begin{center}
\hbox{
\includegraphics[scale=0.52]{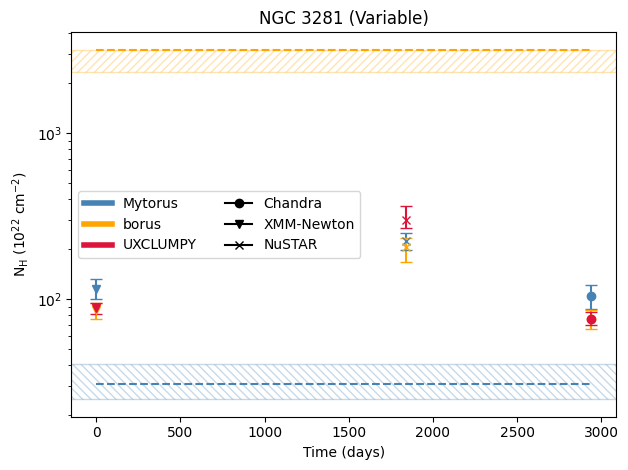}
\hspace{5mm}
\includegraphics[scale=0.52]{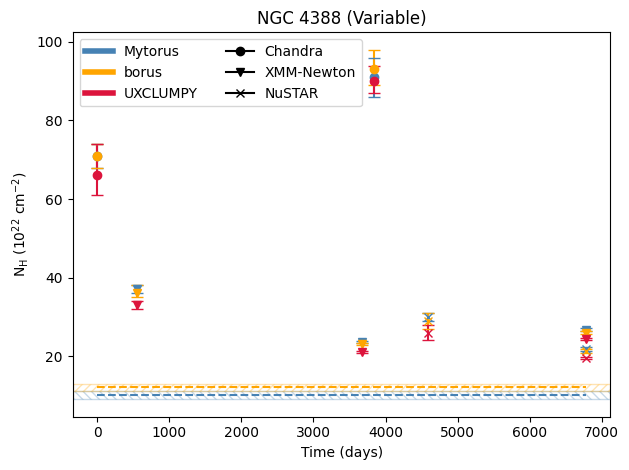}
}
\hbox{
\includegraphics[scale=0.52]{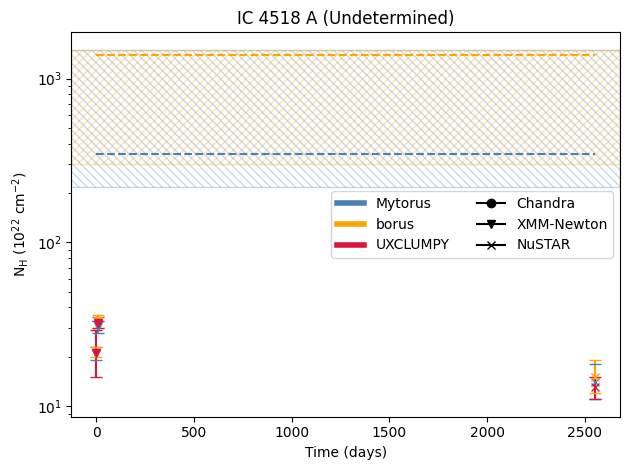}
\hspace{5mm}
\includegraphics[scale=0.52]{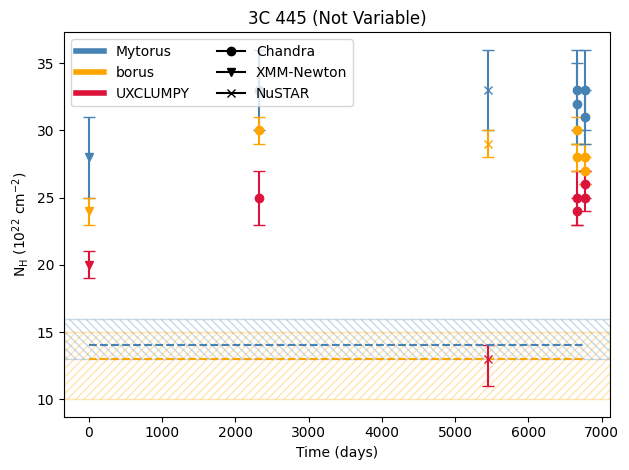}
}
\hbox{
\includegraphics[scale=0.52]{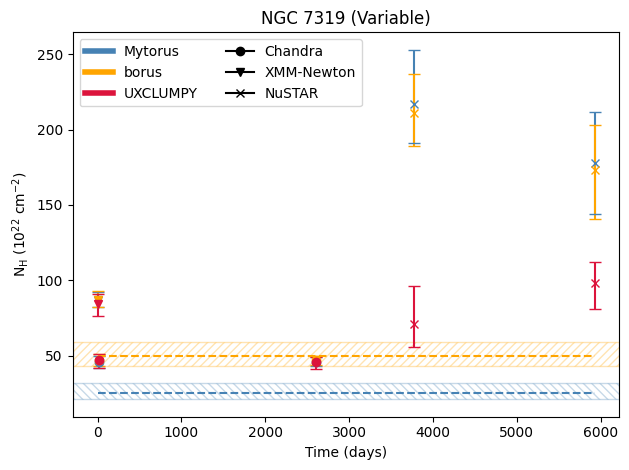}
\hspace{5mm}
\includegraphics[scale=0.52]{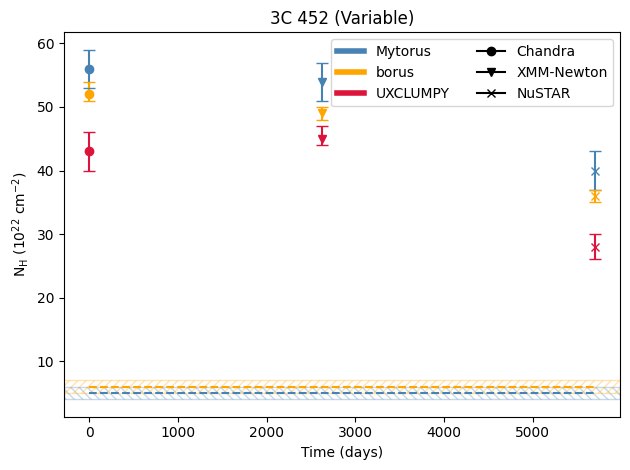}
}
\caption{Same as Fig. \ref{fig:nhvstime1}.}
\label{fig:nhvstime2}
\end{center}
\end{figure*}

\section{Discussion}\label{Discussion}

Using the comparison between $\chi^2_{red}$ in the no-variability scenario and the best-fit scenario, it is easy to see that all sources in the sample require some form of variability in order to fit the data. About $42\%$ of the sample (5/12) presents $N_{\rm H,los}$ variability for certain; a number that could be as high as $\sim58\%$ if all our `Undetermined' cases turned out to be $N_{\rm H,los}$ variable. For 5 sources in the sample we can confidently say no $N_{\rm H,los}$ variability is present between the given observations. 

When analyzing the results, however, one must take into account the following two factors: 1) The sample was intentionally biased toward variable sources, meaning that we expect to detect more $N_{\rm H,los}$ variability than in a blind survey. 2) The fact that we did not detect $N_{\rm H,los}$ variability for any given source does not mean it has never varied in $N_{\rm H,los}$.

For the two `Undetermined' sources, we are not able to claim whether flux variability or $N_{\rm H,los}$ variability is needed to fit the source, but we can claim that at least one of them is required. This showcases the difficulty in disentangling the two types of variability in X-ray datasets, even when dealing with nearby, bright AGN. In particular, this behavior is amplified when fitting \nustar data: for both 3C 445 and NGC 7319 the clumpy model \uxc favors higher flux variability and smaller $N_{\rm H,los}$ variability between other observations and the \nustar one, while the opposite is true for \bor and \myt, the homogeneous models. It is likely that simultaneous \nustar and \xmm observations would allow to properly disentangle the two scenarios.

\begin{figure*}[h]
\begin{center}
\hbox{
\includegraphics[scale=0.52]{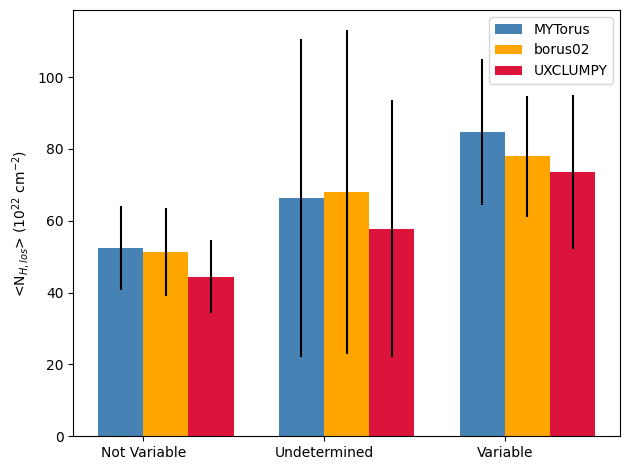}
\hspace{5mm}
\includegraphics[scale=0.52]{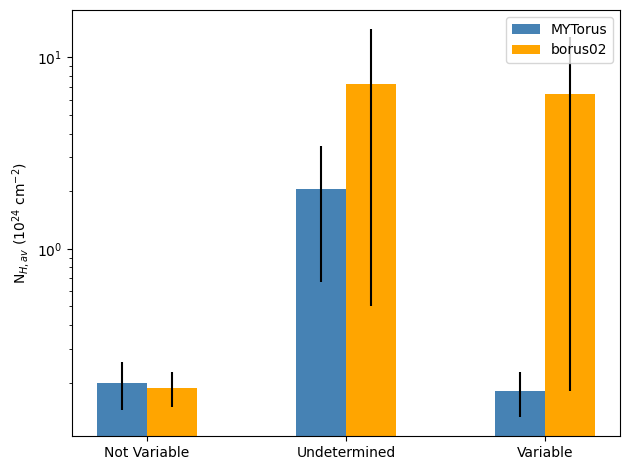}
}
\hbox{
\includegraphics[scale=0.52]{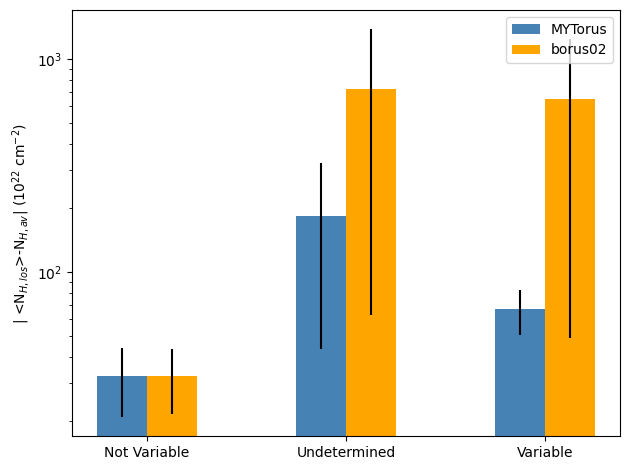}
\hspace{5mm}
\includegraphics[scale=0.52]{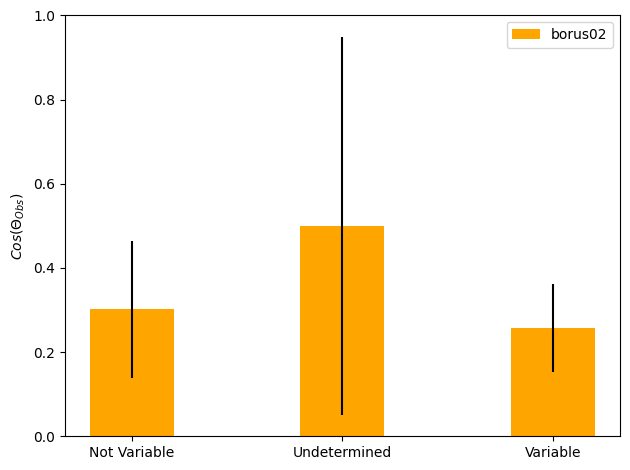}
}
\hbox{
\includegraphics[scale=0.52]{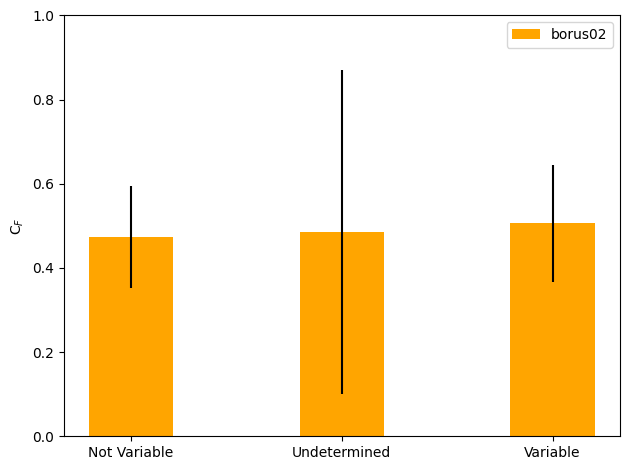}
\hspace{5mm}
\includegraphics[scale=0.52]{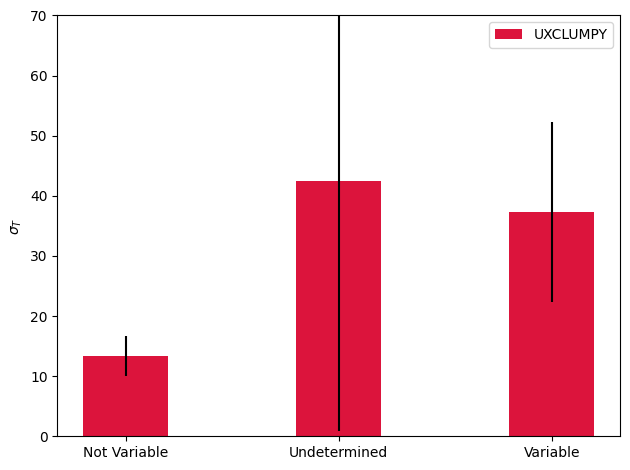}
}
\caption{Histograms containing the averaged best-fit properties of all sources in the sample, grouped by variability class. All models providing the plotted parameter are shown (\myt in blue, \bor in orange, \uxc in red). Source properties are as follows: \textit{Top left,} time average of all $N_{\rm H,los}$ (i.e. average value of the obscurer column density) for each single source. \textit{Top right,} $N_{\rm H,av}$ (i.e. column density of the reflector) considered constant with time. \textit{Middle left,} absolute value of the difference between the two properties plotted above. \textit{Middle right,} cosine of the inclination angle, $\theta_{\rm Obs}$. \textit{Bottom left,} covering factor of the torus. \textit{Bottom right,} dispersion of the torus cloud distribution.}
\label{fig:Sample_Properties}
\end{center}
\end{figure*}

\subsection{Disagreement between average torus $N_{\rm H}$ and l.o.s. $N_{\rm H}$}\label{cloudsizes}

One of the most obvious results of our analysis can be appreciated at first glance when looking at the plots in Figures \ref{fig:nhvstime1} and \ref{fig:nhvstime2}. For the majority of sources, there is a large difference between the column density in the line-of-sight (at all times) and the average column density of the torus. 

If one assumes that the whole (or the majority) of the torus is responsible for both obscuration and reflection, one would expect that the time-averaged value of $N_{\rm H, los}$ (i.e. $\langle N_{\rm H, los} \rangle$) would be similar to the value of $N_{\rm H, av}$. This is because, as the torus rotates, our line-of-sight should intercept a variety of cloud densities, representative of the density of the torus.

To estimate the feasibility that we are probing a significant fraction of the torus, we make some simple calculations. We assume Keplerian velocities, with black hole masses in the range $M_{\rm SMBH}=10^{7}-10^{8} M_\sun$ (representative of the local Universe), distances in the range $1-10$~pc (representative of the torus scales), and timescales in the $8-20$~yr range (representative of our sample). Under these assumptions, we estimate the torus to have rotated between $0.003-0.3\degree$ within the timespan of our observations\footnote{We note that this is a very simplified calculation, given how the torus is composed of individual clouds, with independent orbits, which are not necessarily circular.}. At the mentioned distances, this corresponds to a physical size of $6\times 10^{-4}-6\times10^{-3}$~pc. 

The number of works that place constraints on torus cloud/clump size (hereafter $r_{\rm c}$) is small. For reference, we list here a few determinations and/or commonly used values in the literature. \citet{Maiolino2010} place the most direct lower limit on cloud size, based on their X-ray observations of a whole eclipsing event (i.e. from ingress to egress). They estimate the size of the cloud head (i.e. denser, spherical region) to be $r_{\rm c}>10^{-7}$~pc, while the size of the following `cometary tail' of less-dense material would be $r_{\rm tail}>3\times10^{-6}~$pc. However, one must take into account these estimates correspond to a cloud placed in the broad line region (BLR), which does not necessarily have the same size as clouds orbiting the SMBH at larger distances. 

Infrared emission models of patchy/clumpy tori only require the clouds to be `small enough' in order to reproduce the observed MIR SEDs \citep[e.g.][]{Nenkova2008}. X-ray clumpy models based on the previous work assume cloud sizes of the order of $r_{\rm c}=2\times10^{-3}$~pc \citep{Tanimoto2019}, or $\theta_{\rm c}=0.1'-1\degree$. All of these are larger than the region sizes we estimate. These, however, do not necessarily correspond to observed cloud sizes, but rather to modeling or computational requirements.

The region sizes we obtain from our estimates ($6\times 10^{-4}-6\times10^{-3}$ pc) would not correspond to the size of a single cloud, given how multiple of our sources show variability at shorter timescales. However, in order to explain why we systematically see this $N_{\rm H,los}$ variability at a level incompatible to $N_{\rm H,av}$, this would need to be the size of the underdense/overdense region. 

While this is in principle not unfeasible, one needs to take into consideration the chances of systematically looking through overdense regions (as is the case of at least 6/12 of our sources), while in only 1 (or 2, depending on the model considered for NGC 3281) are observed through underdense ones. Furthermore, one should consider that the overdense regions are so by a factor 2$-$10 with respect to the torus average, while the underdense regions are so by orders of magnitude \citep[see not only IC 4518 A and NGC 3281 in this work, but also NGC 7479 in][]{Pizzetti2022}.

A study of the actual feasibility of this geometry would require: 1) A dynamical model to generate and sustain these underdense/overdense regions within a torus; and 2) An analysis of the probability of systematically observing overdense regions in a sample of 12 sources. Both of these studies are beyond the scope of this paper.

In the sections below we explore other possibilities that could explain the observed disagreement, by assuming that the material responsible for obscuration (characterized by $N_{\rm H,los}$ and, hereafter, the obscurer) and the material responsible for reflection (characterized by $N_{\rm H,av}$ and, hereafter, the reflector) are not the same.

\subsubsection{Inner Reflector Ring}

The need for an additional, thick reflector, disentangled from the rest of the torus material, has been proposed in the past. As already mentioned above, \citet{Pizzetti2022} suggested this possibility to explain the $N_{\rm H,los}$ variability curve in NGC 7479. Furthermore, the only clumpy model used in this work, \uxc, requires the addition of one such thick ring to reproduce the spectrum of sources with strong reflection \citep{Buchner2019}. In fact, both IC 4518 A and NGC 7479 require this inner ring component to model the spectrum when using \uxc, which is in agreement with the large column densities invoked by \myt and \bor. 

This theory could explain the large differences in $N_{\rm H}$ between the two structures in the torus (of factors between $10-100$) without the need to invoke a particularly underdense region of size up to $\sim 0.3\degree$ through which we observe the source. It has been suggested that such a ring could correspond to a launch site for  a Compton-thick cloud wind \citep[e.g.][]{Krolik1988}, an inner wall \citep[e.g.][]{Lightman1988}, the inner rim of a hot disk, as seen in proto-planetary disks \citep[e.g.][]{Dullemond2010}, or a warped disk \citep[e.g.][particularly suitable to explain the spectrum of Circinus]{Buchner2019,Buchner2021}.

\subsubsection{Multiple reflectors}

The majority of sources in our sample have a thin reflector, rather than a thick one. This is of particular interest, given how even if one assumes a disentangled thinner reflector near the SMBH, one needs to explain why then the thicker cloud distribution does not reflect.

\begin{figure}[h]
\begin{center}
\includegraphics[scale=0.35,angle=-90]{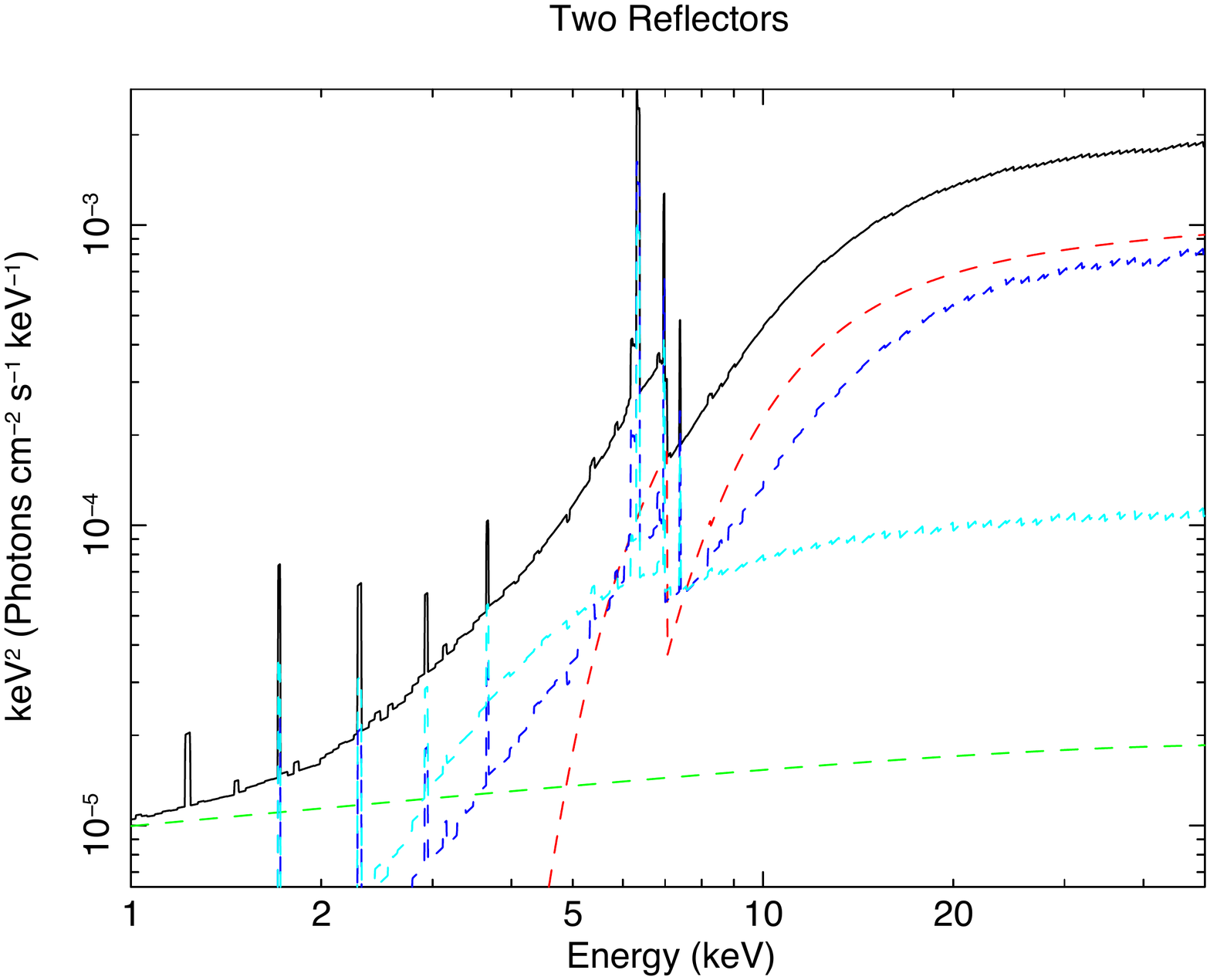}
\caption{\bor AGN X-ray spectrum resulting from an obscured l.o.s. ($N_{\rm H,los}=10^{24}$ cm$^{-2}$, in red), a scattered component ($F_{\rm S}=10^{-2}$, in green), a medium-thick reflector ($N_{\rm H,av}=10^{24}$ cm$^{-2}$, in blue), and a thin reflector ($N_{\rm H,av}=10^{23}$ cm$^{-2}$, in cyan). We use $\Gamma=1.8$, $C_{\rm F}=0.5$ and $\cos(\theta_{\rm Obs})=0.5$.}
\label{fig:tworeflectors}
\end{center}
\end{figure}

Figure \ref{fig:tworeflectors} shows the overall X-ray spectrum in the $1-50$ keV range resulting from an obscured l.o.s. (with $N_{\rm H,los}=10^{24}$ cm$^{-2}$, in red), a scattered component (with $F_{\rm S}=10^{-2}$, in green), a medium-thick reflector (with $N_{\rm H,av}=10^{24}$ cm$^{-2}$, in blue), and a thin reflector (with $N_{\rm H,av}=10^{23}$ cm$^{-2}$, in cyan). 

As can be appreciated in the model, thin reflectors have more significant contributions in the $2-5$ keV range, where the line-of-sight component (in the case of heavily obscured AGN) does not contribute. The medium-thick reflector, while also having a minor contribution in that range, has a shape more similar to that of the line-of-sight component. It is thus possible that when only one reflector is considered, the thin reflector is made necessary by the detected emission in the $2-5$ keV range. However, the medium-thick reflector, if present, could could be more difficult to recognize given the degeneracies with the combined contribution of the line-of-sight component and the thin reflector.

While this possibility is brought forward when observing the spectra in Figure \ref{fig:tworeflectors}, it must be thoroughly tested. We propose to do that in future works, using sources with good quality data, in which we may be able to disentangle the three components.

If such was the case, the idea of a two-phase medium \citep[as propsoed by e.g.,][]{Siebenmorgen2015} could explain the observations: a thinner, inter-cloud medium could act as the thin reflector, while the cloud distribution itself would be the medium-thick reflector. 

\subsection{Torus geometry as a function of Variability}

Figure \ref{fig:Sample_Properties} shows a series of histograms, which showcase how certain torus properties depend on source variability. We computed the plots by averaging a given parameter for sources in each of the three variability categories defined (i.e. Variable, Not Variable, and Undetermined). 

Each of these categories contains a low number of sources (particularly, we only classify 2 sources as 'Undetermined', which results in large error bars), and thus we are unable to make strong claims about torus geometry differences for ($N_{\rm H,los}$-) variable and non-variable sources. However, a few trends are seen in the plots in Figure \ref{fig:Sample_Properties}. 

The top, left panel of the figure shows the histogram for the average value of $N_{\rm H,los}$ across time. Meaning, the average column density of the obscurer. We observe a tendency for $N_{\rm H,los}$-variable sources to have thicker obscurers compared to their non-variable counterparts.

When it comes to the average torus column density, $N_{\rm H,av}$, this trend is not necessarily maintained. When considering the \myt results, we find overall thin reflectors for the whole sample, as already mentioned. However, the results are apparently different when considering \bor. We note that the error bar of the \bor bar for Variable sources is particularly large, and that the high average value is largely due to the \bor model yielding $N_{\rm H,av}>10^{25}$ cm$^{-2}$ for a single source (NGC 3281, but also IC 4518 A for the Undetermined sources data point).

This effect is similarly present in the middle, left plot. In here, we show the absolute value of the difference between the $N_{\rm H}$ of the obscurer and that of the reflector. The large value and large error bar of \bor are again due to the two sources mentioned above. However, \myt also suggests a larger difference between the absorber and the reflector for variable sources. Meaning, non-variable sources are much more consistent with having homogeneous tori.

We see no significant difference between inclination angles for the two different source populations. This means the observed variability (or lack thereof) is not a result of relative orientation.

We again see no difference between the two samples when it comes to $C_{\rm F}$, as determined by \bor. However, a difference is present when considering $\sigma_{\rm T}$, as determined by \uxc. This is interesting, as both parameters are representative of the height of the material responsible for reflection. It is not obvious what could be the cause of such discrepancy, but it likely lays in the different shapes assumed for the reflector: for \bor, a homogeneous sphere with two conical cut-outs; for \uxc, a cloud distribution of different densities. \uxc thus already contains the `multiple reflector' concept, and is perhaps more representative of the whole shape of the torus. If we assume, however, that \bor only models the thin reflector, the actual $C_{\rm F}$ of the medium-thick material is left unknown. In any case, \uxc results suggest that $N_{\rm H,los}$-variable sources have broader cloud distributions.

Previous work by \citet{Marchesi2022} successfully used a small \bor $C_{\rm F}$ to select a variable source, NGC 1358. They argued that, in some cases, as small $C_{\rm F}$ can represent a patchy and broad cloud distribution, rather than a homogeneous and flat one. If the theory is correct, one should expect a difference in the average values for variable and non-variable sources. However, once again, the discrepancy may be due to our inability to model all reflectors in the source. 

We observe no clear difference in average X-ray luminosity among the three different populations. 

\subsection{$\Delta(N_{\rm H,los}) \ {\rm vs}. \ \Delta(t)$}

\begin{figure}[h]
\begin{center}
\includegraphics[scale=0.54]{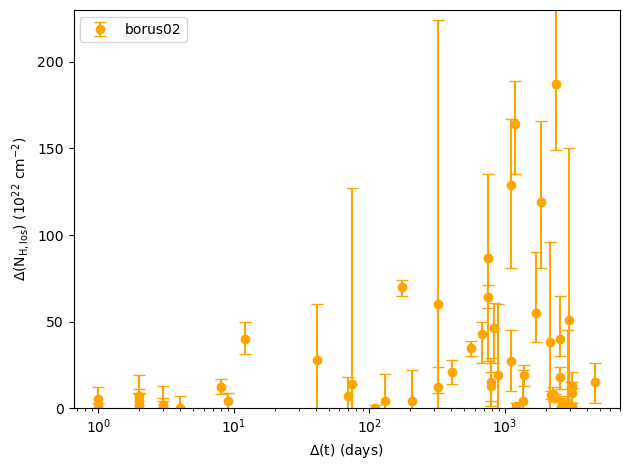}
\caption{\bor-obtained values of $\Delta(N_{\rm H,los})$ between two consecutive observations, as a function of the time difference between said observations, for the whole sample.}
\label{fig:deltanh_deltat}
\end{center}
\end{figure}

Figure \ref{fig:deltanh_deltat} shows the change in $N_{\rm H,los}$ between any two consecutive observations, as a function of the time difference between said observations. We opt to show results of only one model, \bor, in order to make the plot more easily readable. 

As can be appreciated in the figure, while small changes in $N_{\rm H,los}$ can be observed at all given time differences between observations ($\Delta(t)\sim 1-5000$ days), large changes in $N_{\rm H,los}$ ($\Delta(N_{\rm H,los})>50\times 10^{22}$ cm$^{-2}$) are only observed with large $\Delta(t)$ ($>100$~d).

This is likely a consequence of the fact that individual clouds are not homogenous in $N_{\rm H}$ \citep[as already shown for BLR clouds by e.g.][]{Maiolino2010}, but rather present a density gradient toward their centers. Performing calculations similar to those in Sect. \ref{cloudsizes}, imposing that a $\Delta(t)>100$~d is needed for a significant change in $N_{\rm H,los}$ implies clouds are generally larger than $r_{\rm c}>6\times 10^{-6}-2\times 10^{-5}$ pc, depending on underlying assumptions (such as black hole mass and cloud distance to the black hole).

Considering that events with $\Delta(N_{\rm H,los})>50\times 10^{22}$~cm$^{-2}$ are still rare for $\Delta(t)<1000$~d, one could further infer that the majority of clouds have minimum sizes $r_{\rm c}>6\times 10^{-5}-2\times 10^{-4}$~pc. The lower limits we derive are $\sim2-60$ times larger than the ones for the `cometary tails' of BLR clouds obtained by \citep{Maiolino2010}.

However, this estimate is highly dependent on the fact that the majority of timescales probed are at $\Delta(t)>1000$~d. A much larger sample than the one considered in this work is needed to fully populate the plot in Fig. \ref{fig:deltanh_deltat} and derive more reliable constraints on typical torus cloud size. 
\subsection{Constant Parameters and Treatment of Reflection}

In order to fit the data across multiple observations we have assumed that the following parameters remain unchanged across time: $\Gamma$ for all three models, $N_{\rm H,av}$ for \myt and \bor, $\theta_{\rm Obs}$ and $C_{\rm F}$ for \bor and \uxc, and $\sigma_{\rm T}$ for \uxc.

The inclination angle of the torus with respect to the observer, $\theta_{\rm Obs}$ is not a quantity that is expected to change with time. Similarly, due to the large scale of the torus ($\sim1-10$ pc), its overall geometry is not expected to vary significantly in timescales of up to $\sim 20$ yr. Therefore, all parameters associated to the reflection component ($N_{\rm H,av}$,$C_{\rm F}$,$\sigma_{\rm T}$), can be considered constant across different observations. 

A recent work on multiepoch observations of NGC 1358 performed by \cite{Marchesi2022} found that fitting the torus parameters individually at each epoch produced results that were compatible with those of the joint fit, but with much higher uncertainties. This is compatible with our assumption. We note that an equivalent test cannot easily be performed unless one possesses multiple sets of simultaneous \xmm and \nustar observations, which is unlikely to be the case for any other source. 

For a handful of sources in the literature, with extremely good data quality, further tests on the treatment of the reflection component may also be performed. One such example is NGC 4388 in this work, which is not well-fit under our assumptions. While large variations of torus geometry still seem unlikely, other assumptions are present in our treatment of reflection. One of them is the already-discussed assumption of one single reflector. As such, NGC 4388 is a good candidate for a future study including multiple reflectors. Another assumption lays in the relation between the normalization of the line-of-sight component and the reflection component. In the analysis of obscured AGN, the widely-used assumption is that the two components have the same normalization \citep[e.g.][]{Balokovic2018,Marchesi2019,Zhao2020,Torres-Alba2021,Esparza-Arredondo2021,Tanimoto2022}. However, due to the non-simultaneous origin of the intrinsic and the reflected emission, this is not necessarily the case. In sources with very large flux variability, it is possible that the normalization of the reflection component corresponds to a past flux level of the intrinsic emission. We will explore these possibilities for sources with good data quality in the future. 

We also assume that the photon index does not vary between different observations. While some works have suggested variability of $\Gamma$ with strong luminosity variability in AGN \citep[e.g.][]{Connolly2016}\footnote{We note that the mentioned work used \swift-XRT data, which makes the disentanglement of $N_{\rm H,los}$, $\Gamma$ and intrinsic luminosity variability additionally complicated.}, we note that none of the sources for which we had multiple \nustar observations suggested a need for $\Gamma$ variability. Furthermore, we do not observe extreme intrinsic luminosity variability for the sources in this sample\footnote{The largest flux variation observed is of a factor of $\sim4$, and all others are under a factor of 3.}.

\subsection{Agreement with previous results and model comparison}

Our results show satisfactory agreement  with those obtained by \citet{Zhao2020}. However, for 4/12 sources we obtain $N_{\rm H,av}$ values that are incompatible with (and in 3 sources, much lower than) those of their work. This could be a result of introducing the $0.5-2$~keV emission into the fit, which \citet{Zhao2020} did not do. If the hypothesis of the thin reflector is correct, this could result in a different sub-component disentanglement needed to explain the emission at around $\sim 2$ keV. Alternatively, it could also mean that a larger number of observations is needed to break degeneracies between parameters, and obtain reliable values of $N_{\rm H,av}$ (i.e. not pinned at the model hard limit). 

Within our sample, there is reasonable agreement within the three used models. The most notable differences are the following:
\begin{itemize}
\setlength\itemsep{1em}
    \item As already mentioned, \bor has a slight tendency to move to very large values of $N_{\rm H,av}$, sometimes even pegged at the upper limit, in sources for which \myt suggests more moderate densities. 
    \item \uxc may favor scenarios in which, instead of higher obscuration, a combination of lower obscuration and lower intrinsic flux is preferred. This is particularly true for \nustar data (see Fig. \ref{fig:nhvstime2}, sources 3C 445 and NGC 7319).
    \item The three models tend to give slightly different $N_{\rm H,los}$ results. While the agreement is still remarkable, and very often the values stay within errors, Fig. \ref{fig:Sample_Properties} (top, left) shows a systematic trend between the three models. \myt yields the highest $N_{\rm H,los}$ values, followed by \bor and further followed by \uxc, with the lowest values. Interestingly, this is in disagreement with the results obtained by \citet{Saha2022} (see their Fig. 13), who saw large agreement between \myt and \bor while \uxc had a tendency to yield larger $N_{\rm H,los}$ values. Both our results and theirs, however, agree that these differences tend to remain small.
\end{itemize}

%Add a subsection about comparing with previous work (i.e., my work). Talking about adding more soft X-ray observations could provide better constraints on tori properties (by how much). also discuss if the photon index, line-of-sight nh is better constrained or not. You can emphasize if multi-epochs X-ray data are needed to fully constrain the source properties. 

\section{Conclusions}\label{Conclusions}

In this work we have analyzed multiepoch X-ray data for a sample of 12 local Compton-thin AGN, selected from the work of \citet{Zhao2020}. We have derived the amount of obscuring column density in our line-of-sight ($N_{\rm H,los}$) for each source, for each epoch available. We have also obtained values of the average torus column density, $N_{\rm H,av}$, covering factor, $C_{\rm F}$, inclination angle, $\theta_{\rm Obs}$, and cloud dispersion, $\sigma_{\rm T}$, among others. In this section we summarize our main conclusions:

\begin{itemize}
\setlength\itemsep{1em}
    \item At least 42\% (5/12) sources in the sample present $N_{\rm H,los}$ variability (through the available observations). All sources require some form of variability, either in flux, in $N_{\rm H,los}$, or both. This is expected, given how the sample was selected to target variable sources. 
    \item The majority of sources show strong disagreement between the time-average of $N_{\rm H,los}$ (or average density of the obscurer) and $N_{\rm H,av}$ (average density of the reflector). This behavior is particularly strong in $N_{\rm H,los}$-variable sources. The difference between the two oscillates between a factor of $\sim2-100$.
    \item Based on the previous point, if the reflector and the obscurer are the same (and representative of the density of the torus), we must be observing the torus through overdense/underdense regions. We estimate those to have angular sizes between $0.003-0.3\degree$ (i.e. $6\times 10^{-4}-6\times 10^{-3}$~pc). These regions would have to contain a number of clouds of different densities to explain the observed $N_{\rm H,los}$ variability at shorter timescales. Furthermore, it is unclear how statistically feasible it is that we observe 6/12 sources through underdense regions, while observing only 1 (or 2) through an overdense one. It is equally unclear if such structures are dynamically feasible.
    \item We provide alternative explanations to the disagreement between $N_{\rm H,los}$ and $N_{\rm H,av}$. These imply the possibility that the material responsible for reflection and the material responsible for obscuration are not the same. We suggest the possible presence of an inner, thicker ring for sources with $N_{\rm H,av}$>$N_{\rm H,los}$. We suggest the possibility of a two-phase medium (or the presence of multiple reflectors) for sources with $N_{\rm H,los}$>$N_{\rm H,av}$.
    \item We observe a tendency for $N_{\rm H,los}$-variable sources to have, on average, larger obscuring density (i.e. $N_{\rm H,los}$) and broader cloud distributions than their non-variable counterparts.
    \item We observe no difference between inclination angle or torus covering factors for variable and non-variable sources.
    \item We observe small changes in $\Delta(N_{\rm H,los})$ at all timescales, but we only observe large changes ($\Delta(N_{\rm H,los})>50\times 10^{22}$~cm$^{-2}$) at large timescales (>100d). This suggests clouds are extended, with a density profile increasing toward their centers. While this is not unexpected, we use these numbers to place rough constraints on minimum cloud sizes. We obtain that, even in the most rapid variability scenarios, $r_{\rm c}>6\times 10^{-6}-2\times 10^{-5}$~pc for smaller clouds. And, for the majority of cases, $r_{\rm c}>6\times 10^{-5}-2\times 10^{-4}$~pc. However, we note that these estimates are highly dependent on availability of observations spanning smaller timescales. 
    \item We observe a tendency for \uxc to result in systematically lower $N_{\rm H,los}$ values than \myt and \bor. This is in disagreement with behavior observed in previous works.
\end{itemize}

Future work will extend this analysis to include the following: 12 more sources, for which new observations have been taken since 2019 (Pizzetti et al. in prep.); NGC 6300 (Sengupta et al. in prep.), Mrk 477 and NGC 7582 (Torres-Albà et al. in prep.) and NGC 4507 (Cox et al. in prep.). This will result in the completion of the $\sim30$ source sample of variable sources selected from \citet{Zhao2020}. We will further expand the sample by selecting potential $N_{\rm H,los}$-variable galaxies by applying the newly-developed method of Cox et. al 2023.

\section{Acknowledgments}

N.T.A., M.A., R.S., A.P. and I.C. acknowledge funding from NASA under contracts 80NSSC19K0531, 80NSSC20K0045 and, 80NSSC20K834. S.M. acknowledges funding from the INAF ``Progetti di Ricerca di Rilevante Interesse Nazionale'' (PRIN), Bando 2019 (project: ``Piercing through the clouds: a multiwavelength study of obscured accretion in nearby supermassive black holes''). The scientific results reported in this article are based on observations made by the X-ray observatories NuSTAR and XMM-Newton, and has made use of the NASA/IPAC Extragalactic Database (NED), which is operated by the Jet Propulsion Laboratory, California Institute of Technology under contract with NASA. We acknowledge the use of the software packages XMM-SAS and HEASoft.

\bibliographystyle{aa}
\bibliography{bibliography}

\begin{thebibliography}{51}
\expandafter\ifx\csname natexlab\endcsname\relax\def\natexlab#1{#1}\fi

\bibitem[{{Aird} {et~al.}(2015){Aird}, {Coil}, {Georgakakis}, {Nandra},
  {Barro}, \& {P{\'e}rez-Gonz{\'a}lez}}]{Aird2015}
{Aird}, J., {Coil}, A.~L., {Georgakakis}, A., {et~al.} 2015, \mnras, 451, 1892

\bibitem[{{Anders} \& {Grevesse}(1989)}]{Anders1989}
{Anders}, E. \& {Grevesse}, N. 1989, \gca, 53, 197

\bibitem[{{Andrae} {et~al.}(2010){Andrae}, {Schulze-Hartung}, \&
  {Melchior}}]{Andrae2010}
{Andrae}, R., {Schulze-Hartung}, T., \& {Melchior}, P. 2010, arXiv e-prints,
  arXiv:1012.3754

\bibitem[{{Arnaud}(1996)}]{Arnaud1996}
{Arnaud}, K.~A. 1996, in Astronomical Society of the Pacific Conference Series,
  Vol. 101, Astronomical Data Analysis Software and Systems V, ed. G.~H.
  {Jacoby} \& J.~{Barnes}, 17

\bibitem[{Balokovi{\'{c}} {et~al.}(2018)Balokovi{\'{c}}, Brightman, Harrison,
  Comastri, Ricci, Buchner, Gandhi, Farrah, \& Stern}]{Balokovic2018}
Balokovi{\'{c}}, M., Brightman, M., Harrison, F.~A., {et~al.} 2018, The
  Astrophysical Journal, 854, 42

\bibitem[{{Balokovi{\'c}} {et~al.}(2020){Balokovi{\'c}}, {Harrison},
  {Madejski}, {Comastri}, {Ricci}, {Annuar}, {Ballantyne}, {Boorman}, {Brandt},
  {Brightman}, {Gandhi}, {Kamraj}, {Koss}, {Marchesi}, {Marinucci}, {Masini},
  {Matt}, {Stern}, \& {Urry}}]{Balokovic2020}
{Balokovi{\'c}}, M., {Harrison}, F.~A., {Madejski}, G., {et~al.} 2020, \apj,
  905, 41

\bibitem[{{Barlow}(2003)}]{Barlow2003}
{Barlow}, R. 2003, in Statistical Problems in Particle Physics, Astrophysics,
  and Cosmology, ed. L.~{Lyons}, R.~{Mount}, \& R.~{Reitmeyer}, 250

\bibitem[{{Bianchi} {et~al.}(2009){Bianchi}, {Piconcelli}, {Chiaberge},
  {Bail{\'o}n}, {Matt}, \& {Fiore}}]{Bianchi2009}
{Bianchi}, S., {Piconcelli}, E., {Chiaberge}, M., {et~al.} 2009, \apj, 695, 781

\bibitem[{{Buchner} {et~al.}(2021){Buchner}, {Brightman}, {Balokovi{\'c}},
  {Wada}, {Bauer}, \& {Nandra}}]{Buchner2021}
{Buchner}, J., {Brightman}, M., {Balokovi{\'c}}, M., {et~al.} 2021, \aap, 651,
  A58

\bibitem[{{Buchner} {et~al.}(2019){Buchner}, {Brightman}, {Nandra}, {Nikutta},
  \& {Bauer}}]{Buchner2019}
{Buchner}, J., {Brightman}, M., {Nandra}, K., {Nikutta}, R., \& {Bauer}, F.~E.
  2019, \aap, 629, A16

\bibitem[{{Buchner} {et~al.}(2015){Buchner}, {Georgakakis}, {Nandra},
  {Brightman}, {Menzel}, {Liu}, {Hsu}, {Salvato}, {Rangel}, {Aird}, {Merloni},
  \& {Ross}}]{Buchner2015}
{Buchner}, J., {Georgakakis}, A., {Nandra}, K., {et~al.} 2015, \apj, 802, 89

\bibitem[{{Connolly} {et~al.}(2016){Connolly}, {McHardy}, {Skipper}, \&
  {Emmanoulopoulos}}]{Connolly2016}
{Connolly}, S.~D., {McHardy}, I.~M., {Skipper}, C.~J., \& {Emmanoulopoulos}, D.
  2016, \mnras, 459, 3963

\bibitem[{{Dullemond} \& {Monnier}(2010)}]{Dullemond2010}
{Dullemond}, C.~P. \& {Monnier}, J.~D. 2010, \araa, 48, 205

\bibitem[{{Elvis} {et~al.}(2004){Elvis}, {Risaliti}, {Nicastro}, {Miller},
  {Fiore}, \& {Puccetti}}]{Elvis2004}
{Elvis}, M., {Risaliti}, G., {Nicastro}, F., {et~al.} 2004, \apjl, 615, L25

\bibitem[{{Esparza-Arredondo} {et~al.}(2021){Esparza-Arredondo},
  {Gonzalez-Mart{\'\i}n}, {Dultzin}, {Masegosa}, {Ramos-Almeida},
  {Garc{\'\i}a-Bernete}, {Fritz}, \& {Osorio-Clavijo}}]{Esparza-Arredondo2021}
{Esparza-Arredondo}, D., {Gonzalez-Mart{\'\i}n}, O., {Dultzin}, D., {et~al.}
  2021, \aap, 651, A91

\bibitem[{Fruscione {et~al.}(2006)Fruscione, McDowell, Allen, Brickhouse,
  Burke, Davis, Durham, Elvis, Galle, Harris, Huenemoerder, Houck, Ishibashi,
  Karovska, Nicastro, Noble, Nowak, Primini, Siemiginowska, Smith, \&
  Wise}]{Fruscione2006}
Fruscione, A., McDowell, J.~C., Allen, G.~E., {et~al.} 2006, in Observatory
  Operations: Strategies, Processes, and Systems, ed. D.~R. Silva \& R.~E.
  Doxsey, Vol. 6270, International Society for Optics and Photonics (SPIE), 586
  -- 597

\bibitem[{{Harrison} {et~al.}(2013){Harrison}, {Craig}, {Christensen},
  {Hailey}, {Zhang}, {Boggs}, {Stern}, {Cook}, {Forster}, {Giommi},
  {Grefenstette}, {Kim}, {Kitaguchi}, {Koglin}, {Madsen}, {Mao}, {Miyasaka},
  {Mori}, {Perri}, {Pivovaroff}, {Puccetti}, {Rana}, {Westergaard}, {Willis},
  {Zoglauer}, {An}, {Bachetti}, {Barri{\`e}re}, {Bellm}, {Bhalerao},
  {Brejnholt}, {Fuerst}, {Liebe}, {Markwardt}, {Nynka}, {Vogel}, {Walton},
  {Wik}, {Alexander}, {Cominsky}, {Hornschemeier}, {Hornstrup}, {Kaspi},
  {Madejski}, {Matt}, {Molendi}, {Smith}, {Tomsick}, {Ajello}, {Ballantyne},
  {Balokovi{\'c}}, {Barret}, {Bauer}, {Blandford}, {Brandt}, {Brenneman},
  {Chiang}, {Chakrabarty}, {Chenevez}, {Comastri}, {Dufour}, {Elvis}, {Fabian},
  {Farrah}, {Fryer}, {Gotthelf}, {Grindlay}, {Helfand}, {Krivonos}, {Meier},
  {Miller}, {Natalucci}, {Ogle}, {Ofek}, {Ptak}, {Reynolds}, {Rigby},
  {Tagliaferri}, {Thorsett}, {Treister}, \& {Urry}}]{Harrison2013}
{Harrison}, F.~A., {Craig}, W.~W., {Christensen}, F.~E., {et~al.} 2013, \apj,
  770, 103

\bibitem[{{Isobe} {et~al.}(2002){Isobe}, {Tashiro}, {Makishima}, {Iyomoto},
  {Suzuki}, {Murakami}, {Mori}, \& {Abe}}]{Isobe2002}
{Isobe}, N., {Tashiro}, M., {Makishima}, K., {et~al.} 2002, \apjl, 580, L111

\bibitem[{{Jana} {et~al.}(2020){Jana}, {Chatterjee}, {Kumari}, {Nandi}, {Naik},
  \& {Patra}}]{Jana2020}
{Jana}, A., {Chatterjee}, A., {Kumari}, N., {et~al.} 2020, \mnras, 499, 5396

\bibitem[{{Kalberla} {et~al.}(2005){Kalberla}, {Burton}, {Hartmann}, {Arnal},
  {Bajaja}, {Morras}, \& {P{\"o}ppel}}]{Kalberla2005}
{Kalberla}, P.~M.~W., {Burton}, W.~B., {Hartmann}, D., {et~al.} 2005, \aap,
  440, 775

\bibitem[{{Krolik} \& {Begelman}(1988)}]{Krolik1988}
{Krolik}, J.~H. \& {Begelman}, M.~C. 1988, \apj, 329, 702

\bibitem[{{Laha} {et~al.}(2020){Laha}, {Markowitz}, {Krumpe}, {Nikutta},
  {Rothschild}, \& {Saha}}]{Laha2020}
{Laha}, S., {Markowitz}, A.~G., {Krumpe}, M., {et~al.} 2020, \apj, 897, 66

\bibitem[{{Lightman} \& {White}(1988)}]{Lightman1988}
{Lightman}, A.~P. \& {White}, T.~R. 1988, \apj, 335, 57

\bibitem[{{Maiolino} {et~al.}(2010){Maiolino}, {Risaliti}, {Salvati},
  {Pietrini}, {Torricelli-Ciamponi}, {Elvis}, {Fabbiano}, {Braito}, \&
  {Reeves}}]{Maiolino2010}
{Maiolino}, R., {Risaliti}, G., {Salvati}, M., {et~al.} 2010, \aap, 517, A47

\bibitem[{{Marchesi} {et~al.}(2019){Marchesi}, {Ajello}, {Zhao}, {Comastri},
  {La Parola}, \& {Segreto}}]{Marchesi2019}
{Marchesi}, S., {Ajello}, M., {Zhao}, X., {et~al.} 2019, \apj, 882, 162

\bibitem[{{Marchesi} {et~al.}(2022){Marchesi}, {Zhao}, {Torres-Alb{\`a}},
  {Ajello}, {Gaspari}, {Pizzetti}, {Buchner}, {Bertola}, {Comastri}, {Feltre},
  {Gilli}, {Lanzuisi}, {Matzeu}, {Pozzi}, {Salvestrini}, {Sengupta}, {Silver},
  {Tombesi}, {Traina}, {Vignali}, \& {Zappacosta}}]{Marchesi2022}
{Marchesi}, S., {Zhao}, X., {Torres-Alb{\`a}}, N., {et~al.} 2022, arXiv
  e-prints, arXiv:2207.06734

\bibitem[{{Markowitz} {et~al.}(2014){Markowitz}, {Krumpe}, \&
  {Nikutta}}]{Markowitz2014}
{Markowitz}, A.~G., {Krumpe}, M., \& {Nikutta}, R. 2014, \mnras, 439, 1403

\bibitem[{{Murphy} \& {Yaqoob}(2009)}]{Murphy2009}
{Murphy}, K.~D. \& {Yaqoob}, T. 2009, \mnras, 397, 1549

\bibitem[{{Nenkova} {et~al.}(2002){Nenkova}, {Ivezi{\'c}}, \&
  {Elitzur}}]{Nenkova2002}
{Nenkova}, M., {Ivezi{\'c}}, {\v{Z}}., \& {Elitzur}, M. 2002, \apjl, 570, L9

\bibitem[{{Nenkova} {et~al.}(2008){Nenkova}, {Sirocky}, {Nikutta},
  {Ivezi{\'c}}, \& {Elitzur}}]{Nenkova2008}
{Nenkova}, M., {Sirocky}, M.~M., {Nikutta}, R., {Ivezi{\'c}}, {\v{Z}}., \&
  {Elitzur}, M. 2008, \apj, 685, 160

\bibitem[{{Oh} {et~al.}(2018){Oh}, {Koss}, {Markwardt}, {Schawinski},
  {Baumgartner}, {Barthelmy}, {Cenko}, {Gehrels}, {Mushotzky}, {Petulante},
  {Ricci}, {Lien}, \& {Trakhtenbrot}}]{Oh2018}
{Oh}, K., {Koss}, M., {Markwardt}, C.~B., {et~al.} 2018, \apjs, 235, 4

\bibitem[{{Pizzetti} {et~al.}(2022){Pizzetti}, {Torres-Alb{\`a}}, {Marchesi},
  {Ajello}, {Silver}, \& {Zhao}}]{Pizzetti2022}
{Pizzetti}, A., {Torres-Alb{\`a}}, N., {Marchesi}, S., {et~al.} 2022, \apj,
  936, 149

\bibitem[{{Ramos Almeida} {et~al.}(2014){Ramos Almeida}, {Alonso-Herrero},
  {Levenson}, {Asensio Ramos}, {Rodr{\'\i}guez Espinosa},
  {Gonz{\'a}lez-Mart{\'\i}n}, {Packham}, \&
  {Mart{\'\i}nez}}]{Ramos-Almeida2014}
{Ramos Almeida}, C., {Alonso-Herrero}, A., {Levenson}, N.~A., {et~al.} 2014,
  \mnras, 439, 3847

\bibitem[{{Ricci} {et~al.}(2015){Ricci}, {Ueda}, {Koss}, {Trakhtenbrot},
  {Bauer}, \& {Gandhi}}]{Ricci2015}
{Ricci}, C., {Ueda}, Y., {Koss}, M.~J., {et~al.} 2015, \apjl, 815, L13

\bibitem[{{Risaliti} {et~al.}(2005){Risaliti}, {Elvis}, {Fabbiano}, {Baldi}, \&
  {Zezas}}]{Risaliti2005}
{Risaliti}, G., {Elvis}, M., {Fabbiano}, G., {Baldi}, A., \& {Zezas}, A. 2005,
  \apjl, 623, L93

\bibitem[{{Risaliti} {et~al.}(2002){Risaliti}, {Elvis}, \&
  {Nicastro}}]{Risaliti2002}
{Risaliti}, G., {Elvis}, M., \& {Nicastro}, F. 2002, \apj, 571, 234

\bibitem[{{Risaliti} {et~al.}(2009){Risaliti}, {Salvati}, {Elvis}, {Fabbiano},
  {Baldi}, {Bianchi}, {Braito}, {Guainazzi}, {Matt}, {Miniutti}, {Reeves},
  {Soria}, \& {Zezas}}]{Risaliti2009}
{Risaliti}, G., {Salvati}, M., {Elvis}, M., {et~al.} 2009, \mnras, 393, L1

\bibitem[{{Rivers} {et~al.}(2015){Rivers}, {Balokovi{\'c}}, {Ar{\'e}valo},
  {Bauer}, {Boggs}, {Brandt}, {Brightman}, {Christensen}, {Craig}, {Gandhi},
  {Hailey}, {Harrison}, {Koss}, {Ricci}, {Stern}, {Walton}, \&
  {Zhang}}]{Rivers2015}
{Rivers}, E., {Balokovi{\'c}}, M., {Ar{\'e}valo}, P., {et~al.} 2015, \apj, 815,
  55

\bibitem[{{Saha} {et~al.}(2022){Saha}, {Markowitz}, \& {Buchner}}]{Saha2022}
{Saha}, T., {Markowitz}, A.~G., \& {Buchner}, J. 2022, \mnras, 509, 5485

\bibitem[{{Siebenmorgen} {et~al.}(2015){Siebenmorgen}, {Heymann}, \&
  {Efstathiou}}]{Siebenmorgen2015}
{Siebenmorgen}, R., {Heymann}, F., \& {Efstathiou}, A. 2015, \aap, 583, A120

\bibitem[{{Siemiginowska} {et~al.}(2012){Siemiginowska}, {Stawarz}, {Cheung},
  {Aldcroft}, {Bechtold}, {Burke}, {Evans}, {Holt}, {Jamrozy}, \&
  {Migliori}}]{Siemiginowska2012}
{Siemiginowska}, A., {Stawarz}, {\L}., {Cheung}, C.~C., {et~al.} 2012, \apj,
  750, 124

\bibitem[{Smith {et~al.}(2001)Smith, Brickhouse, Liedahl, \&
  Raymond}]{Smith2001}
Smith, R.~K., Brickhouse, N.~S., Liedahl, D.~A., \& Raymond, J.~C. 2001, The
  Astrophysical Journal, 556, L91

\bibitem[{{Sobolewska} {et~al.}(2012){Sobolewska}, {Siemiginowska}, {Migliori},
  {Stawarz}, {Jamrozy}, {Evans}, \& {Cheung}}]{Sobolewska2012}
{Sobolewska}, M.~A., {Siemiginowska}, A., {Migliori}, G., {et~al.} 2012, \apj,
  758, 90

\bibitem[{{Tanimoto} {et~al.}(2019){Tanimoto}, {Ueda}, {Odaka}, {Kawaguchi},
  {Fukazawa}, \& {Kawamuro}}]{Tanimoto2019}
{Tanimoto}, A., {Ueda}, Y., {Odaka}, H., {et~al.} 2019, \apj, 877, 95

\bibitem[{{Tanimoto} {et~al.}(2022){Tanimoto}, {Ueda}, {Odaka}, {Yamada}, \&
  {Ricci}}]{Tanimoto2022}
{Tanimoto}, A., {Ueda}, Y., {Odaka}, H., {Yamada}, S., \& {Ricci}, C. 2022,
  \apjs, 260, 30

\bibitem[{{Torres-Alb{\`a}} {et~al.}(2018){Torres-Alb{\`a}}, {Iwasawa},
  {D{\'\i}az-Santos}, {Charmandaris}, {Ricci}, {Chu}, {Sand ers}, {Armus},
  {Barcos-Mu{\~n}oz}, {Evans}, {Howell}, {Inami}, {Linden}, {Medling},
  {Privon}, {U}, \& {Yoon}}]{Torres-Alba2018}
{Torres-Alb{\`a}}, N., {Iwasawa}, K., {D{\'\i}az-Santos}, T., {et~al.} 2018,
  \aap, 620, A140

\bibitem[{{Torres-Alb{\`a}} {et~al.}(2021){Torres-Alb{\`a}}, {Marchesi},
  {Zhao}, {Ajello}, {Silver}, {Ananna}, {Balokovi{\'c}}, {Boorman}, {Comastri},
  {Gilli}, {Lanzuisi}, {Murphy}, {Urry}, \& {Vignali}}]{Torres-Alba2021}
{Torres-Alb{\`a}}, N., {Marchesi}, S., {Zhao}, X., {et~al.} 2021, \apj, 922,
  252

\bibitem[{{Urry} \& {Padovani}(1995)}]{Urry1995}
{Urry}, C.~M. \& {Padovani}, P. 1995, \pasp, 107, 803

\bibitem[{{Verner} {et~al.}(1996){Verner}, {Ferland}, {Korista}, \&
  {Yakovlev}}]{Verner1996}
{Verner}, D.~A., {Ferland}, G.~J., {Korista}, K.~T., \& {Yakovlev}, D.~G. 1996,
  \apj, 465, 487

\bibitem[{{Yaqoob}(2012)}]{Yaqoob2012}
{Yaqoob}, T. 2012, \mnras, 423, 3360

\bibitem[{{Zhao} {et~al.}(2021){Zhao}, {Marchesi}, {Ajello}, {Cole}, {Hu},
  {Silver}, \& {Torres-Alb{\`a}}}]{Zhao2020}
{Zhao}, X., {Marchesi}, S., {Ajello}, M., {et~al.} 2021, \aap, 650, A57

\end{thebibliography}

\clearpage
\appendix

\section{X-ray Fitting Results}\label{App:x-ray}

This Appendix is a compilation of tables showing the best-fit results for all sources analyzed in this work (except for NGC 612, which can be found in Table \ref{tab:NGC612}, in the main text).

\begin{table*}
\centering
\begin{threeparttable}
\label{tab:NGC788}
\caption{NGC 788 fitting results}
\renewcommand*{\arraystretch}{1.4}
\begin{tabular}{ccccc}
{\bf Model} & {\bf MYTorus} & {\bf borus02} & {\bf borus02} & {\bf UXCLUMPY} \\ \hline
$\chi^2_{\rm red}$ & 1.13 & 1.13 & 1.13 & 1.17  \\
$\chi^2$/d.o.f. & 572/508 & 571/507 & 570/507 & 596/508 \\ 
$T$ & 2.9$\sigma$ & 2.9$\sigma$ & 2.9$\sigma$ & 3.8$\sigma$ \\ \hline
kT & 0.25$^{+0.07}_{-0.05}$ & 0.24$^{+0.04}_{-0.05}$ &  0.24$^{+0.04}_{-0.05}$& 0.24$^{+0.01}_{-0.03}$ \\
$E_1$ & 0.89$^{+0.01}_{-0.01}$ & 0.90$^{+0.01}_{-0.01}$  & 0.90$^{+0.01}_{-0.01}$  & 0.90$^{+0.01}_{-0.01}$  \\
$E_2$ & 1.86$^{+0.04}_{-0.05}$ & 1.86$^{+0.04}_{-0.06}$ & 1.86$^{+0.04}_{-0.06}$ & 1.87$^{+0.03}_{-0.04}$  \\
$E_3$ & 2.38$^{+0.07}_{-0.05}$ & 2.39$^{+0.05}_{-0.05}$ & 2.39$^{+0.04}_{-0.05}$  & 2.39$^{+0.05}_{-0.05}$  \\ \hline
$\Gamma$ & 1.92$^{+0.11}_{-0.12}$ & 1.77$^{+0.04}_{-0.04}$ & 1.88$^{+0.09}_{-0.04}$ & 1.87$^{+0.07}_{-0.09}$  \\
$N_{H,av}$ & 0.19$^{+0.02}_{-0.02}$ & 0.21$^{+0.06}_{-0.03}$ & 31.6$^{-u}_{-18.2}$ & $-$ \\
A$_{S90}$ & 0.92$^{+0.21}_{-0.16}$ & $-$ & $-$ & $-$  \\
A$_{S0}$ & 0* & $-$ & $-$ & $-$ \\
$C_{\rm F}$ & $-$ & 0.34$^{+0.05}_{-0.05}$ & 0.44$^{+0.05}_{-0.23}$ & 0*  \\
Cos ($\theta_{Obs}$) & $-$ & 0.21$^{+0.05}_{-0.13}$ & 0.46$^{+0.13}_{-0.14}$ & 1.00$^{-u}_{-0.47}$  \\ 
$\sigma_{\rm tor}$ & $-$ & $-$ & $-$ & 7.5$^{+12.0}_{-0.5}$  \\
F$_s$ (10$^{-3}$) & 2.96$^{+1.04}_{-0.95}$ & 4.07$^{+2.00}_{-1.31}$ & 5.09$^{+1.18}_{-0.29}$ & 0.15$^{+1.28}_{-u}$ \\
norm (10$^{-2}$) & 1.45$^{+0.74}_{-0.51}$ & 0.906$^{+0.091}_{-0.098}$ & 0.731$^{+0.675}_{-0.282}$ & 43.4$^{+2.8}_{-1.1}$ \\
\hline 
$N_{H,Ch}$ & 0.79$^{+0.08}_{-0.08}$ & 0.73$^{+0.05}_{-0.05}$ & 0.62$^{+0.04}_{-0.03}$ &  0.55$^{+0.05}_{-0.2}$  \\
$N_{H,xmm}$ & 0.82$^{+0.08}_{-0.08}$ & 0.76$^{+0.04}_{-0.04}$ & 0.65$^{+0.02}_{-0.02}$ & 0.59$^{+0.08}_{-0.08}$  \\
$N_{H,nus}$ & 1.10$^{+0.10}_{-0.09}$ & 1.04$^{+0.07}_{-0.07}$ & 0.86$^{+0.05}_{-0.04}$& 0.83$^{+0.03}_{-0.05}$  \\ \hline
$C_{Ch}$ & 1* & 1* & 1* & 1* \\
$C_{xmm}$ & =$C_{Ch}$ & =$C_{Ch}$ & =$C_{Ch}$ & =$C_{Ch}$    \\
$C_{nus}$ &  =$C_{Ch}$ &  =$C_{Ch}$ & =$C_{Ch}$ & =$C_{Ch}$  \\ \hline \hline
$\chi^2_{\rm red}$ No Var. & 1.47 & 1.47 & 1.37 & 1.49 \\
$T$ & 10.6$\sigma$ & 10.6$\sigma$ & 8.4$\sigma$ & 11.1$\sigma$\\ \hline
$\chi^2_{\rm red}$ No C Var. & 1.13 & 1.13 & 1.13 & 1.17\\
$T$ & 2.9$\sigma$ & 2.9$\sigma$ & 2.9$\sigma$ & 3.8$\sigma$\\ \hline
$\chi^2_{\rm red}$ No $N_{\rm H}$ Var. & 1.15 & 1.15 & 1.13 & 1.19\\
$T$ & 3.4$\sigma$ & 3.4$\sigma$ & 2.9$\sigma$ & 4.3$\sigma$ \\ \hline
P-value & 1.4e-1 & 2.0e-1 & & 1.7e-5 \\ \hline
\end{tabular}
\begin{tablenotes}{\textbf{Notes:} Same as Table \ref{tab:NGC612}, with the following additions: \newline
E$_{\rm n}$: Central energy of the added nth Gaussian line, in keV.}
\end{tablenotes}
\end{threeparttable}
\end{table*}

\begin{table*}
\centering
\begin{threeparttable}
\label{tab:NGC833}
\caption{NGC 833 fitting results}
\renewcommand*{\arraystretch}{1.4}
\begin{tabular}{cccc}
{\bf Model} & {\bf MYTorus} & {\bf borus02} & {\bf UXCLUMPY} \\ \hline
$\chi^2_{\rm red}$ & 0.93  & 0.93 &  0.93 \\
$\chi^2$/d.o.f. & 193/208 & 193/206 & 192/207  \\ 
$T$ & 1.0$\sigma$ & 1.0$\sigma$ & 1.0$\sigma$ \\\hline
kT & 0.60$^{+0.05}_{-0.08}$ & 0.59$^{+0.06}_{-0.11}$ & 0.59$^{+0.06}_{-0.11}$ \\ \hline
$\Gamma$ & 1.69$^{+0.26}_{-0.25}$ & 1.58$^{+0.26}_{-u}$ & 1.55$^{+0.37}_{-0.32}$  \\
$N_{H,av}$ & 0.06$^{+0.08}_{-u}$ & 0.08$^{+u}_{-u}$ & $-$ \\
A$_{S90}$ & 1* & $-$ & $-$  \\
A$_{S0}$ & 1* & $-$ & $-$  \\
$C_{\rm F}$ & $-$ & 0.52$^{+0.30}_{-u}$ & 0*  \\
Cos ($\theta_{Obs}$) & $-$ & 0.15$^{+u}_{-u}$ &  0.0$^{+u}_{-u}$  \\ 
$\sigma_{\rm tor}$ & $-$ & $-$ & 3.8$^{+u}_{-u}$  \\ 
F$_s$ (10$^{-2}$) & 0.61$^{+0.59}_{-0.31}$ & 1.24$^{+0.41}_{-0.77}$ & 0.90$^{+7.41}_{-0.86}$ \\
norm (10$^{-4}$) & 4.44$^{+4.62}_{-2.28}$ & 3.19$^{+3.03}_{-1.24}$ & 6.50$^{+6.05}_{-4.75}$ \\
\hline
$N_{H,xmm}$ & 0.34$^{+0.07}_{-0.06}$ & 0.31$^{+0.07}_{-0.07}$ & 0.26$^{+0.04}_{-0.03}$  \\
$N_{H,Ch1}$ & 0.21$^{+0.07}_{-0.06}$ & 0.19$^{+0.05}_{-0.05}$ & 0.16$^{+0.04}_{-0.03}$  \\
$N_{H,Ch2}$ & $-$ & $-$ & $-$  \\
$N_{H,Ch3}$ & 0.33$^{+0.06}_{-0.05}$ & 0.34$^{+0.07}_{-0.06}$ & 0.28$^{+0.05}_{-0.03}$  \\
$N_{H,Ch4}$ & 0.27$^{+0.05}_{-0.05}$ & 0.27$^{+0.05}_{-0.05}$ & 0.22$^{+0.04}_{-0.04}$  \\
$N_{H,Ch5}$ & 0.28$^{+0.05}_{-0.04}$ & 0.29$^{+0.05}_{-0.06}$ & 0.24$^{+0.04}_{-0.04}$  \\
$N_{H,nus}$ & 0.18$^{+0.10}_{-0.10}$ & 0.14$^{+0.08}_{-0.09}$ & 0.10$^{+0.09}_{-0.06}$  \\ \hline
$C_{xmm}$ & 1.20$^{+0.33}_{-0.17}$ & 1.18$^{+0.13}_{-0.14}$ & 1.21$^{+0.29}_{-0.18}$  \\
$C_{Ch1}$ & 1* & 1* & 1* \\
$C_{Ch2}$ & $-$ & $-$ & $-$ \\
$C_{Ch3}$ & 0.55$^{+0.16}_{-0.12}$ & 0.66$^{+0.14}_{-0.10}$ & 0.66$^{+0.16}_{-0.12}$  \\
$C_{Ch4}$ & $=C_{Ch3}$ & $=C_{Ch3}$ & $=C_{Ch3}$  \\
$C_{Ch5}$ & $=C_{Ch3}$ & $=C_{Ch3}$ & $=C_{Ch3}$  \\
$C_{nus}$ & $=C_{Ch1}$ & $=C_{Ch1}$ & $=C_{Ch1}$  \\ \hline \hline
$\chi^2_{\rm red}$ No Var. & 1.98 & 2.00 & 1.69\\
$T$ & 14.3$\sigma$ & 14.6$\sigma$ & 10.1$\sigma$\\ \hline
$\chi^2_{\rm red}$ No C Var. & 1.18 & 1.19 & 1.19\\
$T$ & 2.6$\sigma$ & 2.7$\sigma$ & 2.7$\sigma$\\ \hline
$\chi^2_{\rm red}$ No $N_{\rm H}$ Var. & 0.99 & 1.02 & 1.05\\
$T$ & 0.1$\sigma$ & 0.3$\sigma$ & 0.7$\sigma$\\ \hline
P-value & 9.7e-1 & 9.2e-1 & 8.5e-1 \\ \hline
\end{tabular}
\begin{tablenotes}
{\textbf{Notes:} Same as Table \ref{tab:NGC612}. The second \textit{Chandra} observation of the system formed by NGC 833 and NGC 835 did not include the former, hence the missing parameters corresponding to the observation. See Appendix \ref{App:sources} for details.}
\end{tablenotes}
\end{threeparttable}
\end{table*}

\begin{table*}
\centering
\begin{threeparttable}
\label{tab:NGC835}
\caption{NGC 835 fitting results}
\renewcommand*{\arraystretch}{1.4}
\begin{tabular}{cccc}
{\bf Model} & {\bf MYTorus} & {\bf borus02} & {\bf UXCLUMPY} \\ \hline
$\chi^2_{\rm red}$ & 1.07 & 1.08 & 1.05  \\
$\chi^2$/d.o.f. & 479/446 & 479/445 &  468/446 \\ 
$T$ & 1.5$\sigma$ & 1.7$\sigma$ & 1.1$\sigma$ \\\hline
kT & 0.61$^{+0.02}_{-0.02}$ & 0.61$^{+0.04}_{-0.03}$ & 0.61$^{+0.02}_{-0.02}$ \\
$E_1$ & 0.68$^{+0.03}_{-0.02}$ & 0.68$^{+0.03}_{-0.19}$ & 0.68$^{+0.02}_{-0.03}$ \\
$E_2$ & 1.29$^{+0.06}_{-0.09}$ & 1.29$^{+0.05}_{-0.10}$ & 1.29$^{+0.06}_{-0.06}$ \\ \hline
$\Gamma$ & 1.68$^{+0.13}_{-0.13}$ & 1.63$^{+0.15}_{-0.12}$ & 1.55$^{+0.22}_{-0.25}$  \\
$N_{H,av}$ & 0.19$^{+0.08}_{-0.09}$ & 0.21$^{+0.10}_{-0.10}$ & $-$ \\
A$_{S90}$ & 0.52$^{+0.18}_{-0.18}$ & $-$ & $-$  \\
A$_{S0}$ & 0* & $-$ & $-$  \\
$C_{\rm F}$ & $-$ & 0.18$^{+0.08}_{-0.04}$ &  0* \\
Cos ($\theta_{Obs}$) & $-$ & 0.05$^{+0.17}_{-u}$ & 0.86$^{+0.04}_{-0.45}$  \\ 
$\sigma_{\rm tor}$ & $-$ & $-$ & 6.8$^{+3.8}_{-4.5}$  \\ 
F$_s$ (10$^{-3}$) & 7.06$^{+1.94}_{-1.68}$ & 6.88$^{+1.82}_{-1-38}$ & 4.93$^{+12.16}_{-u}$ \\
norm (10$^{-3}$) & 1.08$^{+0.41}_{-0.29}$ & 0.96$^{+0.38}_{-0.24}$ & 1.90$^{+0.19}_{-0.48}$ \\
\hline
$N_{H,xmm}$ & 1.53$^{+1.07}_{-0.26}$ & 1.48$^{+1.50}_{-0.23}$ & 1.35$^{+0.05}_{-0.02}$  \\
$N_{H,Ch1}$ & 0.89$^{+0.25}_{-0.14}$ & 0.88$^{+0.28}_{-0.14}$ & 1.04$^{+0.18}_{-0.19}$  \\
$N_{H,Ch2}$ & 0.86$^{+0.32}_{-0.14}$ & 0.85$^{+0.33}_{-0.14}$ & 0.94$^{+0.24}_{-0.16}$  \\
$N_{H,Ch3}$ & 0.31$^{+0.02}_{-0.03}$ & 0.30$^{+0.03}_{-0.02}$ & 0.28$^{+0.04}_{-0.03}$  \\
$N_{H,Ch4}$ & 0.32$^{+0.03}_{-0.03}$ & 0.32$^{+0.03}_{-0.03}$ & 0.31$^{+0.04}_{-0.04}$  \\
$N_{H,Ch5}$ & 0.33$^{+0.03}_{-0.03}$ & 0.32$^{+0.03}_{-0.03}$ & 0.32$^{+0.03}_{-0.03}$  \\
$N_{H,nus}$ & 0.46$^{+0.06}_{-0.05}$ & 0.45$^{+0.06}_{-0.05}$ & 0.27$^{+0.16}_{-0.12}$  \\ \hline
$C_{xmm}$ & 1.34$^{+0.10}_{-0.09}$ & 1.25$^{+0.07}_{-0.07}$ & 1.28$^{+0.18}_{-0.16}$  \\
$C_{Ch1}$ & 1* & 1* & 1* \\
$C_{Ch2}$ & $=C_{Ch1}$  & $=C_{Ch1}$ & $=C_{Ch1}$ \\
$C_{Ch3}$ & $=C_{Ch1}$  & $=C_{Ch1}$ & $=C_{Ch1}$ \\
$C_{Ch4}$ & $=C_{Ch1}$  & $=C_{Ch1}$ & $=C_{Ch1}$ \\
$C_{Ch5}$ & $=C_{Ch1}$  & $=C_{Ch1}$ & $=C_{Ch1}$ \\
$C_{nus}$ & $=C_{Ch1}$  & $=C_{Ch1}$ & 0.63$^{+0.12}_{-0.22}$  \\ \hline \hline
$\chi^2_{\rm red}$ No Var. & 4.44 & 4.63 & 4.55 \\ 
$T$ & 73.2$\sigma$ & 77.2$\sigma$ & 75.6$\sigma$\\ \hline
$\chi^2_{\rm red}$ No C Var. & 1.17 & 1.18 & 1.18\\
$T$ & 3.6$\sigma$ & 3.8$\sigma$ & 3.8$\sigma$\\ \hline
$\chi^2_{\rm red}$ No $N_{\rm H}$ Var. & 2.31 & 3.84 & 3.85 \\
$T$ & 27.6$\sigma$ & 59.9$\sigma$ & 60.2$\sigma$\\ \hline
P-value & 4.7e-20 & 3.1e-13 & 5.7e-52 \\ \hline
\end{tabular}
\begin{tablenotes}{\textbf{Notes:} Same as Table \ref{tab:NGC612}, with the following additions: \newline
E$_{\rm n}$: Central energy of the added nth Gaussian line, in keV.}
\end{tablenotes}
\end{threeparttable}
\end{table*}

\begin{table*}
\centering
\begin{threeparttable}
\label{tab:3C105}
\caption{3C 105 fitting results}
\renewcommand*{\arraystretch}{1.4}
\begin{tabular}{cccc}
{\bf Model} & {\bf MYTorus} & {\bf borus02} & {\bf UXCLUMPY} \\ \hline
$\chi^2_{\rm red}$ & 1.01 & 1.01 &  1.01 \\
$\chi^2$/d.o.f. & 240/237 & 240/236 &  240/237 \\ 
$T$ & 0.2$\sigma$ & 0.2$\sigma$ & 0.2$\sigma$ \\\hline
kT & 0.21$^{+0.03}_{-0.03}$ & 0.20$^{+0.03}_{-0.03}$ & 0.20$^{+0.03}_{-0.03}$ \\ \hline
$\Gamma$ & 1.48$^{+0.15}_{-u}$ & 1.44$^{+0.14}_{-u}$ & 1.57$^{+0.17}_{-0.03}$  \\
$N_{H,av}$ & 0.40$^{+0.57}_{-0.21}$ & 0.43$^{+0.24}_{-0.15}$ & $-$ \\
A$_{S90}$ & 0.75$^{+0.48}_{-0.40}$ & $-$ & $-$  \\
A$_{S0}$ & 0* & $-$ & $-$  \\
$C_{\rm F}$ & $-$ & 0.30$^{+0.13}_{-0.12}$ & 0*  \\
Cos ($\theta_{Obs}$) & $-$ & 0.10$^{+0.80}_{-u}$ & 0.00$^{-u}_{-u}$  \\ 
$\sigma_{\rm tor}$ & $-$ & $-$ & 15.9$^{+20.8}_{-6.9}$  \\ 
F$_s$ (10$^{-3}$) & 2.67$^{+1.18}_{-1.13}$ & 2.75$^{+0.95}_{-0.93}$ & 2.93$^{+4.21}_{-1.26}$ \\
norm (10$^{-3}$) & 2.92$^{+1.65}_{-0.84}$ & 2.50$^{+0.06}_{-0.69}$ & 5.09$^{+2.64}_{-1.56}$ \\
\hline 
$N_{H,ch}$ & 0.45$^{+0.08}_{-0.05}$ & 0.46$^{+0.04}_{-0.04}$ & 0.49$^{+0.03}_{-0.09}$  \\
$N_{H,xmm}$ & 0.39$^{+0.05}_{-0.04}$ & 0.39$^{+0.03}_{-0.03}$ & 0.39$^{+0.02}_{-0.03}$  \\
$N_{H,nus1}$ & 0.45$^{+0.08}_{-0.07}$ & 0.45$^{+0.03}_{-0.03}$ & 0.44$^{+0.03}_{-0.08}$  \\
$N_{H,nus2}$ & 0.39$^{+0.06}_{-0.06}$ & 0.39$^{+0.06}_{-0.03}$ & 0.40$^{+0.03}_{-0.07}$  \\ \hline
$C_{ch}$ & 1*& 1* & 1*  \\
$C_{xmm}$ & 0.63$^{+0.16}_{-0.15}$ & 0.62$^{+0.04}_{-0.08}$ & 0.59$^{+0.03}_{-0.13}$  \\
$C_{nus1}$ & 0.28$^{+0.08}_{-0.07}$ & 0.27$^{+0.02}_{-0.06}$ & 0.25$^{+0.08}_{-0.06}$  \\
$C_{nus2}$ & =$C_{nus1}$ & =$C_{nus1}$ & =$C_{nus1}$  \\ \hline \hline
$\chi^2_{\rm red}$ No Var. & 2.66 & 2.67 & 2.65 \\
$T$ & 25.8$\sigma$ & 26.0$\sigma$ & 25.7$\sigma$\\ \hline
$\chi^2_{\rm red}$ No C Var. & 1.20 & 1.21 & 1.23\\
$T$ & 3.1$\sigma$ & 3.3$\sigma$ & 3.6$\sigma$\\ \hline
$\chi^2_{\rm red}$ No $N_{\rm H}$ Var. & 1.05 & 1.02 & 1.01\\
$T$ & 0.8$\sigma$ & 0.3$\sigma$ & 0.2$\sigma$\\ \hline
P-value & 9.2e-1 & 9.2e-1 & 8.0e-1 \\ \hline
\end{tabular}
\begin{tablenotes}
{\textbf{Notes:} Same as Table \ref{tab:NGC612}.}
\end{tablenotes}
\end{threeparttable}
\end{table*}

\begin{table*}
\centering
\begin{threeparttable}
\label{tab:4C+29.30}
\caption{4C+29.30 fitting results}
\renewcommand*{\arraystretch}{1.4}
\begin{tabular}{cccc}
{\bf Model} & {\bf MYTorus} & {\bf borus02} & {\bf UXCLUMPY} \\ \hline
Stat$_{\rm red}$ & 425/432 & 421/431 & 437/433  \\
Stat/d.o.f. & 0.98 & 0.98 & 1.01\\ 
$T$ & 0.4$\sigma$ & 0.4$\sigma$ & 0.2$\sigma$\\ \hline
kT & 0.64$^{0.04}_{-0.04}$ & 0.63$^{+0.04}_{-0.04}$ & 0.64$^{+0.04}_{-0.04}$ \\ \hline
$\Gamma$ & 1.72$^{+0.22}_{-0.20}$ & 1.70$^{+0.19}_{-0.19}$ & 1.90$^{+0.14}_{-0.20}$  \\
$N_{H,av}$ & 0.21$^{+0.04}_{-0.02}$ & 0.22$^{+0.07}_{-0.03}$ & $-$ \\
A$_{S90}$ & 0.81$^{+0.19}_{-0.15}$ & $-$ & $-$  \\
A$_{S0}$ & 0* & $-$ & $-$  \\
$C_{\rm F}$ & $-$ & 0.28$^{+0.06}_{-0.03}$ & 0* \\
Cos ($\theta_{Obs}$) & $-$ & 0.10$^{+0.09}_{-u}$ & 0.16$^{+0.14}_{-u}$  \\ 
$\sigma_{\rm tor}$ & $-$ & $-$ & 17.5$^{+8.6}_{-7.4}$  \\ 
F$_s$ (10$^{-3}$) & 2.07$^{+1.79}_{-0.88}$ & 1.75$^{+0.70}_{-0.68}$ & 2.22$^{+1.58}_{-0.80}$ \\
norm (10$^{-3}$) & 2.66$^{+2.47}_{-1.36}$ & 2.14$^{+1.45}_{-0.56}$ & 3.22$^{+2.18}_{-1.69}$ \\
\hline 
$N_{H,Ch1}$ & 0.72$^{+0.16}_{-0.16}$ & 0.68$^{+0.14}_{-0.06}$ & 0.61$^{+0.10}_{-0.11}$  \\
$N_{H,xmm}$ & 0.87$^{+0.18}_{-0.19}$ & 1.08$^{+0.04}_{-0.11}$ & 0.98$^{+0.08}_{-0.10}$  \\
$N_{H,Ch2}$ & 0.65$^{+0.06}_{-0.06}$ & 0.65$^{+0.06}_{-0.03}$ & 0.61$^{+0.04}_{-0.04}$  \\
$N_{H,Ch3}$ & 0.59$^{+0.05}_{-0.05}$ & 0.60$^{+0.05}_{-0.01}$ & 0.55$^{+0.04}_{-0.02}$  \\
$N_{H,Ch4}$ & 0.60$^{+0.06}_{-0.05}$ & 0.60$^{+0.05}_{-0.02}$ & 0.56$^{+0.04}_{-0.02}$  \\
$N_{H,Ch5}$ & 0.62$^{+0.07}_{-0.06}$ & 0.58$^{+0.05}_{-0.02}$ & 0.54$^{+0.03}_{-0.02}$  \\
$N_{H,nus}$ & 0.61$^{+0.17}_{-0.13}$ & 0.62$^{+0.16}_{-0.13}$ & 0.63$^{+0.09}_{-0.14}$  \\ \hline
$C_{Ch1}$ & 1* & 1* & 1* \\
$C_{xmm}$ & 1.31$^{+0.59}_{-0.35}$ & 1.61$^{+0.70}_{-0.07}$ & 1.82$^{+0.83}_{-0.47}$  \\
$C_{Ch2}$ & 1.15$^{+0.50}_{-0.29}$ & 1.30$^{+0.49}_{-0.29}$ & 1.38$^{+0.37}_{-0.25}$  \\
$C_{Ch3}$ & $=C_{Ch2}$ & $=C_{Ch2}$ & $=C_{Ch2}$  \\
$C_{Ch4}$ & $=C_{Ch2}$ & $=C_{Ch2}$ & $=C_{Ch2}$  \\
$C_{Ch5}$ & $=C_{Ch2}$ & $=C_{Ch2}$ & $=C_{Ch2}$ \\
$C_{nus}$ & 0.73$^{+0.18}_{-0.13}$ & 0.84$^{+0.04}_{-0.28}$ & $=C_{Ch1}$ \\ \hline \hline
Stat$_{\rm red}$ No Var. & 2.40 & 2.41 & 2.41 \\
$T$ & 29.4$\sigma$ & 29.6$\sigma$ & 29.7$\sigma$\\ \hline
Stat$_{\rm red}$ No C Var. & 0.99 & 0.99 & 1.03 \\
$T$ & 0.2$\sigma$ & 0.2$\sigma$ & 0.6$\sigma$\\ \hline
Stat$_{\rm red}$ No $N_{\rm H}$ Var. & 0.98 & 1.16& 1.07 \\
$T$ & 0.4$\sigma$ & 3.3$\sigma$ & 1.5$\sigma$\\ \hline
P-value & 9.9e-1 & 6.7e-1 & 5.4e-1 \\\hline
\end{tabular}
\begin{tablenotes}
{\textbf{Notes:} Same as Table \ref{tab:NGC612}.}
\end{tablenotes}
\end{threeparttable}
\end{table*}

\begin{table*}
\centering
\begin{threeparttable}
\label{tab:NGC3281}
\caption{NGC 3281 fitting results}
\renewcommand*{\arraystretch}{1.4}
\begin{tabular}{cccc}
{\bf Model} & {\bf MYTorus} & {\bf borus02} & {\bf UXCLUMPY} \\ \hline
$\chi^2_{\rm red}$ & 1.10 & 1.04 & 1.07  \\
$\chi^2$/d.o.f. & 469/427 & 444/427 & 460/428 \\ 
$T$ & 2.1$\sigma$ & 0.8$\sigma$ & 1.4$\sigma$\\ \hline
kT & 0.58$^{+0.05}_{-0.09}$ & 0.58$^{+0.04}_{-0.11}$ & 0.57$^{+0.10}_{-0.06}$ \\ \hline
$\Gamma$ & 1.65$^{+0.11}_{-0.12}$ & 1.81$^{+0.14}_{-0.07}$ & 1.75$^{+0.04}_{-0.05}$  \\ 
$N_{H,av}$ & 0.31$^{+0.10}_{-0.06}$ & 31.6$^{-u}_{-8.4}$ & $-$ \\
A$_{S90}$ & 0.21$^{+0.23}_{-u}$ & $-$ & $-$  \\
A$_{S0}$ & 0.31$^{+0.30}_{-0.17}$ & $-$ & $-$  \\
$C_{\rm F}$ & $-$ & 0.52$^{+0.04}_{-0.14}$ & 0*  \\
Cos ($\theta_{Obs}$) & $-$ & 0.53$^{+0.15}_{-0.08}$ & 0.00$^{-u}_{-u}$  \\
$\sigma_{\rm tor}$ & $-$ & $-$ & 28.0$^{+16.5}_{-8.4}$  \\ 
F$_s$ (10$^{-4}$) & 8.17$^{+6.76}_{-3.39}$ & 17.3$^{+6.8}_{-3.2}$ & 51.9$^{+24.6}_{-51.6}$ \\
norm (10$^{-2}$) & 1.65$^{+1.15}_{-0.75}$ & 0.90$^{+0.48}_{-0.24}$ & 1.06$^{+0.45}_{-0.15}$ \\
\hline 
$N_{H,xmm}$ & 1.16$^{+0.17}_{-0.16}$ & 0.86$^{+0.09}_{-0.10}$ & 0.89$^{+0.06}_{-0.07}$  \\
$N_{H,nus}$ & 2.25$^{+0.24}_{-0.26}$ & 2.05$^{+0.28}_{-0.38}$ & 3.01$^{+0.62}_{-0.35}$  \\ 
$N_{H,Ch}$ & 1.04$^{+0.17}_{-0.17}$ & 0.76$^{+0.10}_{-0.10}$ & 0.76$^{+0.08}_{-0.06}$  \\
\hline
$C_{xmm}$ & =$C_{Ch}$ & =$C_{Ch}$ & =$C_{Ch}$  \\
$C_{nus}$ & 1.43$^{+0.22}_{-0.17}$ & 1.53$^{+0.16}_{-0.15}$ & 1.53$^{+0.14}_{-0.15}$  \\
$C_{Ch}$ & 1* & 1* & 1* \\
\hline \hline
$\chi^2_{\rm red}$ No Var.  & 1.53 & 1.43 & 1.99 \\
$T$ & 11.0$\sigma$ & 8.9$\sigma$ & 20.6$\sigma$\\ \hline
$\chi^2_{\rm red}$ No C Var. & 1.16 & 1.10 & 1.18\\
$T$ & 3.3$\sigma$ & 2.1$\sigma$ & 3.7$\sigma$\\ \hline
$\chi^2_{\rm red}$ No $N_{\rm H}$ Var. & 1.43 & 1.25 & 1.48\\
$T$ & 8.9$\sigma$ & 5.2$\sigma$ & 9.9$\sigma$\\ \hline
P-value & 8.3e-3 & 2.4e-5 & 1.2e-27 \\\hline
\end{tabular}
\begin{tablenotes}
{\textbf{Notes:} Same as Table \ref{tab:NGC612}.}
\end{tablenotes}
\end{threeparttable}
\end{table*}

\begin{table*}
\centering
\begin{threeparttable}
\label{tab:NGC4388}
\caption{NGC 4388 fitting results}
\renewcommand*{\arraystretch}{1.4}
\begin{tabular}{cccc}
{\bf Model} & {\bf MYTorus} & {\bf borus02} & {\bf UXCLUMPY} \\ \hline
$\chi^2_{\rm red}$ & 1.28 & 1.25 & 1.31 \\
$\chi^2$/d.o.f. & 6708/5224 & 6532/5224 & 6847/5225 \\
$T$ & 20.2$\sigma$ & 18.0$\sigma$ & 22.4$\sigma$\\\hline
kT & 0.28$^{+0.02}_{-0.02}$ & 0.26$^{+0.02}_{-0.02}$ & 0.27$^{+0.02}_{-0.02}$ \\
kT$_2$ & 0.70$^{+0.03}_{-0.04}$ & 0.68$^{+0.06}_{-0.04}$ & 0.69$^{+0.14}_{-0.06}$ \\
$N_{\rm H,apec}$  &0.59$^{+0.09}_{-0.10}$ & 0.62$^{+0.12}_{-0.17}$ & 0.60$^{+0.09}_{-0.09}$ \\ \hline
$\Gamma$ & 1.58$^{+0.01}_{-0.01}$ & 1.53$^{+0.02}_{-0.02}$ & 1.81$^{+0.03}_{-0.03}$  \\
$N_{H,av}$ & 0.10$^{+0.01}_{-0.01}$ & 0.12$^{+0.01}_{-0.01}$ & $-$ \\
A$_{S90}$ & 1.23$^{+0.20}_{-0.21}$ & $-$ & $-$  \\
A$_{S0}$ & 0.53$^{+0.12}_{-0.12}$ & $-$ & $-$  \\
$C_{\rm F}$ & $-$ & 0.52$^{+0.04}_{-0.04}$ & 0*  \\
Cos ($\theta_{Obs}$) & $-$ & 0.45$^{+0.03}_{-0.03}$ & 0.00$^{+0.14}_{-u}$  \\ 
$\sigma_{\rm tor}$ & $-$ & $-$ & 66.7$^{+8.7}_{-5.0}$  \\ 
F$_s$ (10$^{-3}$) & 1.01$^{+0.59}_{-0.52}$ & 0.84$^{+0.56}_{-0.54}$ & 11.5$^{2.0}_{-0.9}$ \\
norm (10$^{-2}$) & 1.54$^{+0.10}_{-0.10}$ & 1.40$^{+0.05}_{-0.05}$ & 2.41$^{+0.24}_{-0.14}$ \\
\hline 
$N_{H,Ch1}$ & 0.71$^{+0.03}_{-0.03}$ & 0.71$^{+0.04}_{-0.03}$ & 0.66$^{+0.08}_{-0.05}$  \\
$N_{H,xmm1}$ & 0.37$^{+0.01}_{-0.01}$ & 0.36$^{+0.02}_{-0.01}$ & 0.33$^{+0.01}_{-0.01}$  \\
$N_{H,xmm2}$ & 0.235$^{+0.003}_{-0.003}$ & 0.231$^{+0.003}_{-0.003}$ & 0.211$^{+0.002}_{-0.003}$  \\
$N_{H,Ch2}$ & 0.91$^{+0.05}_{-0.05}$ & 0.93$^{+0.05}_{-0.04}$ & 0.90$^{+0.04}_{-0.03}$  \\
$N_{H,nus1}$ & 0.30$^{+0.01}_{-0.01}$ & 0.29$^{+0.02}_{-0.02}$ & 0.26$^{+0.02}_{-0.02}$  \\
$N_{H,xmm3}$ & 0.267$^{+0.004}_{-0.004}$ & 0.260$^{+0.004}_{-0.004}$ & 0.243$^{+0.003}_{-0.003}$  \\
$N_{H,nus2}$ & 0.219$^{+0.004}_{-0.005}$ & 0.214$^{+0.004}_{-0.005}$ & 0.195$^{+0.003}_{-0.003}$  \\\hline
$C_{Ch}$ & 1* & 1* & 1* \\
$C_{xmm1}$ & 1.25$^{+0.07}_{-0.05}$ & 1.20$^{+0.06}_{-0.06}$ & $=C_{Ch2}$  \\
$C_{xmm2}$ & 1.57$^{+0.08}_{-0.07}$ & 1.53$^{+0.09}_{-0.08}$ & 1.55$^{+0.09}_{-0.08}$  \\
$C_{Ch2}$ & 1.11$^{+0.06}_{-0.06}$ & 1.13$^{+0.06}_{-0.05}$ & 1.16$^{+0.07}_{-0.06}$  \\
$C_{nus1}$ & 0.35$^{+0.02}_{-0.02}$ & 0.33$^{+0.02}_{-0.02}$ & 0.33$^{+0.02}_{-0.01}$  \\ 
$C_{xmm3}$ & 1.40$^{+0.07}_{-0.08}$ & 1.36$^{+0.07}_{-0.07}$ & 1.38$^{+0.08}_{-0.07}$  \\
$C_{nus2}$ & $=C_{xmm3}$ & $=C_{xmm3}$ & $=C_{xmm3}$ \\ \hline \hline
$\chi^2_{\rm red}$ No Var.  & 23.9 & 23.9 & 24.1 \\
$T$ & 1727$\sigma$ & 1727$\sigma$ & 1741$\sigma$\\ \hline
$\chi^2_{\rm red}$ No C Var. & 1.61 & 1.62 & 1.80\\
$T$ & 44.1$\sigma$ & 44.8$\sigma$ & 57.8$\sigma$\\ \hline
$\chi^2_{\rm red}$ No $N_{\rm H}$ Var. & 1.84 & 1.71 & 1.73\\
$T$ & 60.7$\sigma$ & 51.3$\sigma$ & 52.7$\sigma$\\ \hline
P-value & 0 & 0 & 0 \\\hline
\end{tabular}
\begin{tablenotes}
{\textbf{Notes:} Same as Table \ref{tab:NGC612}, with the following additions: \newline
kT$_2$: Second (hotter) \texttt{apec} component temperature, in units of keV. \newline
$N_{\rm H, apec}$: Obscuring column density associated to the second \texttt{apec} component, in units of 10$^{22}$ cm$^{-2}$.}
\end{tablenotes}
\end{threeparttable}
\end{table*}

\begin{table*}
\centering
\begin{threeparttable}
\label{tab:IC4518A}
\caption{IC 4518 A fitting results}
\renewcommand*{\arraystretch}{1.4}
\begin{tabular}{cccc}
{\bf Model} & {\bf MYTorus} & {\bf borus02} & {\bf UXCLUMPY} \\ \hline
$\chi^2_{\rm red}$ & 1.07 &  1.06 & 1.16  \\
$\chi^2$/d.o.f. & 413/386 & 408/385 & 448/385 \\ 
$T$ & 1.4$\sigma$ & 1.2$\sigma$ & 3.1$\sigma$ \\\hline
kT & 0.66$^{+0.03}_{-0.03}$ & 0.67$^{+0.03}_{-0.03}$ & 0.67$^{+0.03}_{-0.03}$ \\ \hline
$\Gamma$ & 1.91$^{+0.15}_{-0.14}$ & 1.84$^{+0.09}_{-0.08}$ & 1.76$^{+0.03}_{-0.06}$  \\
$N_{H,av}$ & 3.46$^{-u}_{-1.29}$ & 14.0$^{-u}_{-11.1}$ & $-$ \\
A$_{S90}$ & 0* & $-$ & $-$  \\
A$_{S0}$ & 2.65$^{+0.75}_{-0.58}$ & $-$ & $-$  \\
$C_{\rm F}$ & $-$ & 0.87$^{+0.02}_{-0.19}$ & 0.29$^{+0.03}_{-0.09}$  \\
Cos ($\theta_{Obs}$) & $-$ & 0.95$^{-u}_{-0.57}$ & 0.50$^{+0.42}_{-0.24}$ \\ 
$\sigma_{\rm tor}$ & $-$ & $-$ & 84.0$^{-u}_{-0.14}$  \\ 
F$_s$ (10$^{-2}$) & 1.22$^{+0.46}_{-0.37}$ & 1.26$^{+0.25}_{-0.34}$ & 23.5$^{+0.30}_{-0.62}$ \\
norm (10$^{-3}$) & 2.18$^{+0.85}_{-0.60}$ & 1.85$^{+0.46}_{-0.31}$ & 2.19$^{+0.27}_{-0.13}$ \\
\hline 
$N_{H,xmm1}$ & 0.21$^{+0.02}_{-0.02}$ & 0.21$^{+0.02}_{-0.01}$ & 0.21$^{+0.08}_{-0.06}$  \\
$N_{H,xmm2}$ & 0.31$^{+0.04}_{-0.03}$ & 0.33$^{+0.03}_{-0.03}$ & 0.32$^{+0.01}_{-0.02}$  \\
$N_{H,nus}$ & 0.14$^{+0.04}_{-0.03}$ & 0.15$^{+0.04}_{-0.03}$ & 0.13$^{+0.02}_{-0.02}$  \\ \hline
$C_{xmm1}$ & 1* & 1* & 1* \\
$C_{xmm2}$ & 0.88$^{+0.06}_{-0.06}$ & 0.90$^{+0.06}_{-0.06}$ & 0.93$^{+0.05}_{-0.05}$  \\
$C_{nus}$ & 1.45$^{+0.15}_{-0.13}$ & 1.49$^{+0.15}_{-0.14}$ & 1.44$^{+0.10}_{-0.05}$  \\ 
\hline \hline
$\chi^2_{\rm red}$ No Var.  & 2.66 & 2.94 & 3.04 \\
$T$ & 32.7$\sigma$ & 38.3$\sigma$ & 40.2$\sigma$\\ \hline
$\chi^2_{\rm red}$ No C Var. & 1.25 & 1.24 & 1.27\\
$T$ & 4.9$\sigma$ & 4.7$\sigma$ & 5.3$\sigma$\\ \hline
$\chi^2_{\rm red}$ No $N_{\rm H}$ Var. & 1.33 & 1.32 & 1.43\\
$T$ & 6.5$\sigma$ & 6.3$\sigma$ & 8.5$\sigma$\\ \hline
P-value & 3.6e-2 & 1.8e-2 & 1.7e-5 \\\hline
\end{tabular}
\begin{tablenotes}
{\textbf{Notes:} Same as Table \ref{tab:NGC612}.}
\end{tablenotes}
\end{threeparttable}
\end{table*}

\begin{table*}
\centering
\begin{threeparttable}
\label{tab:3C445}
\caption{3C 445 fitting results}
\renewcommand*{\arraystretch}{1.4}
\begin{tabular}{cccc}
{\bf Model} & {\bf MYTorus} & {\bf borus02} & {\bf UXCLUMPY} \\ \hline
$\chi^2_{\rm red}$ & 1.02 & 1.03 & 1.00  \\
$\chi^2$/d.o.f. & 2220/2178 & 2248/2177 & 2180/2178 \\
$T$ & 0.9$\sigma$ & 1.4$\sigma$ & 0.0$\sigma$\\\hline
kT & 0.62$^{+0.04}_{-0.04}$ & 0.56$^{+0.09}_{-0.08}$ & 0.56$^{+0.10}_{-0.31}$ \\
kT$_2$ & 0.71$^{+0.56}_{-0.24}$ & 1.63$^{+0.09}_{-0.09}$ & 1.29$^{+0.33}_{-0.09}$ \\ 
$N_{\rm H,apec}$ & 26.1$^{+5.7}_{-5.4}$ & 5.14$^{+0.16}_{-0.15}$ & 6.04$^{+0.65}_{-0.74}$ \\ \hline
$\Gamma$ & 1.75$^{+0.07}_{-0.07}$ & 1.62$^{+0.01}_{-0.01}$ & 1.60$^{+0.04}_{-0.03}$  \\
$N_{H,av}$ & 0.14$^{+0.02}_{-0.01}$ & 0.13$^{+0.02}_{-0.03}$ & $-$ \\
A$_{S90}$ & 7.99$^{+5.70}_{-u}$ & $-$ & $-$  \\
A$_{S0}$ & 4.26$^{+7.29}_{-u}$ & $-$ & $-$  \\
$C_{\rm F}$ & $-$ & 0.93$^{+0.04}_{-0.03}$ & 0*  \\
Cos ($\theta_{Obs}$) & $-$ & 0.95$^{-u}_{-0.02}$ & 0.00$^{-u}_{-u}$  \\ 
$\sigma_{\rm tor}$ & $-$ & $-$ & 84.0$^{-u}_{-5.9}$  \\ 
F$_s$ (10$^{-2}$) & 0.60$^{+0.41}_{-0.37}$ & 1.96$^{+0.16}_{-0.06}$ & 21.8$^{+2.3}_{-3.1}$ \\
norm (10$^{-3}$) & 4.36$^{+0.95}_{-1.11}$ & 2.76$^{+0.03}_{-0.03}$ & 3.31$^{+0.24}_{-0.20}$ \\
\hline
$N_{H,xmm}$ & 0.28$^{+0.03}_{-0.03}$ & 0.24$^{+0.01}_{-0.01}$ & 0.20$^{+0.01}_{-0.01}$  \\
$N_{H,Ch1}$ & 0.26$^{+0.03}_{-0.01}$ & 0.23$^{+0.01}_{-0.01}$ & 0.22$^{+0.02}_{-0.01}$  \\
$N_{H,nus}$ & 0.33$^{+0.03}_{-0.03}$ & 0.29$^{+0.01}_{-0.01}$ & 0.13$^{+0.01}_{-0.02}$  \\
$N_{H,Ch2}$ & 0.33$^{+0.03}_{-0.03}$ & 0.30$^{+0.01}_{-0.01}$ & 0.25$^{+0.02}_{-0.02}$  \\
$N_{H,Ch3}$ & 0.32$^{+0.03}_{-0.03}$ & 0.28$^{+0.01}_{-0.01}$ & 0.24$^{+0.01}_{-0.01}$  \\
$N_{H,Ch4}$ & 0.33$^{+0.03}_{-0.03}$ & 0.28$^{+0.01}_{-0.01}$ & 0.25$^{+0.01}_{-0.01}$  \\
$N_{H,Ch5}$ & 0.31$^{+0.02}_{-0.02}$ & 0.27$^{+0.01}_{-0.01}$ & 0.26$^{+0.01}_{-0.01}$  \\
 \hline
$C_{xmm}$ & $=C_{Ch4}$ & $=C_{Ch4}$ & $=C_{Ch4}$  \\
$C_{Ch1}$ & 1* & 1* & 1* \\
$C_{nus}$ & $=C_{Ch2}$ & $=C_{Ch2}$ & 0.77$^{+0.05}_{-0.05}$  \\
$C_{Ch2}$ & 1.16$^{+0.07}_{-0.06}$ & 1.14$^{+0.03}_{-0.03}$ & 1.11$^{+0.05}_{-0.05}$  \\
$C_{Ch3}$ & $=C_{Ch2}$ & $=C_{Ch2}$ & $=C_{Ch2}$  \\
$C_{Ch4}$ & 1.26$^{+0.08}_{-0.05}$ & 1.21$^{+0.02}_{-0.02}$ & 1.21$^{+0.05}_{-0.04}$  \\
$C_{Ch5}$ & $=C_{Ch2}$ & $=C_{Ch2}$ & $=C_{Ch2}$  \\
 \hline \hline
$\chi^2_{\rm red}$ No Var. & 1.16 & 1.18 &  1.18\\
$T$ & 7.5$\sigma$ & 8.4$\sigma$ & 8.4$\sigma$\\ \hline
$\chi^2_{\rm red}$ No C Var. & 1.04 & 1.06 & 1.07\\
$T$ & 1.9$\sigma$ & 2.8$\sigma$ & 3.3$\sigma$\\ \hline
$\chi^2_{\rm red}$ No $N_{\rm H}$ Var. & 1.03 & 1.05 & 1.06\\
$T$ & 1.4$\sigma$ & 2.3$\sigma$ & 2.8$\sigma$\\ \hline
P-value & 9.9e-1 & 6.3e-1 & 2.7e-3 \\\hline
\end{tabular}
\begin{tablenotes}
{\textbf{Notes:} Same as Table \ref{tab:NGC612}, with the following additions: \newline
kT$_2$: Second (hotter) \texttt{apec} component temperature, in units of keV. \newline
$N_{\rm H, apec}$: Obscuring column density associated to the second \texttt{apec} component, in units of 10$^{22}$ cm$^{-2}$.}
\end{tablenotes}
\end{threeparttable}
\end{table*}

\begin{table*}
\centering
\begin{threeparttable}
\label{tab:NGC7319}
\caption{NGC 7319 fitting results}
\renewcommand*{\arraystretch}{1.4}
\begin{tabular}{cccc}
{\bf Model} & {\bf MYTorus} & {\bf borus02} & {\bf UXCLUMPY} \\ \hline
$\chi^2_{\rm red}$ & 1.08 & 1.07 &  1.10 \\
$\chi^2$/d.o.f. & 542.71/501 & 538.10/501 & 553.84/502 \\
$T$ & 1.8$\sigma$ & 1.6$\sigma$ & 2.2$\sigma$ \\ \hline
kT & 0.41$^{+0.11}_{-0.09}$ & 0.35$^{+0.10}_{-0.06}$ & 0.34$^{+0.06}_{-0.06}$ \\
kT$_2$ & 0.73$^{+0.16}_{-0.12}$ & 0.67$^{+0.15}_{-0.07}$ & 0.66$^{+0.14}_{-0.06}$ \\
$N_{\rm H,apec}$ & 0.72$^{+0.14}_{-0.20}$ & 0.71$^{+0.13}_{-0.14}$ & 0.72$^{+0.09}_{-0.09}$ \\ \hline
$\Gamma$ & 1.73$^{+0.15}_{-0.17}$ & 1.75$^{+0.15}_{-0.14}$ & 2.04$^{+0.22}_{-0.13}$  \\
$N_{H,av}$ & 0.25$^{+0.07}_{-0.04}$ & 0.33$^{+0.09}_{-0.07}$ & $-$ \\
A$_{S90}$ & 0.95$^{+0.30}_{-0.44}$ & $-$ & $-$  \\
A$_{S0}$ & 0.15$^{+0.27}_{-u}$ & $-$ & $-$  \\
$C_{\rm F}$ & $-$ & 0.31$^{+0.06}_{-0.04}$ & 0*  \\
Cos ($\theta_{Obs}$) & $-$ & 0.26$^{+0.03}_{-0.04}$ & 0.00$^{-u}_{-u}$  \\ 
$\sigma_{\rm tor}$ & $-$ & $-$ & 77.9$^{-u}_{-10.7}$  \\ 
F$_s$ (10$^{-4}$) & 9.78$^{+10.0}_{-9.61}$ & 3.23$^{+9.88}_{-u}$ & 0* \\
norm (10$^{-3}$) & 3.55$^{+0.15}_{-0.12}$ & 3.70$^{+1.59}_{-1.03}$ & 7.92$^{+2.96}_{-2.50}$ \\
\hline
$N_{H,xmm}$ & 0.87$^{+0.05}_{-0.05}$ & 0.87$^{+0.06}_{-0.05}$ & 0.84$^{+0.07}_{-0.08}$  \\
$N_{H,Ch1}$ & 0.46$^{+0.04}_{-0.04}$ & 0.47$^{+0.04}_{-0.04}$ & 0.47$^{+0.04}_{-0.05}$  \\
$N_{H,Ch2}$ & 0.46$^{+0.03}_{-0.03}$ & 0.47$^{+0.03}_{-0.03}$ & 0.46$^{+0.03}_{-0.05}$  \\
$N_{H,nus1}$ & 2.17$^{+0.36}_{-0.26}$ & 2.11$^{+0.26}_{-0.22}$ & 0.71$^{+0.25}_{-0.15}$  \\ 
$N_{H,nus2}$ & 1.78$^{+0.34}_{-0.34}$ & 1.73$^{+0.30}_{-0.32}$ & 0.98$^{+0.14}_{-0.17}$  \\\hline
$C_{xmm}$ & 1.31$^{+0.08}_{-0.08}$ & 1.32$^{+0.09}_{-0.08}$ & 1.29$^{+0.09}_{-0.08}$  \\
$C_{Ch1}$ & 1* & 1* & 1* \\
$C_{Ch2}$ & $=C_{Ch1}$ & $=C_{Ch1}$ & $=C_{Ch1}$ \\
$C_{nus1}$ & $=C_{Ch1}$ & $=C_{Ch1}$ & 0.32$^{+0.11}_{-0.07}$  \\ 
$C_{nus2}$ & 0.83$^{+0.13}_{-0.16}$ & 0.85$^{+0.13}_{-0.15}$ & 0.44$^{+0.08}_{-0.08}$  \\ \hline \hline
$\chi^2_{\rm red}$ No Var.  & 5.44 & 5.47 & 5.71 \\
$T$ & 99.9$\sigma$ & 100$\sigma$ & 106$\sigma$\\ \hline
$\chi^2_{\rm red}$ No C Var. & 1.19 & 1.19 & 1.20\\
$T$ & 4.3$\sigma$ & 4.3$\sigma$ & 4.5$\sigma$\\ \hline
$\chi^2_{\rm red}$ No $N_{\rm H}$ Var. & 1.91 & 1.88 & 1.92\\
$T$ & 20.4$\sigma$ & 19.7$\sigma$ & 20.6$\sigma$\\ \hline
P-value & 5.3e-46 & 4.5e-42 & 8.0e-5 \\\hline
\end{tabular}
\begin{tablenotes}
{\textbf{Notes:} Same as Table \ref{tab:NGC612}, with the following additions: \newline
kT$_2$: Second (hotter) \texttt{apec} component temperature, in units of keV. \newline
$N_{\rm H, apec}$: Obscuring column density associated to the second \texttt{apec} component, in units of 10$^{22}$ cm$^{-2}$.}
\end{tablenotes}
\end{threeparttable}
\end{table*}

\begin{table*}
\centering
\begin{threeparttable}
\label{tab:3C452}
\caption{3C 452 fitting results}
\renewcommand*{\arraystretch}{1.4}
\begin{tabular}{cccc}
{\bf Model} & {\bf MYTorus} & {\bf borus02} & {\bf UXCLUMPY} \\ \hline
$\chi^2_{\rm red}$  & 1.03 & 1.03 & 1.08 \\
$\chi^2$/d.o.f. & 1394/1353  & 1388/1352 & 1459/1353 \\
$T$ & 1.1$\sigma$ & 1.1$\sigma$ & 2.9$\sigma$ \\\hline
kT & $-$ & $-$ & $-$\\ \hline
$\Gamma$ & 1.53$^{+0.05}_{-0.05}$ & 1.42$^{+0.03}_{-u}$& 1.57$^{+0.01}_{-0.01}$  \\
$N_{H,av}$ & 0.05$^{+0.01}_{-0.01}$ & 0.06$^{+0.01}_{-0.01}$ & $-$ \\
A$_{S90}$ & 2.55$^{+0.46}_{-0.40}$ & $-$ & $-$  \\
A$_{S0}$ & 0* & $-$& $-$  \\
$C_{\rm F}$ & $-$ & 1.00$^{-u}_{-0.10}$  & 0*  \\
Cos ($\theta_{Obs}$) & $-$& 0.00$^{+0.13}_{-u}$ & 1.00$^{-u}_{-0.73}$  \\
$\sigma_{\rm tor}$& $-$ &  $-$ &  7.10$^{+22.41}_{-0.10}$ \\
norm$(10^{-3})$ & 2.24$^{+0.41}_{-0.32}$& 1.72$^{+0.02}_{-0.18}$ & 1.87$^{+0.06}_{-0.06}$  \\
$\Gamma_{\rm jet}$ & 1.40$^{+0.19}_{-0.18}$ & 1.36$^{+0.09}_{-0.09}$ & 0.75$^{+0.06}_{-0.05}$\\
\hline 
$N_{H,ch}$ & 0.55$^{+0.03}_{-0.03}$ & 0.52$^{+0.02}_{-0.03}$ & 0.44$^{+0.03}_{-0.02}$  \\
$N_{H,xmm}$ & 0.52$^{+0.03}_{-0.03}$ & 0.49$^{+0.01}_{-0.03}$ &  0.46$^{+0.02}_{-0.02}$ \\
$N_{H,nus}$ & 0.39$^{+0.03}_{-0.03}$  & 0.36$^{+0.01}_{-0.02}$ & 0.28$^{+0.01}_{-0.01}$\\ \hline
norm $_{jet,ch}(10^{-6})$ & 8.26$^{+1.01}_{-1.01}$ & 7.52$^{+0.82}_{-0.82}$ & 8.13$^{+0.73}_{-0.73}$\\
norm$_{jet,xmm} (10^{-5})$ & 2.46$^{+0.47}_{-0.37}$ & 2.00$^{+0.08}_{-0.08}$ & 2.40$^{+0.63}_{-0.03}$\\
 norm$_{jet,nus}$ & =norm$_{jet,xmm}$ & =norm$_{jet,xmm}$ & =norm$_{jet,xmm}$ \\ \hline \hline
$\chi^2_{\rm red}$ No Var. & 1.50 & 1.49 & 1.55 \\
$T$ & 18.4$\sigma$ & 18.0$\sigma$ & 20.2$\sigma$\\ \hline
$\chi^2_{\rm red}$ No C Var. & 1.25 & 1.25 & 1.31\\
$T$ & 9.2$\sigma$ & 9.2$\sigma$ & 11.4$\sigma$\\ \hline
$\chi^2_{\rm red}$ No $N_{\rm H}$ Var. & 1.25 & 1.26 & 1.33\\
$T$ & 9.2$\sigma$ & 9.6$\sigma$ & 12.1$\sigma$\\ \hline
P-value & 1.4e-3 & 1.9e-16 & 2.5e-8 \\ \hline
\end{tabular}
\begin{tablenotes}
{\textbf{Notes:} Same as Table \ref{tab:NGC612}, with the following additions: \newline
norm$_{jet,instrument}$: Variable normalization on the added jet component required to model the source.}
\end{tablenotes}
\end{threeparttable}
\end{table*}

\clearpage

\section{Source Spectra}\label{App:spectra}

\begin{figure*}[h]
\begin{center}
\hbox{
\includegraphics[scale=0.35,angle=-90]{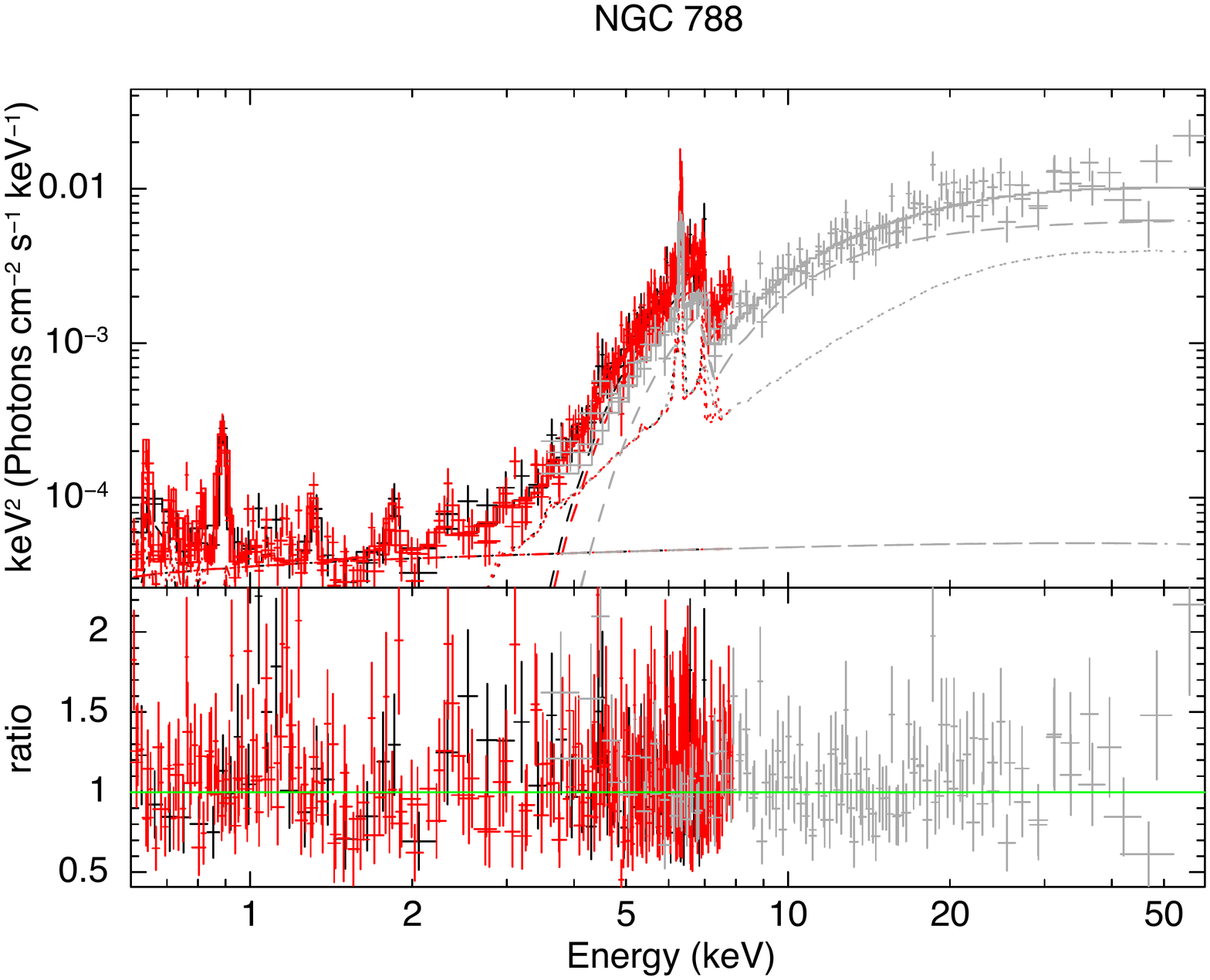}
\hspace{5mm}
\includegraphics[scale=0.35,angle=-90]{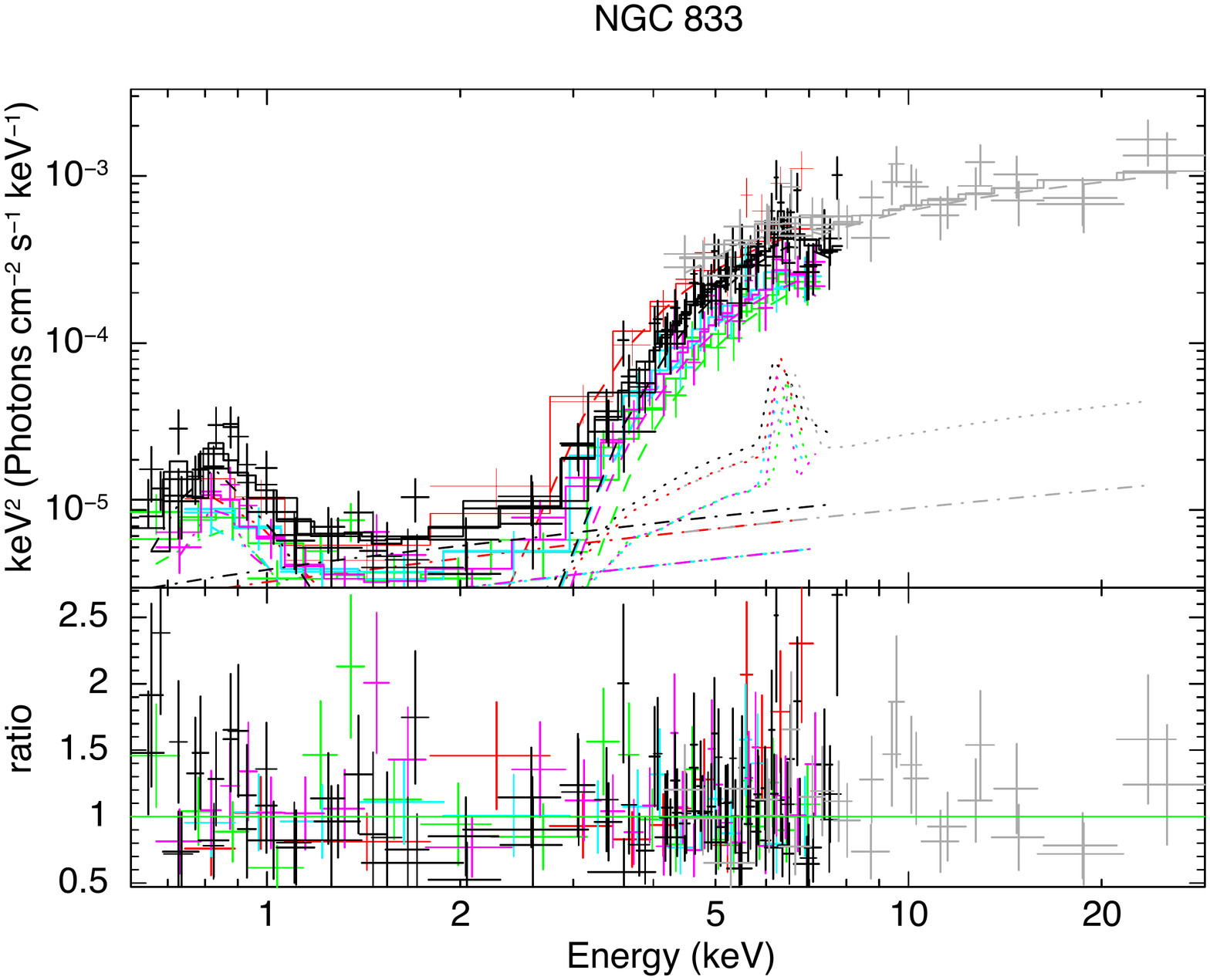}
}
\hbox{
\includegraphics[scale=0.35,angle=-90]{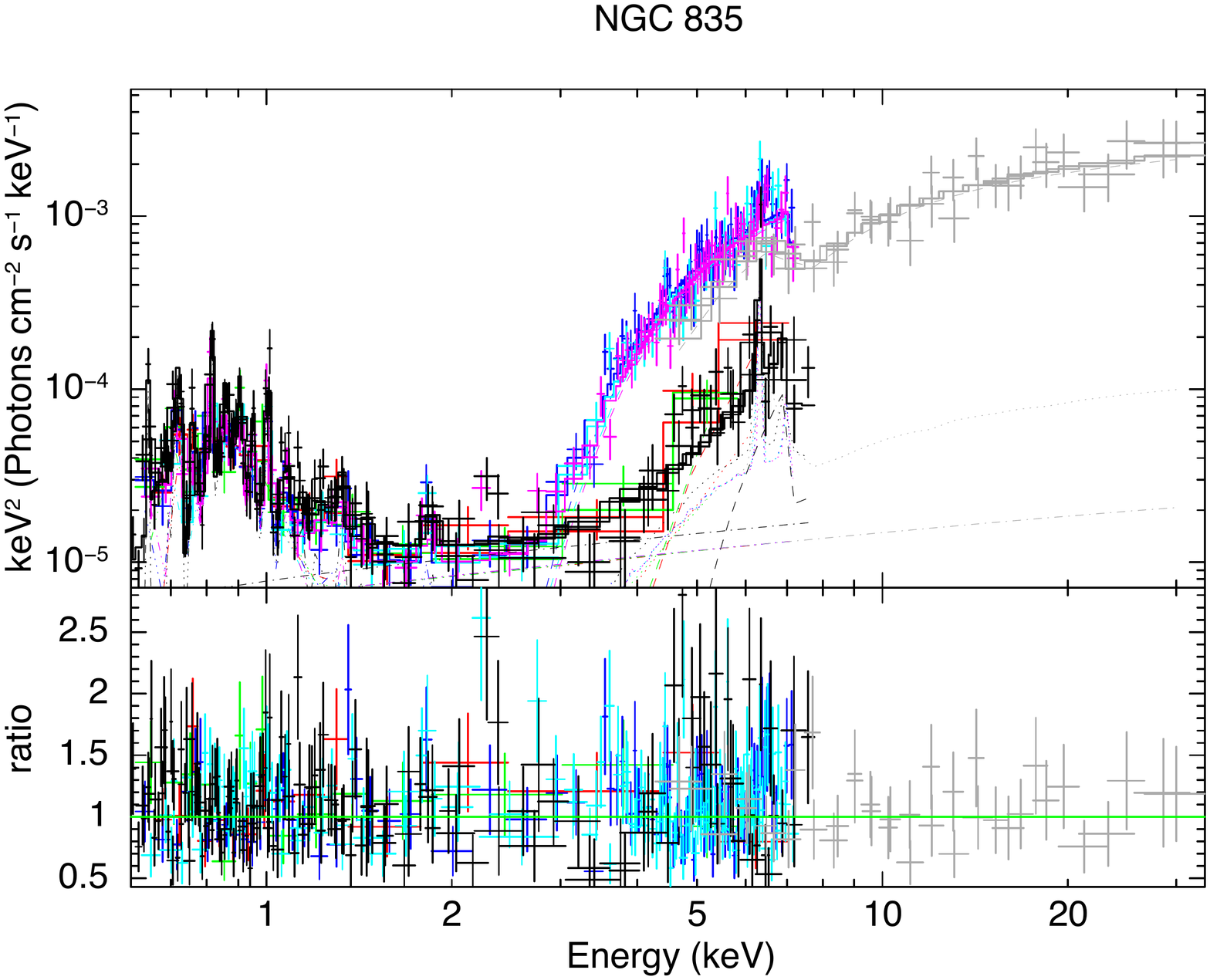}
\hspace{5mm}
\includegraphics[scale=0.35,angle=-90]{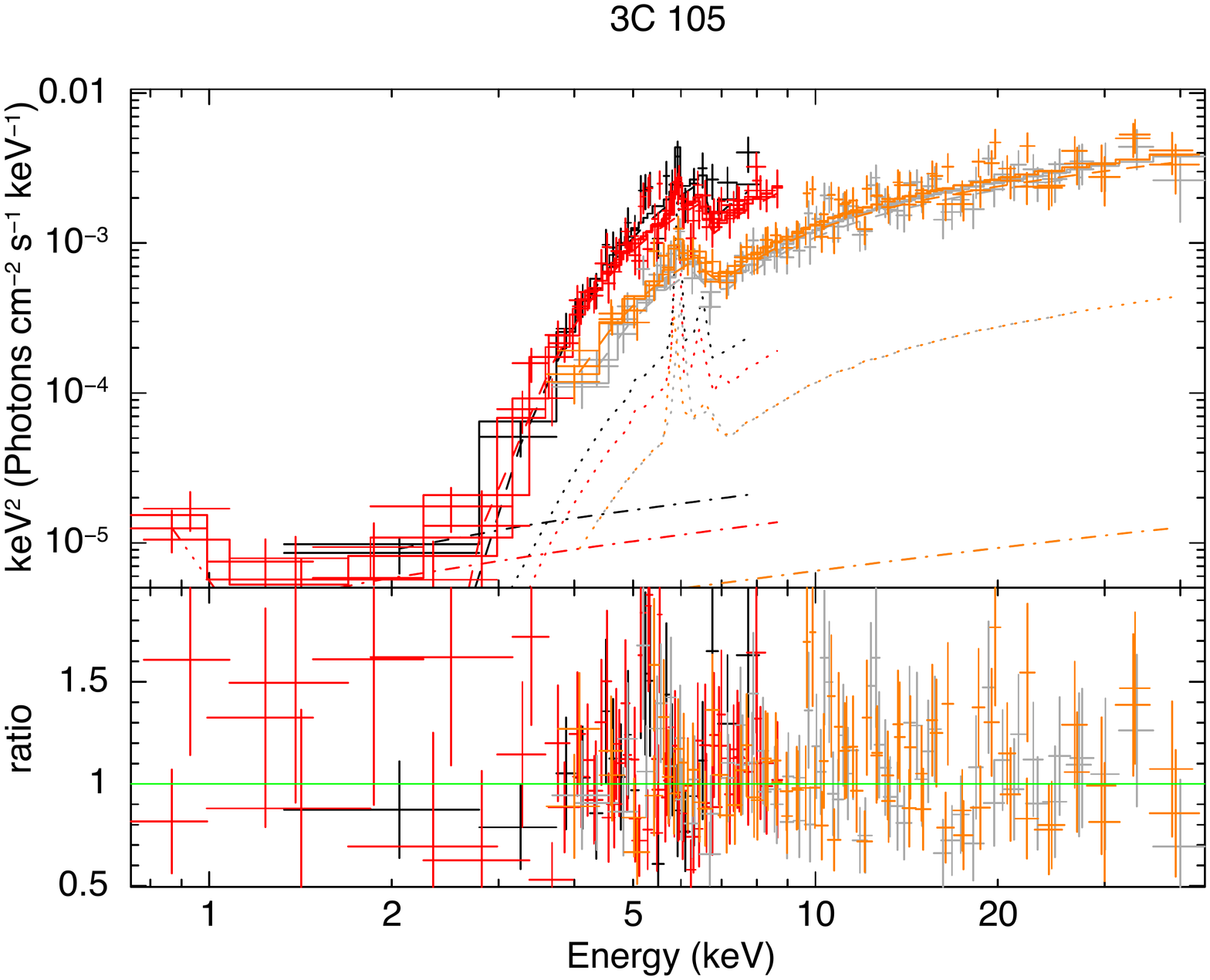}
}
\hbox{
\includegraphics[scale=0.35,angle=-90]{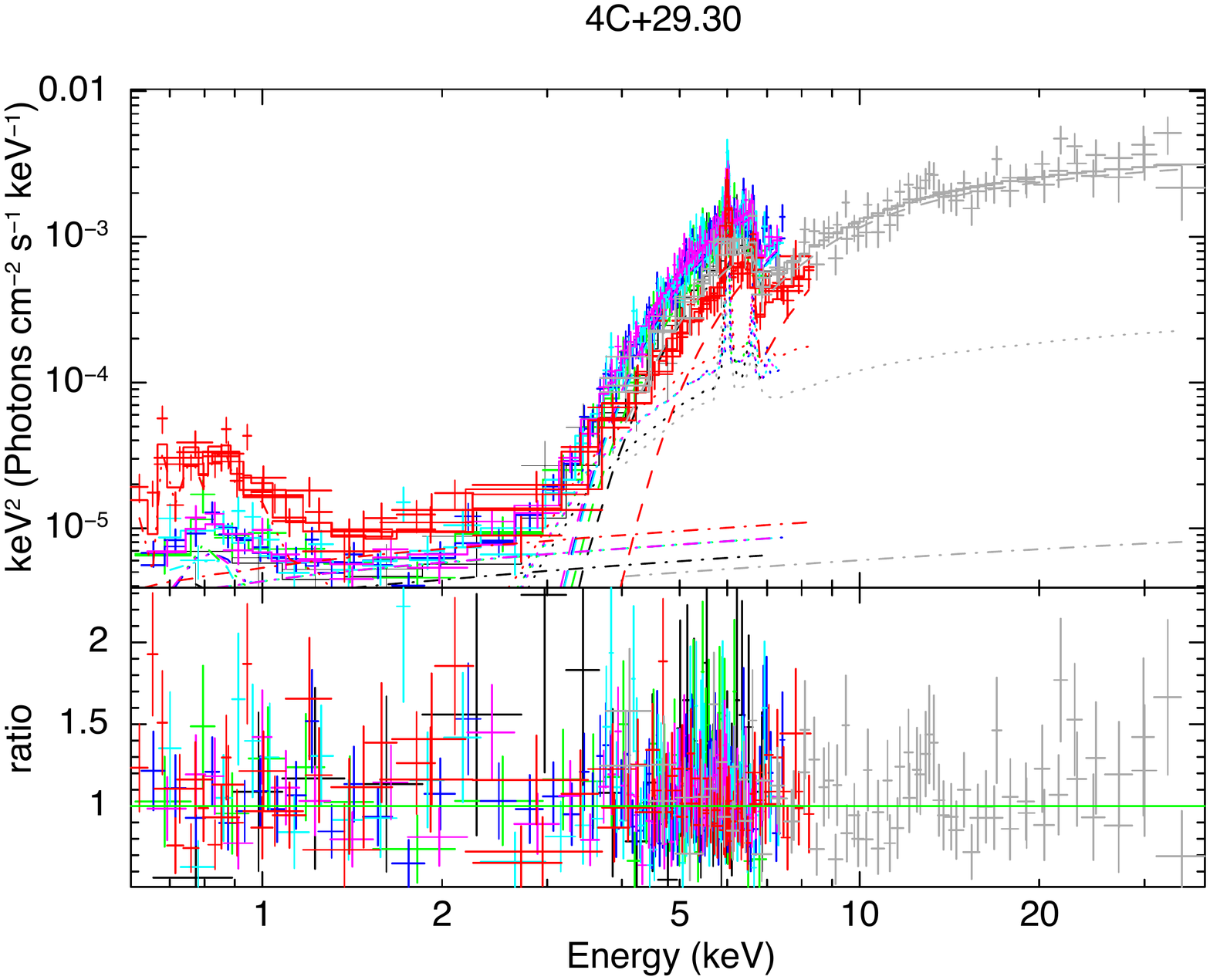}
\hspace{5mm}
\includegraphics[scale=0.35,angle=-90]{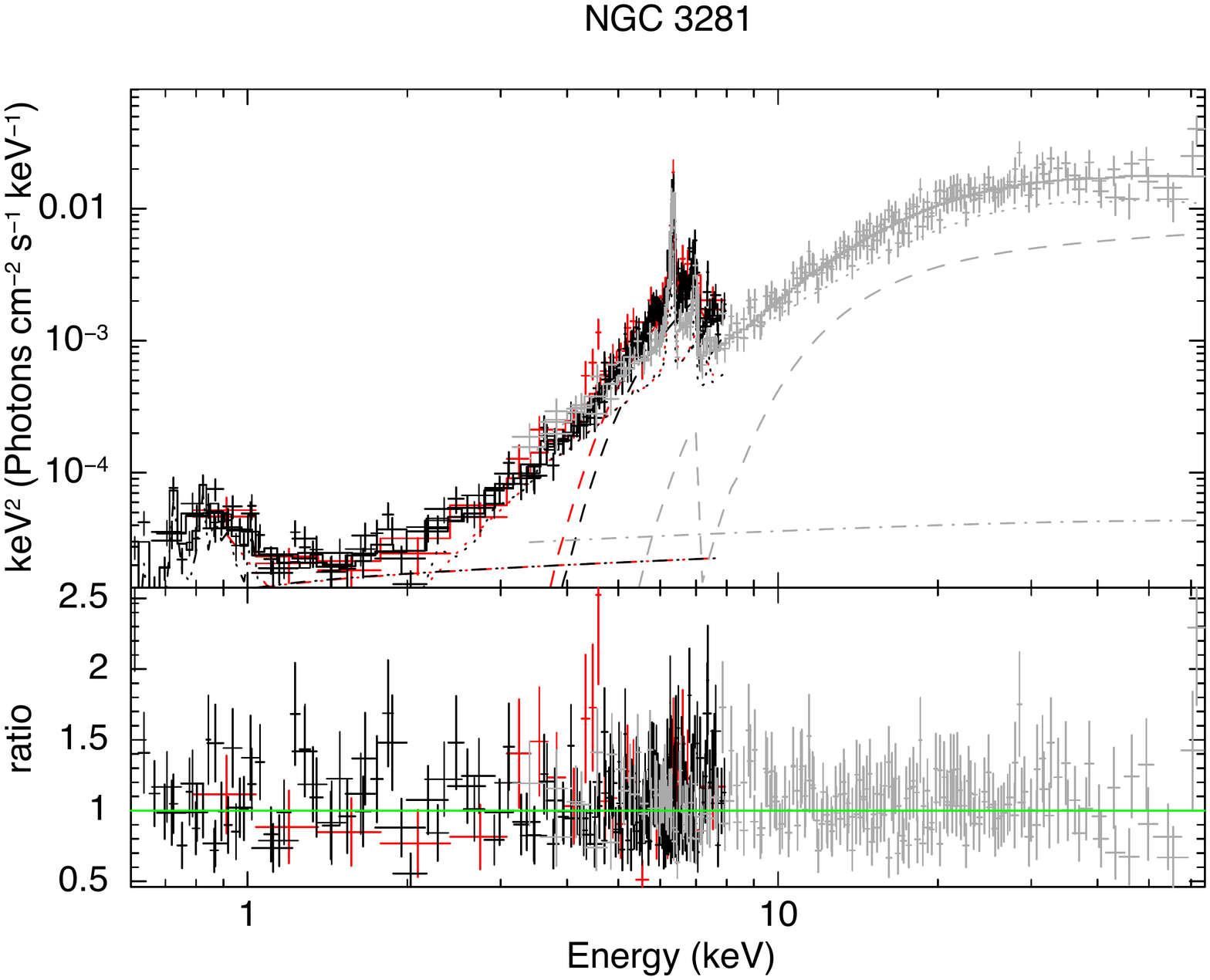}
}
\caption{From left to right, top to bottom: \bor fits to the data for NGC 788, NGC 833, NGC 835, 3C 105, 4C+29.30, NGC 3281. Color code is as explained in Appendix \ref{App:spectra}.}
\label{fig:spectra1}
\end{center}
\end{figure*}

\begin{figure*}[h]
\begin{center}
\hbox{
\includegraphics[scale=0.35,angle=-90]{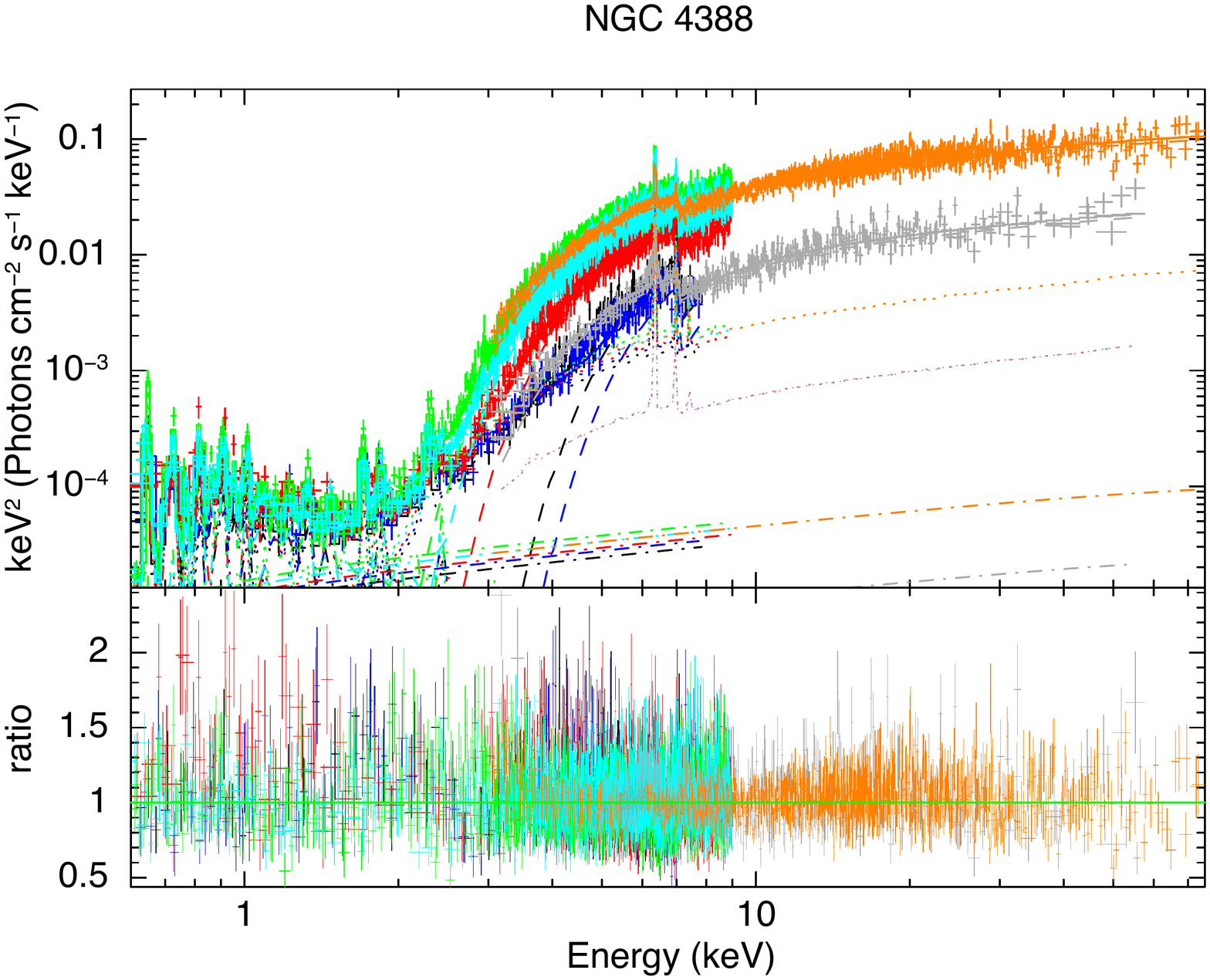}
\hspace{5mm}
\includegraphics[scale=0.35,angle=-90]{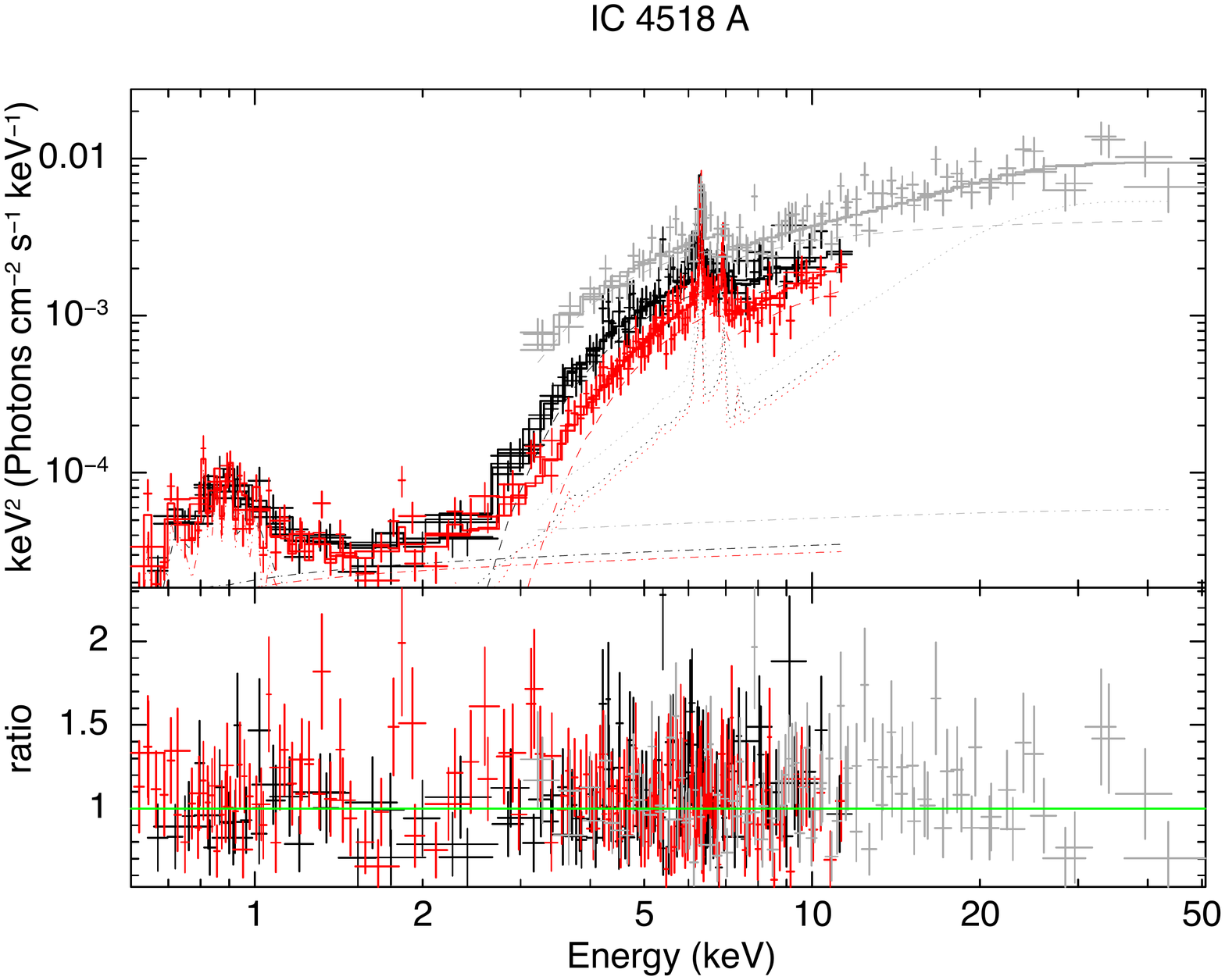}
}
\hbox{
\includegraphics[scale=0.35,angle=-90]{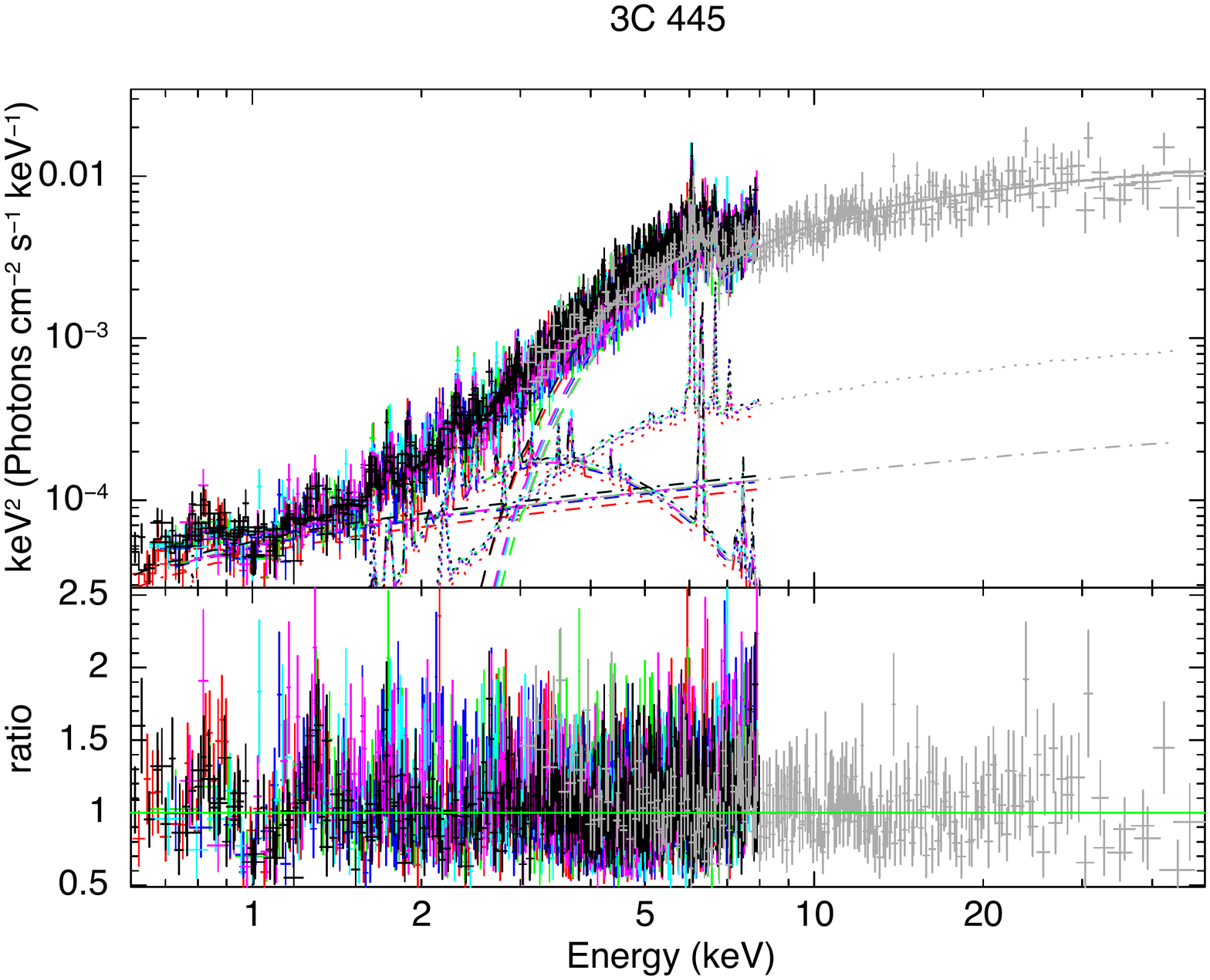}
\hspace{5mm}
\includegraphics[scale=0.35,angle=-90]{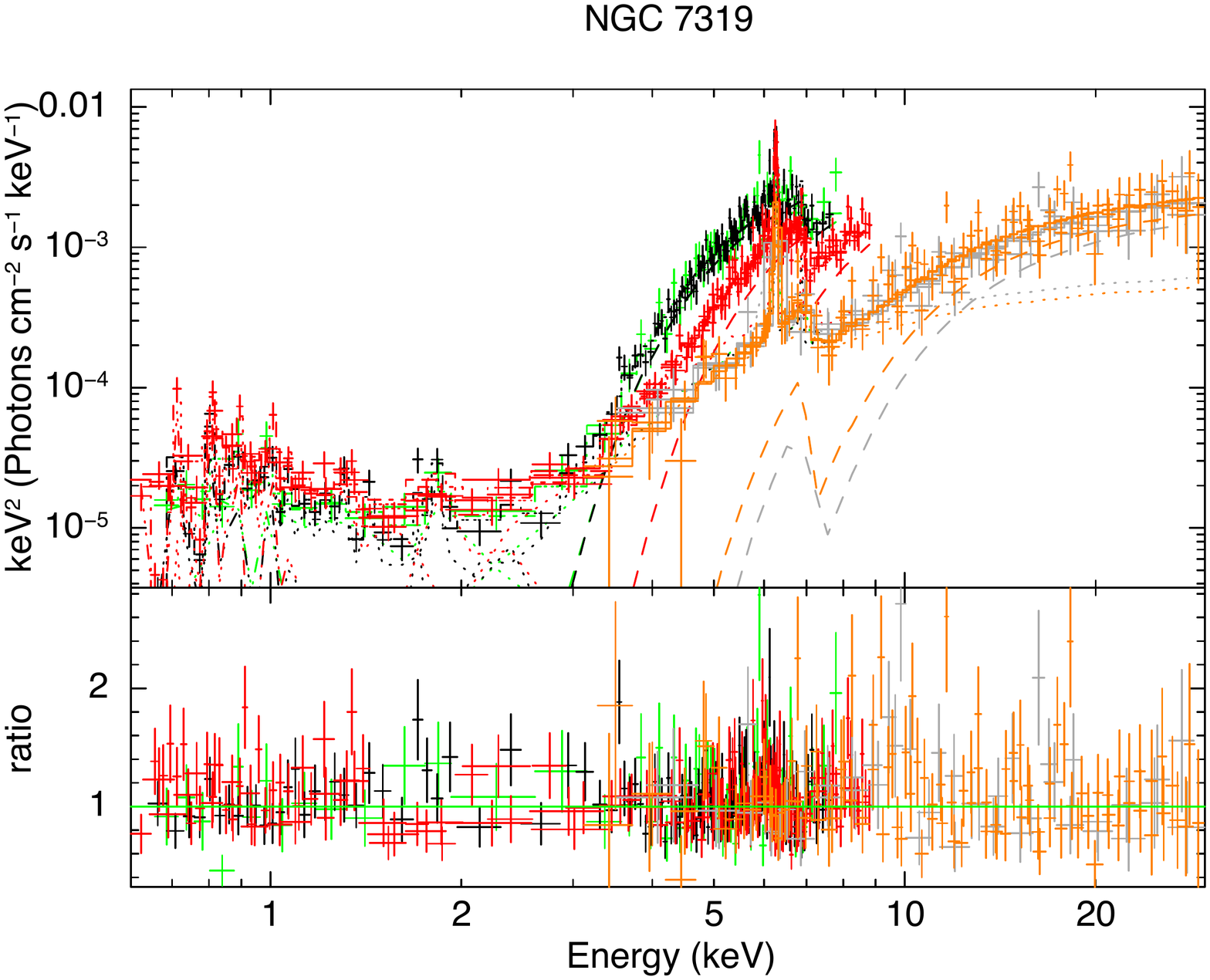}
}
\hbox{
\includegraphics[scale=0.35,angle=-90]{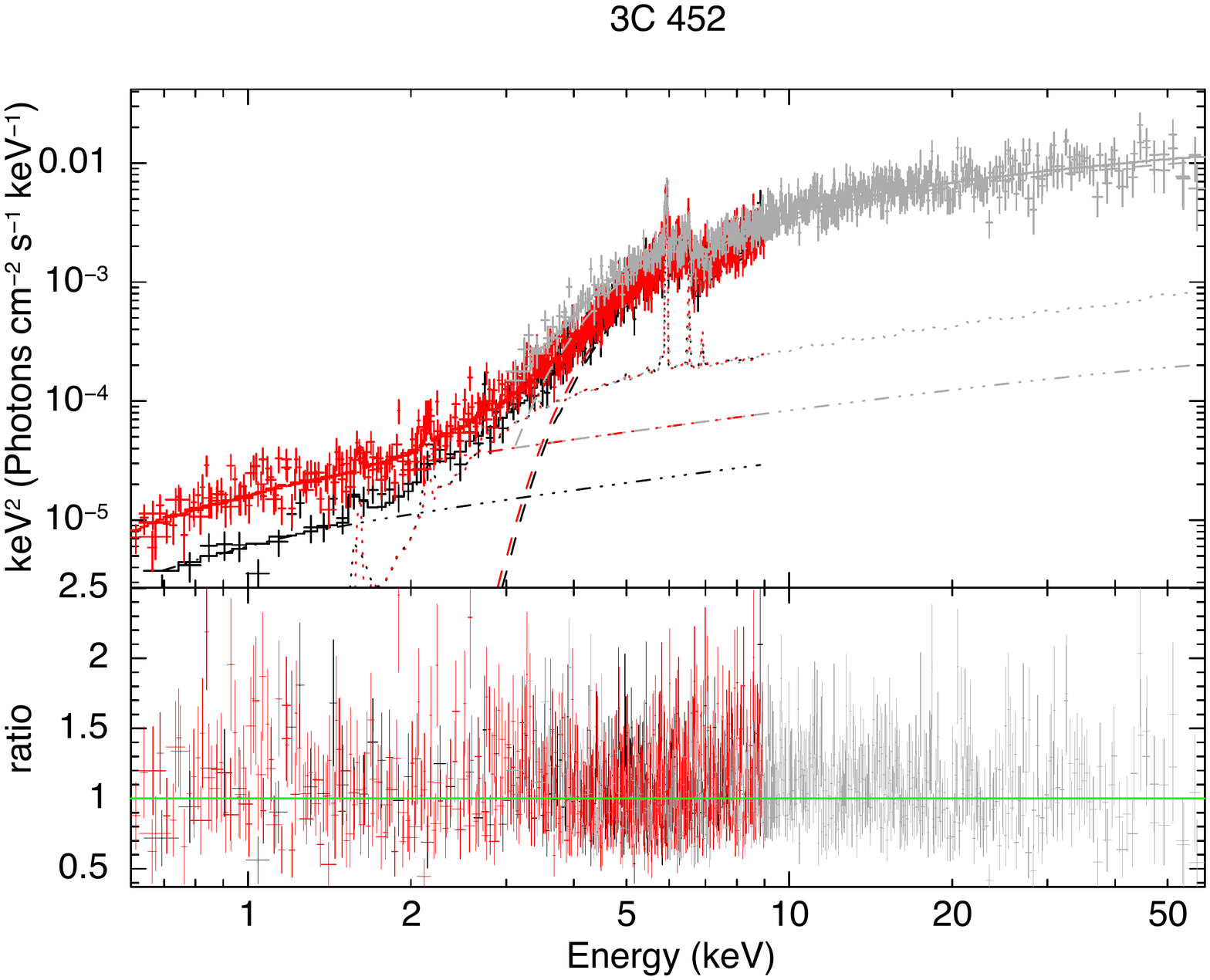}
\hspace{5mm}
}
\caption{From left to right, top to bottom: \bor fits to the data for NGC 4388, IC 4518 A, 3C 445, NGC 7319 and 3C 452. Color code is as explained in Appendix \ref{App:spectra}.}
\label{fig:spectra2}
\end{center}
\end{figure*}

In this section we present the best fit \bor models to the multiepoch spectra of all sources in the sample, shown in Figs. \ref{fig:spectra1} and \ref{fig:spectra2}. We opt to show the \bor fits over those of the other models, since \myt has a reflection component divided into four different individual sub-components, which makes the spectra much more difficult to interpret. \uxc, on the other hand, does not show a distinction between l.o.s. and reflection components, therefore providing less information in the spectral decomposition. The spectra shown in Figs. \ref{fig:spectra1} and \ref{fig:spectra2} should be read as follows: 
\begin{itemize}
\setlength\itemsep{1em}
    \item All observations for a single source are shown together, each one in a different color. Meaning, all detectors in the same telescope are colored the same in each individual observation (i.e. MOS1, MOS2, PN for \xmm, and FPMA, FPMB for \nustar).
    \item Soft band observations (\xmm and \chandra) are colored chronologically, as listed in Tables \ref{tab:NGC612} and \ref{tab:NGC788}-\ref{tab:3C452}. The color order is as follows, from first to last observation: Black, red, green, blue, cyan, magenta.
    \item Hard band observations (i.e. \nustar) are colored, also chronologically, but separated from the soft-band observations. This is done to avoid confusion between different bands. From first to second, the colors are: Grey, orange.
    \item For each individual observation, we plot the overall best-fit model as a solid line, the l.o.s. component as a dashed line, the reflection as a dotted line, the scattering as a dot-dash line, and the soft emission component (single or double \texttt{mekal} and any added lines) as a dash-dot-dot-dot. We note that 3C 452 has a jet component instead of a soft component + scattering, and we use a dash-dot-dot-dot (equivalent to the soft emission component) to represent it.
\end{itemize}

\clearpage

\section{Comments on Individual Sources}\label{App:sources}

In this section we provide a detailed explanation about specific analysis and fitting details for each source, that may deviate (or need clarification) from the methods described in sections \ref{Sample} and \ref{Results}. We also comment on the fitting results for each specific source, add comments on model comparison if discrepancies are present, and compare the obtained fitting parameters to those obtained by \citet{Zhao2020}, from which this sample is selected, and who used \texttt{borus02} on only two observations per source.

\subsection{NGC 612}

\textbf{Data reduction/fitting:} 
C-statistic was used to fit Chandra observations 1 and 2, given how the data quality forced us to bin them with 3 and 5 cts/bin, respectively. Table \ref{tab:NGC612} thus refers to Stat. (total statistic, a mix of $\chi^2$ and C-statistic) instead of $\chi^2$. \texttt{apec} was applied to model solely the \xmm data, as the \chandra data did not show any excess (again, probably due to the lower quality data).
%The XMM and Chandra 2 observations have the same flux according to the model, and linking/un-linking them doesn't affect the fit (doesn't improve it). However, the two chandra emissions look very similar to each other, and so it was interesting to see the error ranges of the un-linked Chandra 2 flux. This way, the two Chandra observations are almost compatible with each other.

\textbf{Analysis of results:} All models fit this source well, and our results are compatible with those derived by \citet{Zhao2020}. The best-fit values for the torus parameters are in good agreement, within errors, for all models. However, that is not the case when it comes to the variability determination. While all models require some form of variability (T$>10\sigma$ for the non-variability scenario), \texttt{MYTorus} is not able to discern between a pure $N_{\rm H,los}$ variability scenario and a pure flux variability with enough significance. \texttt{borus02}, on the other hand, clearly favors an $N_{\rm H,los}$-variable scenario\footnote{We note that, while \texttt{MYTorus} and \texttt{borus02} give practucally identical best-fit parameters, the errors of \texttt{MYTorus} are much larger. This results in the source being compatible with a non $N_{\rm H,los}$-variability scenario}. And finally, \texttt{UXCLUMPY} favors a scenario in which the spectral variability is predominantly caused by intrinsic flux changes, rather than absorption. We thus classify this source as `Undetermined'. 

%For some reason the errors of mytorus are much larger than borus.

\subsection{NGC 788}

\textbf{Data reduction/fitting:} 
Three Gaussian lines (\texttt{zgauss} in \texttt{xspec}) were added to model the source soft emission. The reduced $\chi^2$ showed significant improvement for all models, justifying this decision (1.24 to 1.13 for \myt, 1.27 to 1.13 for \bor and 1.29 to 1.17 for \uxc).

\textbf{Analysis of results:} The models and the data show a more significant tension than for the majority of sources in this sample, at around the $3\sigma$ level. For this source we present two \texttt{borus02} configurations that can explain the data with the same goodness of fit. The two configurations can be described as a low-$N_{\rm H,av}$ scenario and a high-$N_{\rm H,av}$ one. The former is statistically preferred by \texttt{MYTorus}, which cannot reproduce the latter without forcing $N_{\rm H,av}$ to stay at a very high value. \texttt{UXCLUMPY}, while not directly comparable (it does not provide a value for $N_{\rm H,av}$), results in values of $N_{\rm H,los}$ that are more similar to those of the high-$N_{\rm H,av}$ \texttt{borus02} option. Given how the first configuration is practically identical to the \texttt{MYTorus} results, we opt to show the second \texttt{borus02} configuration (the high-$N_{\rm H,av}$ scenario) in all plots regarding the source. The degeneracy between the reflection and line-of-sight component modeling results in different estimates for $N_{\rm H,los}$ for each model, although the upwards trend of $N_{\rm H,los}$ vs time is maintained (see Fig. \ref{fig:nhvstime1}).

The analysis of \citet{Zhao2020} favored the high-$N_{\rm H,av}$ scenario, and preferred pure flux variability over the pure $N_{\rm H,los}$ variability depicted here. However, as shown by our $\chi^2_{\rm red}$ comparisons, either option can explain the data at a similar level for all models. \texttt{UXCLUMPY} is the only model that, when considering the p-value determination, flags this source as variable. This is likely due to the smaller errors and slightly larger differences between $N_{\rm H,los}$ values at different epochs, compared to the \myt and \bor results. However, given how the $\chi^2_{\rm red}$ comparison doesn't show a significant preference for $N_{\rm H,los}$ variability over intrinsic flux variability, we classify this source as `Non-variable in $N_{\rm H,los}$'
% Difference with Xiurui's results: I find a better fit considering NH variation, where Xiurui found a better one when considering flux variability only. Xiurui also had $\Gamma=1.55$, $N_{H,av}=3.16e24$ (compatible with no upper limit) and both $theta_i$ and $C_f$ $\sim 0.20$.

\subsection{NGC 833}

\textbf{Data reduction/fitting:} NGC 833 is part of a closely interacting system with NGC 835 (separation $\sim 1\arcmin$).
The second Chandra observation (Obs. ID: 10394) considered for this merging system does not include NGC 833, but rather only NGC 835. We opted to add this observation to the table (with blank data) to avoid confusion with the epochs shown for NGC 835. Similarly, in the \xmm observation we use data from only the MOS modules, as NGC 833 falls on a prominent CCD line on the PN observation. The \nustar extraction region was limited to 40$\arcsec$ to avoid contamination from NGC 835. For the same reason, the background was extracted from a circular region (instead of the usual annulus) of radius 60$\arcsec$. Nearby source NGC 838, a starburst galaxy at $\sim 3.5\arcmin$ from NGC 835, shows no \nustar emission, and therefore is not contaminating the spectrum. The \chandra spectrum was also extracted from a circular region (15$\arcsec$) radius, to avoid contamination.

\textbf{Analysis of results:} This source is well-fit by all models. The torus parameters are highly unconstrained, likely due to a very subdominant reflection component \citep[see e.g.][]{Torres-Alba2021}. The $\chi^2_{\rm red}$ comparison shows, for all models, that $N_{\rm H,los}$ variability is unnecessary to explain the data. Likewise, the p-value of all $N_{\rm H,los}$ being the same is large enough that one cannot rule out the hypothesis. Thus, we classify this source as `Non-variable in $N_{\rm H,los}$'. 

\subsection{NGC 835}
\textbf{Data reduction/fitting:}
The \nustar extraction region was limited to 40$\arcsec$ to avoid contamination from NGC 833. For the same reason, the background was extracted from a circular region (instead of the usual annulus) of radius 60$\arcsec$. Nearby source NGC 838, a starburst galaxy at $\sim 3.5\arcmin$ from NGC 835, shows no \nustar emission, and therefore is not contaminating the spectrum. The \chandra data was taken using a larger-than-usual 8$\arcsec$ circular region to include all the soft emission (this source is a known Luminous Infrared Galaxy, or LIRG), for easier comparison to the \xmm data. Again, the background was extracted from a circular region (15$\arcsec$) radius, to avoid contamination.
To fit the soft emission in this source we tried both adding Gaussian lines, or adding a second \texttt{apec} component \citep[justified by this source being in a merging system, as well as a known LIRG, see][for details]{Torres-Alba2018}. Adding two lines improved the $\chi^2$ over adding a second \texttt{apec}, and the \texttt{apec} addition resulted in inverted temperatures (i.e. the `cooler' gas was more obscured that the `hotter' gas, which is physically implausible). We thus opted to use the Gaussian lines.
%    \item Tried adding two lines, chi2 went from 500 to 479. Second mekal only improved to 485, and temperatures got inverted.

\textbf{Analysis of results:} The data is well-fitted by all models, which are in reasonable agreement. However, the best-fit values for cos$(\theta_{\rm Obs})$ derived with \texttt{borus02} and \texttt{UXCLUMPY} are incompatible. The former favors an edge-on configuration, while the latter favors an almost face-on one. Our results are compatible with those of \citet{Zhao2020}, whose analysis also favors an edge-on scenario. All models agree that this source shows significant $N_{\rm H,los}$ variability. We classify this source as `variable in $N_{\rm H,los}$'.

\subsection{3C 105}
\textbf{Data reduction/fitting:} No issues to report.

\textbf{Analysis of results:} The data is well-fitted by all models, which are in good agreement. Our results are also consistent with those of \citet{Zhao2020}. Introducing $N_{\rm H,los}$ variability is not necessary to explain the data, and the p-value is also $>0.01$ for all models. We thus classify this source as `Non-variable in $N_{\rm H,los}$'. 

\subsection{4C+29.30}
\textbf{Data reduction/fitting:}
The \chandra data shows a complex morphology in the soft band, including a jet further out from the nucleus \citep[see e.g.][]{Siemiginowska2012}. The usual 5$\arcsec$-radius source region was used, but the background was extracted from a nearby 10$\arcsec$-radius circle, rather than an annulus, in order to avoid contamination. Furthermore, Chandra observation 1 has low quality, forcing us to use 5 cts/bin, and fit with C-statistic. The table shows therefore total Stat. instead of $\chi^2$. The \xmm emission was extracted as usual (avoiding the jet emission), but the larger region (needed to include the \xmm PSF) resulted in including a larger fraction of hot gas. An additional constant was used to weight the normalization of \texttt{apec}, but both \chandra and \xmm data were compatible with having the same exact $kT$. A second \xmm observation exists (Obs. ID: 0504120201) which was not used, at it fell on the same day as the used \xmm observation (Obs. ID: 0504120101) and was much shorter \citep[see e.g.][]{Sobolewska2012}. All emission at $>$2~keV originates in the nucleus, therefore the \nustar data is not affected by the jet presence.

Even though the cross-normalization constants are compatible with 1 within errors, forcing them all to stay equal to 1 resulted in meaningful shifts in $N_{\rm H,los}$. Therefore, we opted to leave the necessary ones free to vary in this case.

\textbf{Analysis of results:} The data is well-fitted by all models, which are in good agreement. We note that \chandra observations 2$-$5 took place within $\sim$1 week, which likely explains the lack of flux/$N_{\rm H,los}$ variability among those observations. While it is clear from the $\chi^2_{\rm red}$ comparison that the data requires some form of variability ($T>20\sigma$), neither intrinsic flux nor $N_{\rm H,los}$ variability is preferred over the other. The one exception to this is perhaps \texttt{borus02}, which shows a tension of $>3\sigma$ between model and data when no $N_{\rm H,los}$ variability is allowed. This is likely due to the high obscuration the model predicts for the \xmm observation. In any case, the tension is not significant enough, and we classify this source as `Non-variable in $N_{\rm H,los}$'.

\subsection{NGC 3281}

\textbf{Data reduction/fitting:}
The \chandra data was extracted using a circle of radius 10$\arcsec$ (background region, annulus 11$-$20$\arcsec$) to include all the extended emission (thus, making the comparison with the \xmm data easier). An additional \nustar observation exists that was not public at the moment this analysis took place. 

\textbf{Analysis of results:} The data is well-fitted by all models, although they are not in strong agreement: \myt favors a low-$N_{\rm H,av}$ scenario, while \bor favors a high-$N_{\rm H,av}$ one. Both models are able to find an equivalent scenario to the best fit of the other, although with worse statistics ($\chi^2_{\rm red}$=1.14 for a \myt configuration with high $N_{\rm H,av}$, and $\chi^2_{\rm red}$=1.09 for a \bor one with low $N_{\rm H,av}$). Our \bor best-fit is consistent with the results of \citet{Zhao2020}.

The models show significant disagreement in the best-fit values of $N_{\rm H,los}$, probably arising from different disentanglements of the degeneracy with $\Gamma$ and $N_{\rm H,av}$. \bor and \uxc show the best agreement, although the \nustar observation is significantly more obscured in the \uxc best fit. \myt, on the other hand, generally prefers higher obscuration. However, the \nustar observation is compatible with the \bor determination. Overall, this results in \uxc painting a much more variable picture of the source. In any case, all models agree that the source is indeed `$N_{\rm H,los}$ variable', and we thus classify it as such.

\subsection{NGC 4388}

\textbf{Data reduction/fitting:}
\chandra observations with Obs. ID 9276, 9277 and 2983 were not considered because they used HETG/LETG grating. This galaxy has a prominent extended emission, likely the result of star formation. We used a 12$\arcsec$-radius region (background annulus at radii 20$-$30$\arcsec$) to include it all in the analysis of \chandra data. The brightness and closeness of this galaxy results in great data quality, and therefore more substructure is appreciated in the soft emission. We used a two-\texttt{apec} model to describe it. %Again used two mekals due to high data quality. For some reason uxclumpy favours 3 lines intead (one of which broad enough to sort of replace the second mekal main peak)

Note that the third \xmm and the second \nustar observations took place simultaneously. Another \xmm observation (Obs. ID: 0110930301) was not included, as it was completely affected by flares.

\textbf{Analysis of results:} The best-fit of all models to the data shows significant tension ($T\sim$20$\sigma$). This may be a result of the large number of counts available for this source, compared to that of the rest of the sample. It may be that our model is too simple to adequately fit it. However, no obvious problem is seen in the fit residuals that may point toward any specific issues. This source is likely a good candidate to implement a more complex treatment of the reflection component, such as the scenarios mentioned in Sect. \ref{Discussion}.

Despite the poorer fit, the models show remarkable agreement, particularly in the $N_{\rm H,los}$ determinations. The largest discrepancy is in the photon index obtained by \uxc, which is largely incompatible with those of \bor and \myt. The values of $\theta_{\rm obs}$ obtained via \uxc and \bor are also incompatible, with \uxc favoring an edge-on scenario, while \bor suggests a much more inclined viewing angle. Our \bor results are mostly in agreement with those of \citet{Zhao2020}, although they obtain much higher $N_{\rm H,av}$, on the order of $10^{24}$ cm$^{-2}$.

Even if the fit to the data might be improved by using more complex models, it is clear that allowing both intrinsic flux and $N_{\rm H,los}$ variability significantly improves the fit. Taking this into account, as well as the derived p-values, we classify this source as `$N_{\rm H,los}$ variable'.

\subsection{IC 4518 A}

\textbf{Data reduction/fitting:}
For the second \xmm observation, MOS2 was not used as it was corrupted.

\textbf{Analysis of results:} The data is well-fitted by \myt and \bor, with \uxc showing poorer statistics. This may be a result of the strong reflection seemingly needed to fit the data. In fact, this is the only source in our sample that requires the addition of an inner, CT reflection ring in \uxc. This component was introduced into the \uxc model precisely because of difficulty fitting sources with strong reflection with only a cloud distribution \citep[see][]{Buchner2019}. \myt  and \bor also yield large values of $N_{\rm H,av}$, which agrees with this interpretation. This scenario is remarkably similar to that described in \citet{Pizzetti2022}.

The results obtained from our best fit are consistent with those of \citet{Zhao2020}, although we obtain higher values for $N_{\rm H,av}$ ($\sim2 \times 10^{24}$ cm$^{-2}$ in the mentioned work).

While both the $\chi^2_{\rm red}$ comparison for all models and the p-value obtained for \uxc suggest the need for $N_{\rm H,los}$ variability, the p-values for \myt and \bor remain above the threshold. Therefore, we classify this source as `Undetermined'.

\subsection{3C 445}
\textbf{Data reduction/fitting:}
We used an extraction region of 7$\arcsec$ for \chandra, as some extended emission is present. The background was taken from an annulus, of radii 10$-$20$\arcsec$. 
%Some extended emission present (the source is elongated). Jet?
The source spectra shows a prominent excess at around 2 keV that is best-fit with a second, very hot \texttt{apec} component. It is not obvious whether star-formation, or perhaps the presence of a jet, could result in such very hot gas.  \citet{Torres-Alba2018} used the two-\texttt{apec} model to explain the soft emisson of a large sample of U/LIRGs, and obtained a T$_{2}$ distribution of median 0.97$\pm$0.18 keV, with a long tail extending up to 4.5 keV. However, this galaxy is not classified as a U/LIRG, nor does it show obvious morphological signs of a merger (that could explain the dense star formation required). The detection of radio emission points toward the presence of a jet, as does the slightly elongated \chandra morphology. However, it is not obvious if the jet presence could justify the addition of the second \texttt{apec} component, from a physical point of view. We still opt to use it in the model, given how it is required to explain the data.

\textbf{Analysis of results:}  The data is well-fitted by all models, although \myt requires unusually large reflection constants ($A_{\rm s90}$ and $A_{\rm s0}$). It also results in a larger $\Gamma$ than the other models. Furthermore, \myt and \bor are barely in agreement in their $N_{\rm H,los}$ determinations, while \uxc results in systematically lower values (incompatible with the other models in 3/5 observations). The most remarkable difference is in the \nustar observation, in which \uxc models the observed flux with lower obscuration than the other models, and compensates this with a lower intrinsic flux value. Precisely because of this, \uxc is the only model that classifies the source as `$N_{\rm H,los}$ variable', according to the p-value. However, the $\chi^2_{\rm red}$ comparison shows that, even for \uxc, an alternative fit exists when imposing no $N_{\rm H,los}$ variability, with $T<3\sigma$. Therefore, we opt to classify this source as `Non-variable in $N_{\rm H,los}$'.

Our \bor results are in good agreement with those of \citet{Zhao2020}, with the exception of the $N_{\rm H,av}$, for which they obtain a much higher value of $10^{24}$ cm$^{-2}$.

%MYTorus has significant differences. I believe the flexibility given by the two reflection component normalizations (which are both abnoramlly high) results in a switch in the soft band emissions, to accomodate for the unusual excess. Indeed, reflection has significant contribution down to 1-2 keV, which is unusual. MYTorus was the only model capable of reproducing the emission without the additional mekal, but with even more excessive constants for A0,90.

\subsection{NGC 7319}
\textbf{Data reduction/fitting:}
We used an annulus of radii 10$-$20$\arcsec$ to extract the \chandra background, in order to avoid a nearby source. Similarly, we used a circular source extraction region of only 15$\arcsec$ for \xmm, to avoid both extreme soft excesses and CCD lines present around the source. No source was detected in \xmm Obs. ID 0021140401, and therefore it is not used in this analysis.

A double-\texttt{apec} model was used to fit the soft emission of this galaxy, since it is part of a closely-interacting system, which is known to increase star forming activity.

\textbf{Analysis of results:} The data is well-fitted by all models. However, \uxc yields significantly different values for $N_{\rm H,los}$ for the \nustar observations. Similarly to the case of 3C 445, it models the \nustar observed flux by using both lower $N_{\rm H,los}$ and lower intrinsic flux values. This scenario is more similar to the best-fit \citet{Zhao2020} found for the source using \bor. They detected no significant $N_{\rm H,los}$ variability between \chandra and \nustar, while needing a much lower intrinsic flux for the \nustar observation. We recovered this solution with \myt and \bor, with worse statistics. Interestingly, the \citet{Zhao2020}/\uxc solution is statistically the best when not accounting for the soft X-ray emission (for \bor and \myt). However, this solution always has $N_{\rm H,av}$ at the maximum value allowed by the models.

Despite the mentioned differences, all models agree that $N_{\rm H,los}$ variability is required to explain the data, although this effect is larger for \myt and \bor. We thus classify this source as `$N_{\rm H,los}$ variable'.

%Xiurui finds very different best-fit scenario: no $N_H$ variability wrt Chandra, but $C_{nus}=0.3$. MyTorus has an unstable solution around this local minimum, but with red $\chi^2=1.15$ instead of 1.08. Also, $N_{\rm av}$ is capped at max value for Xiurui. {\bf Interesting Note:} This solution appears as best after taking care of a better fit for the soft-band emission (i.e. Added lines or second mekal to deal with visible soft excess). Solution with 2 zgauss has $red \chi^2=1.11$, and visible excesses in the soft band. So opted to go for the second mekal.
%Uxclumpy falls a lot closer to Xiurui's results, and favours flux variability as the main reason to explain the change in spectral shape (although with some column density variability, too)

\subsection{3C 452}
\textbf{Data reduction/fitting:}
We used an annulus of radii 12$-$20$\arcsec$ to extract the \chandra background, in order to avoid a nearby source. 3C 452 also shows diffuse, soft, (very) extended emission, which is coming from a jet \citep[see][]{Isobe2002}. Given how the extraction region used by \chandra is smaller than that of \nustar and \xmm, when including the jet emission it is necessary to use different jet normalization (i.e. the variation of the parameter norm$_{jet}$ does not imply that the jet is varying in flux).
This source did not require any cross-normalization for the AGN emission, with the exception of that associated to the jet.

\textbf{Analysis of results:} The data is well-fitted by all models, even if they are not in perfect agreement. \bor yields a significantly smaller $\Gamma$ value, and the determinations of $\theta_{\rm Obs}$ by \uxc and \bor are incompatible within errors. \bor favors an edge-on scenario, while \uxc favors a face-on one, although with very large errors. Additionally, \uxc results in a much harder spectrum for the jet emission, when compared to \myt and \bor. Our results for \bor are compatible with those obtained by \citet{Zhao2020}, with the exception of $\theta_{\rm Obs}$, which in their case results in a face-on scenario. Additionally, \citet{Zhao2020} introduce some AGN flux variability, which in our case is modeled via changes in the normalization of jet flux.

All models agree that $N_{\rm H,los}$ variability is required to explain the data, but \uxc yields smaller values for all observations. We thus classify this source as `$N_{\rm H,los}$ variable'.
    
\end{document}